\documentclass[epj]{svjour}
\usepackage{amsfonts,amssymb,amsmath,amscd}
\usepackage{dsfont}
\usepackage{framed}
\def\bigone{ {\mathds{1}}}
\usepackage{mathrsfs}
\usepackage{wasysym}
\usepackage{stmaryrd}
\journalname{epjp}
\begin{document}


%
%

\title{On magnetic monopoles, the anomalous $g$-factor of the electron and the spin-orbit coupling in the Dirac theory}

\author{Gerrit Coddens 
}                     
\institute{Laboratoire des Solides Irradi\'es,\\ Ecole Polytechnique, CNRS, CEA,\\
Universit\'e Paris-Saclay,\\ 91128-Palaiseau CEDEX, FRANCE\\
}
%


\date{}

\authorrunning{Gerrit Coddens}
\titlerunning{Magnetic monopoles, anomalous $g$-factor and spin-orbit coupling in Dirac theory}

\abstract{
We discuss the algebra and the interpretation of the anomalous Zeeman effect and the spin-orbit coupling 
within the Dirac theory.
Whereas the algebra for the anomalous Zeeman effect is impeccable and therefore in excellent agreement  with experiment, 
the {\em physical  interpretation of that algebra} uses images that are based on macroscopic  intuition but do not correspond to the meaning of this algebra.
The interpretation violates the Lorentz symmetry.
We therefore reconsider the interpretation to see if we can render it consistent also with the symmetry. 
The results confirm 
clearly that the traditional physical interpretation of the anomalous Zeeman effect 
is not correct.
We give an alternative intuitive description of the meaning of  this effect, which respects the symmetry and is exact. 
It can be summarized by stating that a magnetic field makes any charged particle spin. This is even true for charged particles ``without spin''.
Particles ``with spin'' acquire additional spin in a magnetic field. This additional spin must  be combined algebraically with  the pre-existing spin.
We show also that the traditional discussion about magnetic monopoles confuses two issues, {\em viz.} the symmetry of the Maxwell equations
and the quantization of charge. These two issues define each a different concept of magnetic monopole. They  cannot be merged together
into a unique all-encompassing issue. 
We also generalize the minimal substitution for a charged particle, and provide some intuition for the 
magnetic vector potential. We finally explore the algebra of the spin-orbit coupling, which turns out to be badly wrong. The traditional theory
that is claimed to reproduce the Thomas half is based on a number of errors. An error-free  application of the Dirac theory cannot account for the Thomas precession,
because it only accounts for the instantaneous local boosts, not for the rotational component of the Lorentz transformation.
This runs contrary to established beliefs, but Thomas precession can be understood in terms of the Berry phase on a path through the velocity space of the  Lorentz group manifold. 
These results clearly reveal the limitations of the prevailing
working philosophy
to ``shut up and calculate''.
} 

\maketitle


\section{Introduction} \label{introduction}

When Uhlenbeck and Goudsmit  presented the concept of spin, Lorentz pointed out that the
idea  could not account for the magnetic dipole moment
of the electron. Even if one were to put all
the charge of a spherical  electron on its equator, the current produced by the spinning motion would not be enough to match
the magnitude of the anomalous Zeeman splitting  observed. 
The algebra of the Dirac theory accounts very well for the measured values
of this anomalous Zeeman splitting,
but in the present paper we point
out that the traditional physical  interpretation of the mathematical formalism in terms of a magnetic dipole moment
associated with the spin is at variance with the meaning of the algebra itself as it violates the built-in Lorentz symmetry. 
It interprets a vector as a scalar, which is a transgression that is similar to interpreting
a tensor as a vector. The importance of such distinctions based on symmetry is well known. In relativistic quantum mechanics
one discusses e.g. that the only bilinear Lorentz covariants that exist are  one-component scalars, four-component vectors, six-component
tensors, four-component axial vectors and one-component pseudo-scalars. Confusing a covariant of one type
with a covariant of an other type is violating its symmetry. Using  group theory we will propose an approach that
respects the symmetry.  

Section \ref{intro} contains an introduction to some aspects of
the representation SU(2) for the rotation group  and of the Dirac representation for the homogeneous Lorentz group,
which will be used in the paper. This is a subject matter that is considered to be ``well known''.
But we cover it from a very different, geometrical perspective than in its traditional treatment given 
 in textbooks, which is far more  abstract and algebraic. The insight gained from this different perspective will permit us to discern the
 error in the interpretation of the anomalous $g$-factor we mentioned above, and which we
 discuss in Subsection \ref{anomaly-g}.  Apparently  this problem has escaped  attention for more than 80 years,
suggesting that the group theory is not as well understood as routinely assumed.

\section{Group Representation Theory} \label{intro}

\subsection{SU(2) and the rotations of ${\mathbb{R}}^{3}$} \label{SU2-group}

\subsubsection{Principal idea} \label{princ-idea}

The general form of a SU(2) rotation matrix is:

\begin{equation}\label{SU(2)}
{\mathbf{R}} = \left (
\begin{tabular}{cr}
$\alpha_{0}$ & $-\alpha_{1}^{*}$\\
$\alpha_{1}$ & $\alpha_{0}^{*}$\\
\end{tabular}
\right ), \quad {\text{ with $\det {\mathbf{R}} =\alpha_{0}\alpha_{0}^{*} + \alpha_{1}\alpha_{1}^{*} =1$}}.
\end{equation}

\noindent  These matrices work on spinors:

\begin{equation}\label{SU(2) spinor}
{\boldsymbol{\xi}} = \left (
\begin{tabular}{c}
$\xi_{0}$\\
$\xi_{1}$\\
\end{tabular}
\right ), \quad {\text{ with $\xi_{0}\xi_{0}^{*} + \xi_{1}\xi_{1}^{*} =1$}},
\end{equation}

\noindent according to:

\begin{equation}\label{SU(2) formalism}
{\boldsymbol{\xi}}' = {\mathbf{R}} {\boldsymbol{\xi}}.
\end{equation}

\noindent The natural question is of course what the meaning of such a spinor is.
In SO(3), the $3\times 3$ rotation matrices of ${\mathbb{R}}^{3}$ are working on $3\times 1$ matrices that represent
vectors of  ${\mathbb{R}}^{3}$, but in SU(2) the situation is different. Here the $2\times 1$ matrices, i.e. the spinors,
represent rotations. The idea can be illustrated by considering the
group multiplication table for an arbitrary group $(G, \circ)$:

\begin{equation} \label{diagram-shift}
\begin{tabular}{|c|cccccc|l}
\hline
${\circ}$         & $ g_{1}$                       & $g_{2}$                         & $g_{3}$                        & $\cdots$ & $g_{j}$                         & $\cdots$ \\
\hline
$\,g_{1}\,$  & $\,g_{1}\,{\circ}\,g_{1}$\, & $\,g_{1}\,{\circ}\,g_{2}$\, & $\,g_{1}\,{\circ}\,g_{3}$\, & $\,\cdots$\, & $\,g_{1}\,{\circ}\,g_{j}$\, & $\,\cdots$\,\\  
$\,g_{2}\,$  & $g_{2}\,{\circ}\,g_{1}$ & $g_{2}\,{\circ}\,g_{2}$ & $g_{2}\,{\circ}\,g_{3}$ & $\cdots$ & $g_{2}\,{\circ}\,g_{j}$ & $\cdots$\\  
$\,\vdots\,$ & $\vdots$               & $\vdots$               & $\vdots$               &          & $\vdots$               &  \\
\hline
\hline
$\,g_{k}\,$  & $g_{k}\,{\circ}\,g_{1}$ & $g_{k}\,{\circ}\,g_{2}$ & $g_{k}\,{\circ}\,g_{3}$ & $\cdots$ & $g_{k}\,{\circ}\,g_{j}$ & $\cdots$  \\  
\hline
\hline
$\,\vdots\,$ & $\vdots$               & $\vdots$               & $\vdots$               &          & $\vdots$               & \\
\hline
\end{tabular}
\begin{tabular}{ll}
$ $&\\
$ $&\\
$ $&\\
$\leftarrow$ $g_{k}\circ,$ 
\end{tabular}
\end{equation}

\noindent which illustrates that the group element $g_{k}$ defines a function $g_{k}\circ: G \rightarrow G; \, g_{j} \rightarrow g_{k}\,{\circ}\,g_{j}$. 
The notation  $g_{k}\circ$ for this function is somewhat 
arcane, but it has the advantage that the way it acts on a group element is obtained by mere  juxtaposition of the symbols.
In the specific case of the rotation group,
we define then a rotation $g_{k}$ no longer by all its function values $g_{k}({\mathbf{r}}), \forall {\mathbf{r}} \in {\mathbb{R}}^{3}$,
but by all function values  $g_{k}\circ g_{j}, \forall g_{j} \in G$.
More rigorously, an arbitrary  group element $g_{k} \in G$ is identified  with the function 
$T_{g_{k}} \in F(G,G)$ that maps $G$ to $G$ according to $T_{g_{k}}: g_{j} \in G \rightarrow g_{j}' = T_{g_{k}}(g_{j}) = g_{k} \,{\circ} \,g_{j}$,
where $T_{g_{k}}$ is just a more standard notation for the function $g_{k}\circ$ (We use here $F(S_{1},S_{2})$ as a general notation for the set of functions from
 $S_{1}$ to $S_{2}$).  
This identification implies that $T_{g_{k}} \in F(G,G)$ represents $g_{k}\in G$. Let us call this representation  $T_{g_{k}}$ of $g_{k}$ the automorphism representation.
The non-standard notation  $g_{k}\circ$ permits 
 writing $g_{k}\circ : g_{j} \in G \rightarrow g'_{j}  = g_{k} \,{\circ} \,g_{j}$ and   grasping more easily the idea of 
 interpreting a rotation as a function that works on 
other rotations rather than on vectors. 
 If we represent $ g_{j}$ by the SU(2) matrix ${\mathbf{X}}$,  $ g_{k}$ by the SU(2) matrix ${\mathbf{R}}$,
and $ g'_{j}$ by the SU(2) matrix ${\mathbf{X}}'$ then
we have:

\begin{equation}  \label{SU(2) diagram 2}
\begin{tabular} {ccccccccccc}
$  g'_{j} $           &        $= $               & $g_{k}$             &     $\circ$        & $ g_{j} $                                    &                                                                                  &  $ g_{j}'  $                      &  $= $    &  $T_{g_{k}} $       & $ (\,g_{j}\,)$ \\
$\left\downarrow\rule{0cm}{0.5cm}\right. $                            &                                                       &$\left\downarrow\rule{0cm}{0.5cm}\right. $              &                     & $\left\downarrow\rule{0cm}{0.5cm}\right. $                                     & \quad\quad\quad\quad {\text{or}} \quad\quad\quad\quad&  $\left\downarrow\rule{0cm}{0.5cm}\right. $                          &                                  & $\left\downarrow\rule{0cm}{0.5cm}\right. $                   &$\left\downarrow\rule{0cm}{0.5cm}\right. $     \\
${\mathbf{X}}' $     &      $= $                   & ${\mathbf{R}}$ &                     &  ${\mathbf{X}}$             &                                                                                  & ${\mathbf{X}}' $  &  $=$    & ${\mathbf{R}}$   & $ {\mathbf{X}}$\\
\end{tabular}
\end{equation}

\noindent We see thus that we can represent $T_{g_{k}}$ also
by  ${\mathbf{R}}$:
 In other words, the SU(2) matrices represent from this viewpoint two types of mathematical objects,
{\em viz.} group elements $g_{j} \in G$ and group automorphisms $T_{g_{k}} \in F(G,G)$. This translates the fact that the 
automorphism group $(F(G,G), ^{\circ})$  (where $^{\circ}$ is the composition of functions)
of a group  $(G, \circ)$, is isomorphic to  $(G, \circ)$ under the mapping $T\in F(G,F(G,G)): g_{k} \rightarrow T_{g_{k}}$.
In SU(2) it is possible to remove this ambiguity between the representations of
group elements and automorphisms by rewriting the second diagram in Eq. \ref{SU(2) diagram 2}  as:

\begin{equation}  \label{SU(2) diagram 3}
\begin{tabular}{cccc}
 $g_{k}\circ g_{j}$ & $ =$ & $T_{g_{k}}$ &  $(g_{j})$\\
$\left\downarrow\rule{0cm}{0.5cm}\right. $   & & $\left\downarrow\rule{0cm}{0.5cm}\right. $    & $\left\downarrow\rule{0cm}{0.5cm}\right. $     \\
${\boldsymbol{\xi}}'$  & $=$ & ${\mathbf{R}}$ & $ {\boldsymbol{\xi}}$\\
\end{tabular}
\quad {\text{by using the substitution:}} \quad
\begin{tabular}{cccc}
${\mathbf{X}}'$ & $=$ & $ {\mathbf{R}}$ & ${\mathbf{X}}$\\
$\left\downarrow\rule{0cm}{0.4cm}\right. $ & &$\left\downarrow\rule{0cm}{0.4cm}\right. $  & $\left\downarrow\rule{0cm}{0.4cm}\right. $   \\
${\boldsymbol{\xi}}'$  & $ = $ & ${\mathbf{R}}$ & ${\boldsymbol{\xi}}$\\
\end{tabular}\,\,,
\end{equation}

\noindent where we recover Eq. \ref{SU(2) formalism} by defining the  spinor ${\boldsymbol{\xi}}$ as 
just a short-hand for the SU(2) rotation matrix:

\begin{equation}\label{SU(2) matrix Xi}
{\mathbf{X}} = \left (
\begin{tabular}{cr}
$\xi_{0}$ & $-\xi_{1}^{*}$\\
$\xi_{1}$ & $\xi_{0}^{*}$\\
\end{tabular}
\right ), \quad  {\text{ with $\xi_{0}\xi_{0}^{*} + \xi_{1}\xi_{1}^{*} =1$}},
\end{equation}

\noindent by taking its first column.
In fact, all the information of the matrix ${\mathbf{X}} $  is already given by its first column. When we know the first
column we know everything we need to know to write the second column. 
Moreover the first column of
${\mathbf{R}} {\mathbf{X}}$ will be ${\mathbf{R}} {\boldsymbol{\xi}}$. If we note the second column
of ${\mathbf{X}} $ as ${\boldsymbol{\eta}}$, then the second column of ${\mathbf{R}} {\mathbf{X}}$
will be ${\mathbf{R}} {\boldsymbol{\eta}}$ and it will be possible to derive
${\mathbf{R}} {\boldsymbol{\eta}}$ from ${\mathbf{R}} {\boldsymbol{\xi}}$ in the same way
as we could derive $ {\boldsymbol{\eta}}$ from $ {\boldsymbol{\xi}}$, 
viz. $({\mathbf{R}}{\boldsymbol{\eta}})_{0} = -(( {\mathbf{R}}{\boldsymbol{\xi}})_{1})^{*}$ and
 $({\mathbf{R}}{\boldsymbol{\eta}})_{1} = (( {\mathbf{R}}{\boldsymbol{\xi}})_{0})^{*}$,
as the SU(2) matrices constitute a group. 
A spinor ${\boldsymbol{\xi}}$ in SU(2) can thus be considered as a set of parameters  that define a rotation, i.e. a set of coordinates for a rotation. 
This rises the question how the information about the rotations
occurs inside the parameter set ${\boldsymbol{\xi}}$. 

\subsubsection{Constructing the representation} \label{constr-rep}

The answer is that it is done by using the fact that any rotation can be obtained  as
a product of two spatial reflections in ${\mathbb{R}}^{3}$. The reflections with respect to the planes through the origin of ${\mathbb{R}}^{3}$ generate a group
 of rotations and reversals, of which the rotation group is a subgroup.
  It is easy to figure out how we write a $2\times 2$ reflection matrix, and once we know the matrices
 for the reflections
 we can calculate the matrices for the  rotations and reversals by making products.
A reflection $A$ is defined by a unit vector ${\mathbf{a}}$
that is normal to its reflection plane. The coordinates of ${\mathbf{a}}$ can be expected to occur as parameters in the $2 \times 2$ 
matrix ${\mathbf{A}}$ that defines the reflection $A$  but we do not know where. We therefore write the reflection matrix ${\mathbf{A}}$ heuristically as
${\mathbf{A}} = a_{x} \sigma_{x} + a_{y} \sigma_{y} + a_{z} \sigma_{z}$. The matrix $\sigma_{x}$ will tell us where and with which coefficients
$a_{x}$ appears in ${\mathbf{A}}$. The same is true, {\em mutatis mutandis} for $\sigma_{y}$ and
$\sigma_{z}$. To find the matrices $\sigma_{x}$, $\sigma_{y}$,  $\sigma_{z}$, we express that ${\mathbf{A}}^{2} = \bigone$.
We find  that we can meet this requirement when the  matrices $\sigma_{j}$ satisfy the conditions $\sigma_{j}\sigma_{k} + \sigma_{k}\sigma_{j} = 2 \delta_{jk} \bigone$.
In other words, identifying them with the Pauli matrices will give us the representation searched for. By expressing a rotation as the product
of two reflections, one can then derive the well-known Rodrigues formula:

\begin{equation} \label{Rodrigues}
 {\mathbf{R}}({\mathbf{n}},\varphi) = \cos(\varphi/2)\,\bigone - \imath
\sin(\varphi/2)\,[\,{\mathbf{n}}{\boldsymbol{\cdot\sigma}}\,],
\end{equation}

\noindent for a rotation by an angle $\varphi$ around an axis defined by the unit vector ${\mathbf{n}}$.
To derive this result it suffices to consider two reflections
 $A$  (with matrix $[{\mathbf{a}}{\boldsymbol{\cdot\sigma}}]$) and  $B$ (with matrix $[{\mathbf{b}}{\boldsymbol{\cdot\sigma}}]$) 
 whose planes contain ${\mathbf{n}}$, and which have an angle $\varphi/2$ between them, and 
 to use
the algebraic identity $[{\mathbf{b}}{\boldsymbol{\cdot\sigma}}]\,[{\mathbf{a}}{\boldsymbol{\cdot\sigma}}] = ({\mathbf{b\cdot a}})\,\bigone + \imath
({\mathbf{b}}\wedge {\mathbf{a}}){\boldsymbol{\cdot\sigma}}$.
There is an infinite set of such pairs of planes, and which precise pair one chooses from this set does not matter. 

\subsubsection{A parallel formalism for vectors} \label{rot-vect}

By construction, this representation contains for the moment only group elements. Of course, it would be convenient if we were
also able to calculate the action of the group elements on  vectors. This can be done by developing a {\em parallel formalism for the matrices ${\mathbf{A}}$,
wherein ${\mathbf{A}}$ takes a different meaning and obeys a different kind of algebra.}
As the matrix ${\mathbf{A}}$ contains the components of the vector ${\mathbf{a}}$
we can conceive the idea of taking the matrix ${\mathbf{A}}$ also as the representation
of the unit vector ${\mathbf{a}}$. This idea can  be generalized to a vector ${\mathbf{v}}$ of arbitrary length, which
is then represented by ${\mathbf{V}} = v_{x}\sigma_{x} + v_{y}\sigma_{y}  + v_{z}\sigma_{z} $. We have then
${\mathbf{V}}^{2} = - (\det {\mathbf{V}}) \bigone = v^{2}\bigone$.
This idea that within SU(2)
a vector ${\mathbf{v}} \in {\mathbb{R}}^{3}$ is represented  by a matrix $ {\mathbf{v}}{\boldsymbol{\cdot\sigma}}$ according to the isomorphism:

\begin{equation}\label{Cartan}
{\mathbf{v}} = v_{x}{\mathbf{e}}_{x} + v_{y}{\mathbf{e}}_{y} + v_{z}{\mathbf{e}}_{z}   \longleftrightarrow 
 v_{x} \sigma_{x} + v_{y} \sigma_{y} + v_{z} \sigma_{z} =
\left (
\begin{tabular}{cc}
$v_{z}$ & $v_{x} - \imath v_{y}$\\
 $v_{x} + \imath v_{y}$ & $-v_{z}$ 
 \end{tabular} 
 \right ) {\hat{=}} \, {\mathbf{v}}{\boldsymbol{\cdot\sigma}}.
\end{equation}

\noindent was introduced by Cartan \cite{Cartan}. From ${\mathbf{V}}^{2}  = v^{2}\bigone$ it follows that:
 ${\mathbf{V}}_{1}{\mathbf{V}}_{2} + {\mathbf{V}}_{2}{\mathbf{V}}_{1} = 2\,({\mathbf{v}}_{1} {\mathbf{\cdot v}}_{2}) \,\bigone$.
To find out how the group acts on these representations of vectors, it suffices to observe that the reflection $A$,
defined by the unit vector ${\mathbf{a}}$, transforms ${\mathbf{v}}$ into $A({\mathbf{v}}) = {\mathbf{v}} - 2( {\mathbf{v \cdot a}})\,{\mathbf{a}}$.
Expressed in the matrices this yields: ${\mathbf{V}} \rightarrow - {\mathbf{AVA}}$. We see that this transformation law for vectors ${\mathbf{v}}$ is quadratic in ${\mathbf{A}}$
in contrast with the transformation law for group elements $g$, which is linear: ${\mathbf{G}} \rightarrow {\mathbf{AG}}$.
{\em Vectors transform thus quadratically as rank-2 tensor products of spinors, whereas spinors transform linearly.}

Both in the representation matrices ${\mathbf{A}} = {\mathbf{a}}{\boldsymbol{\cdot\sigma}}$  for reflections $A$
and ${\mathbf{V}} = {\mathbf{v}}{\boldsymbol{\cdot\sigma}}$ for vectors ${\mathbf{v}}$, $\sigma_{x}$, $\sigma_{y}$ and $\sigma_{z}$ are thus the Pauli matrices, and the symbol ${\hat{=}}$
serves to flag the introduction of a (rather confusing) stenographic notation 
${\boldsymbol{\sigma}} = (\sigma_{x},\sigma_{y}, \sigma_{z})$. 
The Pauli matrices are thus the images of the basis vectors ${\mathbf{e}}_{x}$, ${\mathbf{e}}_{y}$, ${\mathbf{e}}_{z}$
in the isomorphism  (${\mathbf{e}}_{j} \leftrightarrow \sigma_{j}$)  defined by Eq. \ref{Cartan}. 
The drawback of the convenient convention to use the shorthand
${\boldsymbol{\sigma}}$  for $(\sigma_{x},\sigma_{y}, \sigma_{z})$  is that  it may create the misleading impression
that ${\mathbf{v}}{\boldsymbol{\cdot\sigma}}$ represents a scalar, which it does not. It just represents  the counterpart in the isomorphism of  what would be a pedantic notation 
$(v_{x},v_{y},v_{z}){\mathbf{\cdot}}({\mathbf{e}}_{x}, {\mathbf{e}}_{y}, {\mathbf{e}}_{z})$ for
 $v_{x}{\mathbf{e}}_{x} + v_{y}{\mathbf{e}}_{y} + v_{z}{\mathbf{e}}_{z} = {\mathbf{v}}$. 
 
The reader will notice that the definition ${\mathbf{V}} = {\mathbf{v}}{\boldsymbol{\cdot\sigma}}$ with ${\mathbf{V}}^{2} = v^{2} \bigone$
is analogous to Dirac's way of introducing the gamma matrices to write   the energy-momentum
four-vector as $E \gamma_{t} + c{\mathbf{p}}{\boldsymbol{\cdot\gamma}}$ 
and postulating $(E \gamma_{t} + c{\mathbf{p}}{\boldsymbol{\cdot\gamma}})^{2} = (E^{2} - c^{2}p^{2})\bigone$.
In other words, it is the metric that defines the whole formalism, because we are considering  groups of metric-conserving
transformations (as the definition of a geometry in the philosophy of Felix Klein's Erlangen program).
For more information about the calculus on the rotation and reversal matrices, we refer the reader to reference \cite{Coddens}.
Let us just mention that as a reflection $A$ works on a vector ${\mathbf{v}}$ according to ${\mathbf{V}} \rightarrow - {\mathbf{AVA}} = - {\mathbf{AVA}}^{-1}$,
a rotation $R = BA$ will work on it according to ${\mathbf{V}} \rightarrow  {\mathbf{BAVAB}} = {\mathbf{RVR}}^{-1} = 
{\mathbf{RVR}}^{\dagger}$. The identity ${\mathbf{R}}^{-1} = {\mathbf{R}}^{\dagger}$ explains why we end up with SU(2).

In summary, there are two parallel formalisms in SU(2), one for the vectors and one for the group elements.
In both formalisms a matrix ${\mathbf{V}} ={\mathbf{v}}{\boldsymbol{\cdot\sigma}}$ can occur but with different meanings.
In a formalism for group elements, ${\mathbf{v}}$ fulfills the r\^ole of the unit vector ${\mathbf{a}}$ that defines the reflection $A$, such that we must have $|{\mathbf{v}}| =1$, and then the reflection matrix ${\mathbf{V}} = {\mathbf{A}}$ transforms according to: ${\mathbf{A}} \rightarrow {\mathbf{G}}{\mathbf{A}}$ under a group element $g$ with matrix representation ${\mathbf{G}}$. The new group element represented by ${\mathbf{GA}}$ will then no longer be a reflection that can be associated with a unit vector like it was the case for ${\mathbf{A}}$.
 In a formalism of vectors,  $|{\mathbf{v}}| $ can be different from $1$
and the matrix ${\mathbf{V}}$ (that represents now a vector) transforms according to:   ${\mathbf{V}} \rightarrow {\mathbf{G}}{\mathbf{V}}{\mathbf{G}}^{-1} = {\mathbf{G}}{\mathbf{V}}{\mathbf{G}}^{\dagger}$.
Here  $ {\mathbf{G}}{\mathbf{V}}{\mathbf{G}}^{\dagger}$ can be associated again with a vector.

\subsubsection{Other approaches} \label{oth-app}

The approach outlined above is non-standard. 
The standard treatment follows in general a linearization procedure.
One starts  the development by establishing  the quadratic formalism
${\mathbf{V}} \rightarrow - {\mathbf{A}}{\mathbf{V}}{\mathbf{A}}$ and
${\mathbf{V}} \rightarrow {\mathbf{R}}{\mathbf{V}}{\mathbf{R}}^{\dagger}$
for vectors.
The way back to a linear formalism  ${\mathbf{X}} \rightarrow {\mathbf{R}}{\mathbf{X}}$ (for group elements) or
${\boldsymbol{\xi}} \rightarrow {\mathbf{R}} {\boldsymbol{\xi}}$ (for their spinors) is then tricky and
shrouded in mystery. It amounts so to say to defining
a spinor as a kind of a square root of an isotropic vector. 

This runs for instance as follows.
One first considers a triad of normalized, mutually orthogonal basis vectors $({\mathbf{e}}_{x}, {\mathbf{e}}_{y}, {\mathbf{e}}_{z})$. One then observes that 
$({\mathbf{e}}'_{x}, {\mathbf{e}}'_{y}, {\mathbf{e}}'_{z}) =$
$(R({\mathbf{e}}_{x}), R({\mathbf{e}}_{y}), R({\mathbf{e}}_{z}))$ defines $R$ unambiguously. There is a one-to-one correspondence
between the rotated triads 
$({\mathbf{e}}'_{x}, {\mathbf{e}}'_{y}, {\mathbf{e}}'_{z}) $
and the rotations $R$ that produced them by acting on the chosen reference triad $({\mathbf{e}}_{x}, {\mathbf{e}}_{y}, {\mathbf{e}}_{z})$.
In a second stage, one considers that there is also a one-to-one correspondence between isotropic vectors
${\mathbf{e}}'_{x} + \imath {\mathbf{e}}'_{y}$ and triads $({\mathbf{e}}'_{x}, {\mathbf{e}}'_{y}, {\mathbf{e}}'_{z})$. By separating
the real and imaginary parts in ${\mathbf{e}}'_{x} + \imath {\mathbf{e}}'_{y}$ one can reconstruct
${\mathbf{e}}'_{x}$ and  $ {\mathbf{e}}'_{y}$, while ${\mathbf{e}}'_{z} = {\mathbf{e}}'_{x} \wedge {\mathbf{e}}'_{y}$.

The isotropic vector is thus a parameter set that defines a rotation in a one-to-one fashion. 
It is thus a set of complex coordinates for a rotation. The coordinates of the isotropic
vector $(x',y',z') = {\mathbf{e}}'_{x} + \imath {\mathbf{e}}'_{y}$ are thus not position coordinates but rotation coordinates.
They do not define a position in ${\mathbb{R}}^{3}$ because they were not introduced to do so. The only real point  $(0,0,0) \in {\mathbb{R}}^{3}$ that belongs to the isotropic cone and could define
a position  does not  define a triad.
The complex coordinates define nevertheless  an object in real Euclidean space, viz. the triad.
Therefore  spinors, which (as we will see) represent the information about these isotropic vectors and the corresponding triads, do turn in Euclidean space,
 despite the widespread opinion
that they should be considered as defined in some abstract internal space like in the example of isospin 
\cite{Biedenharn}\footnote{This viewpoint was criticized by Cartan who stated \cite{Cartan}:
   {\em ``Certain physicists regard spinors as entities which are, in a sense, unaffected by the rotations which classical 
geometrical entities (vectors etc.) can undergo, and of which the components in a given reference frame
are susceptible to undergo linear transformations which  are in a sense autonomous.}
Cartan  qualified this idea, which amounts to a parallel interpretation of the mathematics, as ``startling''. 
In fact, the claim that spinors do not turn in physical space could be 
attributed to a lack of understanding of the geometrical meaning of spinors, such that
the isospin-inspired viewpoint is then {\em ad hoc} and neither compelling nor unique. 
Based on the principle of Occam's razor, we should not introduce the isospin-inspired assumption unless it is strictly necessary.
  The viewpoint based on the analogy with isospin contains even more exceptional assumptions than the isospin model itself as in the spin operator ${\hat{S}}_{z}$,
  the index $z$ refers to physical space despite the denial that the spinor would not turn
in physical space. In the isospin operator ${\hat{I}}_{z}$ the index $z$ does not refer to physical space,
  such that the postulates of this  formalism are less demanding.
As will transpire from the main text, the approximation to describe the electron as a point charge leads to the introduction
of many mathematical quantities that should permit one to keep treating the physics of  a finite-size object that one has shrunk to a point
in a mathematically self-consistent way.
In this mathematical procedure, one ``pinches'' a part of the true physical reality out of existence. One should however not interpret 
the apparent consequences of these procedures too literally, which has apparently been the case in the viewpoint voiced by Biedenharn.
\label{Cartan-critics}}.

As  ${\mathbf{e}}'_{x} + \imath {\mathbf{e}}'_{y}$ is a vector
it transforms according to the rule ${\mathbf{V}} \rightarrow {\mathbf{R}}{\mathbf{V}}{\mathbf{R}}^{\dagger}$. One can then discover
spinors by noting that $\det\, [\, ({\mathbf{e}}'_{x} + \imath {\mathbf{e}}'_{y}){\boldsymbol{\cdot\sigma}}\,] = 0$ because
$({\mathbf{e}}'_{x} + \imath {\mathbf{e}}'_{y})^{2} = 0$. The rows and columns of $[\, ({\mathbf{e}}'_{x} + \imath {\mathbf{e}}'_{y}){\boldsymbol{\cdot\sigma}}\,] = 0$
must therefore be proportional. This permits us to write:

\begin{equation} \label{isotropic1}
[\, ({\mathbf{e}}'_{x} + \imath {\mathbf{e}}'_{y}){\boldsymbol{\cdot\sigma}}\,] = \sqrt{2}
\left (
\begin{tabular}{c}
$\xi'_{0}$\\
$\xi'_{1}$\\
\end{tabular}
\right ) \otimes (-\xi'_{1}\quad \xi'_{0}) \,\sqrt{2},
\end{equation}

\noindent where

\begin{equation} \label{isotropic2}
 \left (
\begin{tabular}{c}
$\xi'_{0}$\\
$\xi'_{1}$\\
\end{tabular}
\right )  = {\mathbf{R}} \left (
\begin{tabular}{c}
$\xi_{0}$\\
$\xi_{1}$\\
\end{tabular}
\right ) \quad {\text{and}} \quad (-\xi'_{1}\quad \xi'_{0}) = (-\xi_{1}\quad \xi_{0})\,{\mathbf{R}}^{\dagger}.
\end{equation}

\noindent The numbers $\sqrt{2}$ in Eq. \ref{isotropic1} are introduced to satisfy the normalization condition 
 $\xi_{0} \xi_{0}^{*} + \xi_{1} \xi_{1}^{*} =1$. 
 This way one linearizes the quadratic formalism  ${\mathbf{V}} \rightarrow {\mathbf{R}}{\mathbf{V}}{\mathbf{R}}^{\dagger}$ for  vectors
in terms of a linear formalism ${\boldsymbol{\xi}} \rightarrow {\mathbf{R}}{\boldsymbol{\xi}}$ for spinors. This is then analogous
to the way Dirac linearized the Klein-Gordon equation. 
The approach enhances our understanding of the formalism, as it permits us to see how the information about the rotated basis
that defines the rotation is hidden inside the spinor.
But by using it as the starting point for deriving the formalism, a spinor in SU(2)
remains a mysterious object, a kind of square root of an isotropic vector, while the essential point, that it is just a rotation,
remains hidden. It is conceptually much easier to understand the idea that a vector is a tensor quantity of rank 2 in terms of spinors according to our approach,
than to grasp the idea  that a spinor would be a kind of square root of an isotropic vector, according to the standard approach.
There are several other instances where the standard approach  keeps the reader at bay in puzzlement on a sidetrack.
This renders the presentation  abstract and purely algebraic, while the simple underlying   geometrical ideas are lost.
This unsatisfactory situation is just copied into quantum mechanics, which relies on the group theory.   As we will see,
in quantum mechanics we  pay cash for the loss of geometrical insight that results from using the group theory as a black
box of abstract algebra. 
We may mention that there is yet a third approach to spinors, based on the stereographic projection. As discussed 
 in reference \cite{Coddens}, this derivation also tends to conceal the geometrical ideas by
 installing a confusion between spinors and vectors, as it may make the reader  believe that the information content of a spinor
would be that of a position vector
 of a point of the unit sphere.

\subsection{The homogeneous Lorentz group} \label{Lorentz-group}

Also here the basic idea is that a spinor should be a set of coordinates for a group element.
The conditions the analogues of the Pauli matrices will have to satisfy are now $\gamma_{\mu}\gamma_{\nu} + \gamma_{\nu}\gamma_{\mu} = 2 g_{\mu\nu}\bigone$.
There is no fourth $2\times 2$ matrix that would anti-commute with all the Pauli matrices and therefore could be used to 
represent all reflections in Minkowski
space-time and to generate in a second stage all Lorentz transformations.
This problem 
can be overcome  in the $4\times 4$ representation based on the Dirac matrices, 
where $a_{\mu}\gamma^{\mu}$ represents  the four-vector $(a_{t},{\mathbf{a}})$ 
and $\gamma^{t} \neq \bigone$. We have then to postulate $(\sum a_{\mu}\gamma^{\mu})^{2} = (a_{t}^{2} - {\mathbf{a\cdot a}})\,\bigone$.
The simplest representation of the
 Dirac matrices
is the Weyl presentation:

\begin{equation*}
 (a_{t}, {\mathbf{a}})   \longleftrightarrow 
\left (
\begin{tabular}{cc}
& $a_{t}\bigone + {\mathbf{a\cdot}}{\boldsymbol{\sigma}}$ \\
 $a_{t}\bigone - {\mathbf{a\cdot}}{\boldsymbol{\sigma}}$ & \\
 \end{tabular} 
 \right )  
\end{equation*}
\begin{equation}\label{Weyl}
 = a_{t} \gamma^{t} + a_{x} \gamma^{x} + a_{y} \gamma^{y} + a_{z} \gamma^{z} = a_{\mu}\gamma^{\mu}\,  {\hat{=}}  \, (a_{t},{\mathbf{a}}){\boldsymbol{\cdot}} (\gamma^{t}, {\boldsymbol{\gamma}}).
\end{equation}

\noindent This representation is much more easy to manipulate than the traditional text book representation, as 
due to the 
block structure of the Weyl representation
the formalism  reduces to two sets of calculations with $2\times 2$ matrices. We can write them as ${\mathbf{A}} = a_{t}\bigone + {\mathbf{a\cdot}}{\boldsymbol{\sigma}}$
and ${\mathbf{A}}^{\star} = a_{t}\bigone - {\mathbf{a\cdot}}{\boldsymbol{\sigma}}$.
These matrices occur as blocks
 on the secondary diagonal. They are
 both  matrices that represent four-vectors in a SL(2,${\mathbb{C}}$) representation, but in two different types of SL(2,${\mathbb{C}}$) representation.
Each of the two vector matrices can be used as starting point  to set up a representation SL(2,${\mathbb{C}}$) 
 of the Lorentz group \cite{Jones}. The matrix ${\mathbf{A}}^{\star}$ is obtained  from ${\mathbf{A}}$ 
by the parity transformation ${\mathbf{a}} \,|\, -{\mathbf{a}}$. The SL(2,${\mathbb{C}}$)  representations  that are working on the vector matrices are tricky. 
The formalism does no longer permit using a unit four-vector $(a_{t},{\mathbf{a}})$ to define a general reflection in SL(2,${\mathbb{C}}$)
as there is no fourth Pauli matrix to represent reflections with respect to ${\mathbf{e}}_{t}$. Instead of that $a_{t}$ is associated with $\bigone$.

The SL(2,${\mathbb{C}}$) representations do thus not permit a clear  distinction between ${\mathbf{e}}_{t}$ and the identity element $\bigone$ of the Lorentz group, 
which are both represented by $\bigone$. This difficulty is removed by the introduction of the gamma matrices where clearly $\gamma_{t} \neq \bigone$. Nevertheless, if one contents oneself with describing only true Lorentz transformations which are
products of an even number of space-time reflections, we can see by following the faith of the matrices within the Weyl representation
that the $2\times 2$ formalism builds a representation, whereby the four-vector ${\mathbf{A}} = a_{t}\bigone - {\mathbf{a\cdot}}{\boldsymbol{\sigma}}$ transforms
according to ${\mathbf{A}} \rightarrow {\mathbf{LAL}}^{\dagger}$, where ${\mathbf{L}}^{\dagger} \neq {\mathbf{L}}^{-1}$.
In the other   SL(2,${\mathbb{C}}$)  representation, ${\mathbf{A}}^{\star}$ transforms according to ${\mathbf{A}}^{\star} \rightarrow {\mathbf{L}}^{-1}{\mathbf{A}}^{\star} {\mathbf{L}}^{-1 \dagger}$.
We see that in the Weyl formalism the $2\times 2$ blocks are just sequences where the presence and absence of the symbol $\star$ alternates, 
e.g. ${\mathbf{V}}_{2n}^{\star}{\mathbf{V}}_{2n-1}\cdots {\mathbf{V}}_{2}^{\star}  {\mathbf{V}}_{1}$. The algebra in the other block is just given by inverting the presences and absences of the
$\star$ symbol. Everything that happens in one $2\times 2$ block is thus defined by what happens in the other $2\times 2$  block, such that we can use the 
$2\times 2$ blocks as a shorthand for what happens in the $4\times 4$ formalism.
We may note that ${\mathbf{V}}^{\star}$ has the meaning
of $(v_{t}, -{\mathbf{v}})$ in the representation without stars. This justifies the use we will make of the $2\times 2$ matrices in the following sections.

It is no longer possible to cram all the information about a general Lorentz transformation that is coded in a 
one-to-one fashion within a
SL(2,${\mathbb{C}}$) matrix into a single $2\times 1$ spinor like it was the case in SU(2). Fortunately, will not have to bother about
this technicality in this paper.
Once again, we refer the reader to reference \cite{Coddens} for more details about the solution of this problem and the group calculus.

We must finally point out that a representation has always its own internal self-consistent logic, such that there can be no ground
 to question
any result correctly derived within a given representation by drawing in considerations  from outside the context of that representation.


\section{Lorentz Symmetry of Electromagnetism} \label{Lorentz-symmetry}

\subsection{Some simple algebra in SL(2,${\mathbb{C}}$)} \label{algebra}

\subsubsection{The fields} \label{fields}

In view of the facts outlined in Subsection \ref{Lorentz-group}, in SL(2,${\mathbb{C}}$) the four-gradient $({\partial\over{\partial ct}}, {\mathbf{\nabla}})$  is represented by ${\partial\over{\partial ct}}\bigone - {\mathbf{\nabla\cdot}}{\boldsymbol{\sigma}}$.
Analogously, the four-potential $(V,c{\mathbf{A}})$ is represented by $V\bigone -c {\mathbf{A\cdot}}{\boldsymbol{\sigma}}$. We can now check what will happen if we ``multiply'' these two matrices.
Using the identity $[\,{\mathbf{a}}{\boldsymbol{\cdot\sigma}}\,]\, [\,{\mathbf{b}}{\boldsymbol{\cdot\sigma}}\,] 
= ({\mathbf{a\cdot b}})\bigone + \imath\,[\ ({\mathbf{a}}\wedge{\mathbf{b}}){\boldsymbol{\cdot\sigma}}\,]$
we find:

\begin{equation*}
[\, {\partial\over{\partial ct}}\bigone - {\mathbf{\nabla\cdot}}{\boldsymbol{\sigma}}\,]\,
[\,{V\over{c}}\bigone 
- {\mathbf{A\cdot}}{\boldsymbol{\sigma}}\,] =
\end{equation*}
\begin{equation}\label{definitions}
\begin{tabular}{ccccc}
${\underbrace{
[\, {1\over{c^{2}}}{\partial V\over{\partial t}} + {\mathbf{\nabla\cdot A}}\,]\,\bigone }}$ 
&
$-$ 
&
${\underbrace{ {1\over{c}} [\,({\mathbf{\nabla}} V + {\partialÊ{\mathbf{A}}\over{\partial t}}){\boldsymbol{\cdot\sigma}} \,]}}$
& 
$+$ 
& 
$ {\underbrace{\imath [\,  ({\mathbf{\nabla}}\wedge {\mathbf{A}}){\boldsymbol{\cdot\sigma}} \,]}} $
\\
Lorentz gauge  & & ${1\over{c}} {\mathbf{E\cdot}}{\boldsymbol{\sigma}}$ & & $\imath {\mathbf{B\cdot}}{\boldsymbol{\sigma}}$ 
\end{tabular}
\end{equation}

\noindent With the Lorentz gauge condition ${1\over{c^{2}}}{\partial V\over{\partial t}} + {\mathbf{\nabla\cdot A}} =0$, we obtain thus:

\begin{equation}\label{definitions1}
[\, {\partial\over{\partial ct}}\bigone - {\mathbf{\nabla\cdot}}{\boldsymbol{\sigma}}\,]\,
[\,{V\over{c}}\bigone 
- {\mathbf{A\cdot}}{\boldsymbol{\sigma}}\,] =  {1\over{c}} [\,({\mathbf{E}} + \imath c{\mathbf{B}}){\boldsymbol{\cdot\sigma}}\,].
\end{equation}

\noindent We recover thus automatically the expressions for the Lorentz gauge condition, and for
the electric  and magnetic fields in terms of the potentials.
The term ${\mathbf{E}} + \imath c{\mathbf{B}}$ is the electromagnetic field tensor. 
The presence  of $\imath$ in an expression can be seen to  signal that it is a pseudo-vector or a pseudo-scalar\footnote{
This is general and follows from the identity $[\,{\mathbf{a}}{\boldsymbol{\cdot\sigma}}\,]\, [\,{\mathbf{b}}{\boldsymbol{\cdot\sigma}}\,] 
= ({\mathbf{a\cdot b}})\bigone + \imath\,[\ ({\mathbf{a}}\wedge{\mathbf{b}}){\boldsymbol{\cdot\sigma}}\,]$.
Let us construct a general expression 
$[\, a^{(1)}_{t}\bigone + {\mathbf{a}}^{(1)}{\boldsymbol{\cdot\sigma}}\,]\,[\, a^{(2)}_{t}\bigone + {\mathbf{a}}^{(2)}{\boldsymbol{\cdot\sigma}}\,]\,
\cdots [\, a^{(n)}_{t}\bigone + {\mathbf{a}}^{(n)}{\boldsymbol{\cdot\sigma}}\,]$.
First consider the case that $n$ is even.
All quantities in the  product are rank $n$ by construction. Some of them are one-component quantities. 
One can recognize them by the fact that they are multiples of the unit matrix.
When these quantities contain
an even number of vector terms $ {\mathbf{a}}^{(j)}{\boldsymbol{\cdot\sigma}}$, they will be real like $ ({\mathbf{a\cdot b}})\bigone$
and they will not change sign under parity
transformation. They are scalars. When they contain an odd number of vector terms, they will be imaginary
like $\imath\, {\mathbf{a\cdot}}({\mathbf{b}} \wedge {\mathbf{c}})\,\bigone$, and change sign under a
parity transformation. They are pseudo-scalars. The other quantities are three-component quantities. One can recognize them
by the fact that they contain ${\boldsymbol{\sigma}}$ in their expression.
When they contain an even number of vector terms, they will be imaginary like $ \imath\,({\mathbf{a}}\wedge{\mathbf{b}}){\boldsymbol{\cdot\sigma}}$
and they will not change sign under a parity transformation.
They are pseudo-vectors. When they contain an odd number of vector terms, they are real 
like $a_{t}\,{\mathbf{b}}{\boldsymbol{\cdot\sigma}}$,
and they will change sign under a parity transformation.
They are vectors.\label{note0} }.
The vector ${\mathbf{E}}$ and pseudo-vector ${\mathbf{B}}$ are the symmetric and  anti-symmetric
 three-component parts of the six-component field  tensor.  We see thus that symmetry is enough
to recover all the definitions. It summarizes in a sense the reason why we need the theory of relativity by showing
that Lorentz symmetry is the symmetry that is compatible with the structure of the Maxwell equations.
 A whole text book development is here elegantly summarized in one line of calculation.
With this formalism, one can also write the four Maxwell equations jointly in one, very simple, matrix equation.
It seems that this approach was first discovered by Majorana, but most of the time
the presentation is less concise than here.

\subsubsection{The interactions} \label{all-effects}

The charge-current four-vector $(\rho, {\mathbf{j}}/c)$ for a moving point charge $q$ with velocity ${\mathbf{v}}$ is (up to the Lorentz factor $\gamma = (1 - v^{2}/c^{2})^{-1/2}$) given by 
$(q, -q{\mathbf{v}}/c)$, 
which is represented by
 $q\bigone - {q\over{c}} {\mathbf{v\cdot}}{\boldsymbol{\sigma}}$. 
 Let us now couple this quantity with the electromagnetic-field tensor and 
 calculate\footnote{Of course, in Eq. \ref{forces} we have  dropped out 
$\gamma$ and  $[\, q\bigone - {q\over{c}} {\mathbf{v\cdot}}{\boldsymbol{\sigma}}\,]$ is not a four-vector.
The true four-vector  is actually  $\gamma[\, q\bigone - {q\over{c}} {\mathbf{v\cdot}}{\boldsymbol{\sigma}}\,]$ rather than  $[\, q\bigone - {q\over{c}} {\mathbf{v\cdot}}{\boldsymbol{\sigma}}\,]$. Nevertheless, what we have done is correct, because
the terms we are identifying in the Lorentz force equation, like e.g. ${d{\mathbf{p}}\over{dt}} $ and 
$ q({\mathbf{E}} + {\mathbf{v}} \wedge {\mathbf{B}})$,
are both not covariant, while 
the equation ${d{\mathbf{p}}\over{dt}} = q({\mathbf{E}} + {\mathbf{v}} \wedge {\mathbf{B}})$
is covariant. In fact, it corresponds to rearranging  ${d{\mathbf{p}}\over{ d\tau}} = \gamma ({\mathbf{E}} + {\mathbf{v}} \wedge {\mathbf{B}})$
by swapping  $\gamma$ from the right-hand side to the left-hand side, where we can use it to replace $\gamma d\tau$ by $dt$.
Here $\tau$ is the proper time.
The quantities  $d\tau$, $({1\over{c}}\,dE,d{\mathbf{p}})$ and $ (\gamma {\mathbf{v\cdot E}}/c, \gamma ({\mathbf{E}} + {\mathbf{v}} \wedge {\mathbf{B}}))$ that occur in the equation before the swapping are Lorentz covariant  (See e.g. J.D.~Jackson, in {\em Classical Electrodynamics}, (Wiley, New York, 1963), page 405).
Once we know the equation is covariant we can swap sides with any quantity in it.
Now that we know that the equation is covariant, further covariant operations on them
will yield also covariant equations.\label{note1}}:

\begin{equation}\label{forces}
 [\, q\bigone - {q\over{c}} {\mathbf{v\cdot}}{\boldsymbol{\sigma}}\,]\,
[\,( {\mathbf{E}} + \imath c{\mathbf{B}}){\boldsymbol{\cdot\sigma}}\,].
\end{equation}

\noindent We obtain then:

\begin{equation}\label{forces2-split1}
\boxed{
-\,[\,{q\over{c}} {\mathbf{v\cdot E}}\,]\,\bigone + q({\mathbf{E}} + {\mathbf{v}}\wedge {\mathbf{B}}){\boldsymbol{\cdot\sigma}}
- \imath \,[\,q{\mathbf{v\cdot B}}\,]\,\bigone +  \imath cq\,[\,{\mathbf{B\cdot}}{\boldsymbol{\sigma}}\,]  -  \imath {q\over{c}}[\, ({\mathbf{v}}\wedge {\mathbf{E}}){\boldsymbol{\cdot\sigma}}\,]\quad}
\end{equation}

\noindent  The whole  paper is devoted to the meaning and the consequences of this single equation.
 Again, the presence of $\imath$ signals here pseudo-scalars and pseudo-vectors, while the real terms correspond to
scalars and vectors, as can be checked from the behaviour
of the various terms under a parity transformation.
We recognize here the Lorentz force  ${\mathbf{F}} = q({\mathbf{E}} + {\mathbf{v}}\wedge {\mathbf{B}})$  and 
the power-related term $q{\mathbf{E\cdot v}}/c$. In fact, the term ${\mathbf{F\cdot v}}= q{\mathbf{E\cdot v}}$ represents the power
corresponding to the work ${\mathbf{F\cdot}}d{\mathbf{r}}$ done against the force ${\mathbf{F}}$ during
an infinitesimal displacement $d{\mathbf{r}}$ over a time interval $dt$. 
It is  well known that the four-vector generalization of the force three-vector ${\mathbf{F}}$ is $({\mathbf{F\cdot v}} /c, {\mathbf{F}})$,
which contains this additional power-related term (up to a constant $c$).  As the term  $q{\mathbf{E\cdot v}}$ is here divided by $c$,
the result has again the dimension of a force. We will call such terms therefore scalar force terms.
The other terms in Eq. \ref{forces2-split1}
are all imaginary and they may at first sight look less familiar.

\section{Magnetic Monopoles} \label{monopole-fuss}

\subsection{Symmetry issue}\label{note-symm}

There is a surprise in Eq. \ref{forces2-split1} in that it is seen  to exhibit a complete symmetry 
between the electric and magnetic force terms. 
Each of the imaginary terms in Eq. \ref{forces2-split1} corresponds to a term that is a relativistic counterpart 
of a term in $({\mathbf{F\cdot v}} /c, {\mathbf{F}})$ obtained by
using the substitution ${\mathbf{E}} \rightarrow  c{\mathbf{B}}$, $c{\mathbf{B}} \rightarrow  -{\mathbf{E}}$. 
The addition of these terms is necessary
to obtain full relativistic symmetry for the total result, just like adding $\imath c {\mathbf{B}}$ to ${\mathbf{E}}$ 
is necessary to obtain an expression with full relativistic symmetry.
Such a perfect symmetry in the forces is something that is believed to occur only if  magnetic monopoles were to exist. In such an overall symmetry, the magnetic monopole would  be the
symmetric counterpart of the electric monopole.  It will also not have escaped the attention of the reader that
the imaginary three-component quantities that occur in Eq. \ref{forces2-split1}
describe exactly the force exerted by an electromagnetic field on a magnetic monopole $q_{m}$:

\begin{equation} \label{monopole-force}
{\mathbf{F}}_{m} =   q_{m}\,[\,{\mathbf{B\cdot}}{\boldsymbol{\sigma}}\,]       -  {q_{m}\over{c^{2}}}[\, ({\mathbf{v}}\wedge {\mathbf{E}}){\boldsymbol{\cdot\sigma}}\,],
\end{equation}

\noindent provided we take $q_{m} = cq$.\footnote{With the identities
$q_{m} = cq$,  $ \varepsilon_{0}  \mu_{0} = {1\over{c^{2}}}$, and the substitution
${\mathbf{E}} \rightarrow  c{\mathbf{B}}$,
the equation ${\mathbf{E}} = {q\over{4\pi\varepsilon_{0} r^{2}}} {\mathbf{e}}_{r}$ can be transformed into ${\mathbf{B}} = {q_{m} \mu_{0}\over{4\pi r^{2}}} {\mathbf{e}}_{r}$. 
Applying the theorem of Gauss to the magnetic monopole yields $\oiint \, {\mathbf{B\cdot}}d{\mathbf{S}} = q_{m} \mu_{0}$.\label{magnetic-charge}} 
The one-component quantity $- \imath \,[\,q{\mathbf{v\cdot B}}\,]\,\bigone$ is the corresponding power-related term
that completes the force four-vector.
It looks therefore  as though we  get magnetic monopoles out of Eq. \ref{forces2-split1}, while at face value
we have not introduced any magnetic monopoles in Eq. \ref{forces}. {\em All we have introduced 
is the charge-current four-vector of one single moving point charge within a formalism that automatically accounts for Lorentz symmetry.}
 The terms in Eq. \ref{forces2-split1}  are all forces as Eq. \ref{forces} actually generalizes a term $[\,q\,\bigone\,] \, [\, {\mathbf{E}}{\boldsymbol{\cdot\sigma}}\,]$ in a rest frame
to a moving frame by Lorentz covariance. Eq. \ref{forces2-split1} gives us thus the most general possible expression for an electromagnetic force.
It is obvious that one can find a frame wherein such a most general situation is  realized.
It therefore looks as though invoking magnetic monopoles as a mechanism to obtain the symmetry exhibited by Eq. \ref{forces2-split1}
could be a bit far-fetched. All terms in Eq. \ref{forces2-split1} are referring to phenomena that are occurring to a single particle,
not to various different particles.

\subsection{Point-like rotational motion}\label{intuition}

We will discuss in Subsubsection \ref{q-D}
how the whole theory of electromagnetism just resumes to the description of interactions of charges and currents with other charges and currents.
A magnetic field is produced by moving charges. A constant magnetic field is thus a mathematical expedient to describe
moving charges as a static non-moving phenomenon. It will be argued along similar lines that a magnetic monopole is just a mathematical construct to
treat  a  moving electric charge as a static non-moving quantity. 

This may look an absurd statement
if we think about the charges as  performing  a uniform rectilinear motion, that can be described by an
appropriate Lorentz transformation. 
However a charge that is in uniform circular motion in the laboratory frame, whereby the radius of the circle is very small such
that it would look like a point to the naked eye, could indeed correspond intuitively 
to a phenomenon at rest with respect to the laboratory frame. As we will argue, this is the seed of the idea behind identifying
the force terms that occur in Eq. \ref{monopole-force} as forces acting on
 magnetic monopoles. 
Pushing the idea of a motion that cannot be detected to the extreme, a monopole at rest can be imagined as the limit of a charge traveling on a circular orbit
whose radius tends to zero such that the orbit shrinks to a point. It will then just become  a  charge that is rotating in a fixed position (because the precession
does not vanish when we take the limit). 
In considering such a limit we are defining
a new mathematical object, somewhat in the same way as Laurent Schwartz \cite{Schwartz} defined the mathematical distribution that would correspond to a point-like dipole. This idea of shrinking the orbit to a point will be further elaborated  below in Subsection \ref{comb-issues} and in Subsection \ref{A-pot}.

From this point of view, magnetic monopoles and magnetic fields are both just 
theoretical constructions to describe confined motion as a static phenomenon. 
 Coining such mathematical quantities might look like a stroke of genius, but  history shows that this was done unwittingly, as magnetic fields  were  just 
 introduced as a phenomenological tool to describe  the observations in terms of static quantities before their physics were truly understood in terms of motion.
 They were thus defined based on a visual illusion of rest. Of course, the idea to treat motion as a phenomenon at rest  is a kind of tricky and beyond guessing. 
If not clearly spelled out, it may thus easily lead to confusion.

 \subsection{Quantization issue}\label{charge-quant}

We may note that Dirac's argument that the existence of monopoles would lead to the quantization of charge amounts to postulating:

\begin{equation}\label{Dirac-quantization}
{qq_{m}\over{2\pi \epsilon_{0} \hbar c^{2}}} \in {\mathbb{Z}}.
\end{equation}

\noindent  For an electron with charge $q$ and its associated magnetic monopole charge $q_{m} = cq$, the quantity that is required to be an integer becomes then:

\begin{equation}\label{Dirac-quantization-against-alpha}
{qq_{m}\over{2\pi \epsilon_{0} \hbar c^{2}}} = 2\alpha, \quad {\text{where:}}\quad \alpha = {q^{2}\over{4\pi\epsilon_{0}\hbar c}} \quad {\text{is~the~fine-structure~constant}}.
\end{equation}

\noindent As $\alpha \approx 1/137$, the prediction $2\alpha \in {\mathbb{Z}}$  is way off, but by rewriting Eq. \ref{Dirac-quantization-against-alpha} as:

\begin{equation}\label{Dirac-quantization-corrected}
{qq_{m}\over{2\pi \alpha \epsilon_{0}c^{2}}} = \hbar,
\end{equation}

\noindent and considering $\alpha$ as a constant of nature, Dirac's argument could perhaps be saved.
In fact, as orbital angular momentum occurs in multiples of $\hbar$, charge would have to come in multiples of $q$.
However, in particle physics, $\alpha$ is not considered to be a constant. 

\subsection{Refining the concepts}\label{refine}

There are several attitudes one can adopt with respect to these considerations. The first one would be to reject them all together
with contempt, on the grounds that they seem to question the official viewpoint of hard-won ``established science''.
However, an alternative attitude is possible. Before we can address it we must first further develop our  {\em Weltbild} for the  magnetic
monopole.

In the Appendix we  show
that an electron traveling at a velocity ${\mathbf{v}}$ can be associated with a singular current  (that has not to flow along a closed loop). It will become obvious from this approach that a single moving charge can indeed be considered as a ``magnetic charge''
with a  magnetic moment. This magnetic moment could be called a {\em monopole moment}  because there is only one charge in  motion, 
in contrast with
the macroscopic situation in e.g. a circular current loop where there are many such charges and we talk about a dipole moment. The main difference is that one cannot
really claim that there exists a torque pulling on a single charge on a circular orbit described as a loop, while one can when there is more than one charge
traveling around the loop.
In this respect, the magnetic moment of a single magnetic charge can be called a ``magnetic monopole moment'', where it has to be emphasized 
that there is {\em no hyphen} between ``magnetic'' and ``monopole'' because ``magnetic'' refers to a magnetic moment, not to a magnetic monopole.

A magnetic monopole is an all-together different and unrelated concept.
Not all single  ``magnetic charges'' can be considered as magnetic monopoles.
Only the magnetic charges that correspond to point-like rotational motion will be considered as true magnetic monopoles.
Magnetic monopoles will thus not  be new physical particles one would have to search for to confirm Dirac's predictions.
They will appear   to be  just a mathematical means to deal with
moving charges whose motion remains hidden to the eye by remaining confined 
``inside a point'', with the effect that we think that we are dealing with a purely static situation.
There is then a confusion that  has to be avoided. The ``magnetic monopole moment'' of a magnetic monopole could be
called  a magnetic-monopole moment {\em (with a hyphen)}, where the latter would be an ellipsis for a magnetic-monopole's magnetic monopole moment.

There is an important reason for making the distinction between 
 ``magnetic charges'' corresponding to  currents $q{\mathbf{v}}$ and  true point-like magnetic monopoles. If we associated
magnetic monopoles with the visible, not point-like currents $q{\mathbf{v}}$, then the classification
of the interaction terms would be wrong. The criterion to classify the terms would then be if  some term ${\mathbf{j}} = q{\mathbf{v}}$ occurs in them or otherwise.
The part $q ({\mathbf{v}}\wedge {\mathbf{B}})$ would then be due to the interaction
of the magnetic monopole with the magnetic field, while it is conventionally attributed to the interaction of the moving electric charge with the magnetic field.
On the other hand, the part $cq  {\mathbf{B}}$ would be due to the interaction of the electric charge with the magnetic field, while
it is conventionally attributed to the interaction of the magnetic monopole with the magnetic field.

The classification must  rather be based on a distinction between point-like hidden and not point-like visible currents.
We will  show in Subsection \ref{A-pot} that the terms in Eq. \ref{monopole-force} are related to precession. Such a precession can be mentally visualized as
 a kind of point-like hidden motion and it  has hidden  energy we must account for if we want to get our
 calculations right. The terms in Eq. \ref{monopole-force} correspond thus to hidden rotational motion rather than to
 translational motion (which is conceptually always visible). These rotational effects
 can be described also in terms of a vorticity (as will be discussed in  Subsubsection \ref{Larmor-vortex}). The duality between the force in Eq. \ref{monopole-force} and the Lorentz force is thus rather based
on a duality between non point-like visible (rotational and translational) and
point-like  invisible (rotational) motion than on a duality between electric charges and magnetic charges.

\subsection{Disentangling the  two unrelated issues of symmetry and quantization}\label{comb-issues}

On the basis of all this, we can and {\em must} now also make a further distinction between two  concepts of magnetic monopoles. They should  not be confused
because they address two completely unrelated issues.
The first issue is symmetry in the equations describing electromagnetism.
As Eq. \ref{monopole-force} shows, we get it for free.
We do not need to postulate the existence of a true magnetic  monopole to obtain the symmetrical counterpart of the Lorentz force
 in the form of Eq. \ref{monopole-force}. These terms are already there and we will show that they have already been experimentally observed 
 in the form of the anomalous Zeeman effect and the spin-orbit coupling.
 Our first concept of magnetic monopole is thus only a mathematical hype.
 We rewrite $cq$ as $q_{m}$
just to enhance the symmetry, rendering it more evident. We obtain then
 a shiny  interpretation for the symmetry revealed by the
existence of the terms in Eq. \ref{monopole-force}, but as the derivations show, it is also possible
to describe everything in a less glittering way that  only  calls for electric monopoles.
There is no new physical quantity, the only truly
existing physical quantity is $q$. 

The second issue is the quantization of charge, which we do not get for free at all, which is
why Dirac introduced his magnetic-monopole concept. Dirac's argument  does not explain everything,
as it hinges on the quantization of angular momentum which is also just an empirical fact. We 
know very well to {\em describe} the quantization of angular momentum  by quantum mechanics but  our intuition about it
is not any better than our intuition about the quantization of charge. In fact, quantization of charge
is  conceptually a less difficult  concept than quantization of angular momentum because charge is a fundamental quantity. Its definition does not rely
on the definition of  other quantities 
which are themselves not quantized like ${\mathbf{r}}$  and ${\mathbf{p}}$ in the definition ${\mathbf{r}} \wedge {\mathbf{p}}$ of angular momentum.
Of course,  it is eventually  just experimental evidence that can tell if Dirac's construction  is justified.
If Dirac's monopole existed then it would lead to a second equation that is completely analogous to Eq. \ref{forces2-split1}, with $q$  replaced by $Q_{m}/c$
for some value of $Q_{m} \neq q_{m}$. 

The two different issues lead to two distinct concepts of magnetic monopoles, as we do not need Dirac's monopole to get the symmetry between 
 the electric and the magnetic force terms, while the monopole $q_{m}$, which accounts for  that symmetry
  does not tally with the prediction based on Dirac's construction needed to obtain quantization.  
  Traditionally the two issues are mentioned in one breath. 
If in following this tradition however, we  merge the two {\em a priori  completely unrelated issues} into a single one, then it would appear as though Dirac's 
construction misses the point underlying the introduction
of $q_{m} = cq$ in the first issue, which is that a magnetic monopole
 just serves to describe rotational motion confined to a point. In fact, Dirac's monopole is a current (called a string) 
 that stretches from some point to infinity,
 which is all but confined (Why this is wrong will be further discussed in Subsubsection \ref{two-merry}).
 If we blend the two issues, the existence of two equations, one with
$q_{m}$ and one with $Q_{m}$ will also raise   the question why we use only $Q_{m}$ and not also $q_{m}$ in Dirac's argument.
If we do not confuse them, then one might perhaps think of an argument why we only use $Q_{m}$, based on the idea that $q_{m}$ is not a ``true
monopole'' in Dirac's sense.

\section{Anomalous Zeeman Effect and Spin-Orbit Coupling as Purely Classical Phenomena} \label{worse}

\subsection{A striking similarity} \label{similar}

There is an alluring way to interpret  Eq. \ref{forces2-split1} completely differently as follows:

\begin{framed}
\begin{equation*}
\begin{tabular}{ccccc}
${\underbrace{-\,[\,{q\over{c}} {\mathbf{v\cdot E}}\,]\,\bigone }}$
&\quad\quad&
${\underbrace{ - \imath \,[\,q{\mathbf{v\cdot B}}\,]\,\bigone}}$
&\quad\quad &
$ {\underbrace{+ q({\mathbf{E}} + {\mathbf{v}}\wedge {\mathbf{B}}){\boldsymbol{\cdot\sigma}} }}$ \\
power term & \quad\quad& dual power term  &\quad\quad &  Lorentz force ${\mathbf{F}}$\\
\end{tabular}
\end{equation*}

\begin{equation}\label{forces2}
\begin{tabular}{ccc}
$ {\underbrace{  +  \imath cq\,[\,{\mathbf{B\cdot}}{\boldsymbol{\sigma}}\,]      }}$
&\quad\quad 
$ {\underbrace{  -  \imath {q\over{c}}[\, ({\mathbf{v}}\wedge {\mathbf{E}}){\boldsymbol{\cdot\sigma}}\,]    }}$\\
looks similar to &\quad\quad  looks similar to\\
anomalous Zeeman effect
&\quad\quad
spin-orbit coupling\\
\end{tabular}
\end{equation}
\end{framed}

\noindent  We have labeled here the magnetic-monopole terms from Eq. \ref{monopole-force} 
as ``looking similar'' to the anomalous Zeeman effect and to the spin-orbit coupling.
What we want to refer to with the terminology ``looks similar'',  is that
the imaginary terms on the second line of Eq. \ref{forces2} correspond to force terms which are 
up to the proportionality factor $- {\hbar\over{2m_{0}c}}$
equal to the energy terms  derived from the Dirac theory for the anomalous Zeeman effect and for the spin-orbit coupling, {\em but without
the correction for the Thomas precession.}\footnote{Due to this absence of the term that corresponds to the Thomas precession,
the reader could be inclined to conclude that the approach that leads to Eq. \ref{forces2}
is inferior to the traditional approach wherein it is possible to derive the correct expression for the spin-orbit coupling from the Dirac equation.
However, as will be discussed later on in this  subsection, this derivation is not correct and based on a fundamental error in the algebra.\label{apparences}}

As will be explained in Subsection \ref{correct-minimal-applied}, Eq. \ref{forces2} is obtained from Eq. \ref{forces}  
which just expresses  the product of the three terms
  $[\,q\bigone - {q\over{c}} {\mathbf{v\cdot}}{\boldsymbol{\sigma}}\,]\,[\, {\partial\over{\partial ct}}\bigone - {\mathbf{\nabla\cdot}}{\boldsymbol{\sigma}}\,]\,
[\,{V\over{c}}\bigone  - {\mathbf{A\cdot}}{\boldsymbol{\sigma}}\,] $, 
while the correct physics for the anomalous Zeeman effect and the spin-orbit coupling
are obtained by considering  a variant
$-{\hbar\over{2m_{0}c}}\,[\, {\partial\over{\partial ct}}\bigone - {\mathbf{\nabla\cdot}}{\boldsymbol{\sigma}}\,]\,
[\,q\bigone + {q\over{c}} {\mathbf{v\cdot}}{\boldsymbol{\sigma}}\,]\,
[\,{V\over{c}}\bigone  + {\mathbf{A\cdot}}{\boldsymbol{\sigma}}\,] $, wherein the three types of terms are occurring in a different order.

For the derivation of the anomalous Zeeman term  $ -{\hbar q \over{2m_{0}}}\,[\,{\mathbf{B\cdot}}{\boldsymbol{\sigma}}\,]$,
 the change of order has no incidence, such that the algebra that leads to
the term $ \imath cq\,[\,{\mathbf{B\cdot}}{\boldsymbol{\sigma}}\,]$ in Eq. \ref{forces2} is exactly the same as the one that occurs in the derivation of the anomalous Zeeman effect from the Dirac equation. That we obtain the expression for the anomalous Zeeman effect up to the factor  $-{\hbar\over{2m_{0}c}}$ in Eq. \ref{forces2} 
is thus not a coincidence.
Also for the calculation of the spin-orbit term the order of the three terms in the product does not matter.
For both orders of the three terms the calculation does  not account for the correction
due to the Thomas precession. 

There is a classical rationale for the correct calculation of the spin-orbit coupling that runs as follows. 
As can be seen e.g. from Eq. \ref{relativity},
 the total magnetic
field experienced by the electron in its co-moving frame is (to first approximation, whereby one neglects the factors $\gamma$) given by
${\mathbf{B}}' = {\mathbf{B}} - {1\over{c^{2}}} {\mathbf{v}} \wedge {\mathbf{E}} = {\mathbf{B}} + {\mathbf{B}}_{n}$.
The electric field of the nucleus gives thus  rise to a magnetic field ${\mathbf{B}}_{n} = - {1\over{c^{2}}} ({\mathbf{v}}\wedge {\mathbf{E}}){\boldsymbol{\cdot\sigma}}$ in a frame that is co-moving with the traveling electron. 
The interaction of the electron
with the magnetic field ${\mathbf{B}}_{n}$ gives rise to an anomalous Zeeman term  
$-{q\hbar\over{2m_{0}}} [\,{\mathbf{B}}_{n}{\boldsymbol{\cdot\sigma}}\,]=  {\hbar\over{2m_{0}c}} \times {q\over{c}} [\, ({\mathbf{v}}\wedge {\mathbf{E}}){\boldsymbol{\cdot\sigma}}\,]$. This leads finally to an interaction energy ${\hbar\over{2m_{0}^{2} c^{2}}} {1\over{r}} {\partial U\over{\partial r}}$.
This has to be corrected for the Thomas precession ${\boldsymbol{\omega}}_{T} = {1\over{2 c^{2}}} {\mathbf{v}} \wedge {\mathbf{a}} =
{q\over{2m_{0}c^{2}}} {\mathbf{v}} \wedge {\mathbf{E}}$ 
of the electron in its orbit around the nucleus (see Eq. \ref{Wigner}). According to  Subsection \ref{precession}, the corresponding energy  is  $\hbar\omega_{T}/2 = 
{\hbar \over{4m_{0}^{2} c^{2}}}{1\over{r}} {\partial U\over{\partial r}}$ which has to be subtracted from the Zeeman term.
The absolute value of the correction for the Thomas precession is half that of the Zeeman term, which is the reason why it is referred to as the Thomas half.
Both terms on the second line of Eq. \ref{forces2} are thus anomalous Zeeman terms due to magnetic fields in the rest frame of the electron,
while the Thomas precession is a relativistic correction in the laboratory frame. 

As mentioned above in Footnote 
\ref{apparences}, it is claimed in text books that the exact spin-orbit coupling term including the correction for Thomas precession
can be derived from the Dirac theory (see e.g. \cite{Itzykson}). We will show that this is a falsehood.
The calculations claimed to derive the spin-orbit coupling with its correction for Thomas precession from the Dirac equation are
wrong algebra leading to a correct physical result.
The traditional Dirac theory is in  reality unable to derive any of the two terms containing 
$ {\mathbf{v}} \wedge {\mathbf{E}}$ described above, from the Dirac equation with minimal coupling.
It only succeeds in deriving  the correct result by introducing logical  errors. 
Our approach is thus superior to the traditional approach.
By  using the Dirac equation with the correct  physical coupling, it is able
to derive one of the two terms. The term it fails to derive is the correction for Thomas precession.

We may note that even  an electron at rest within a magnetic field is subjected to
the anomalous Zeeman effect. But in the absence of motion there is no Thomas precession,
such that the anomalous Zeeman effect does then not require
a correction for Thomas precession. Of course, Thomas precession occurs in any type of motion.
There exists thus also a correction  for Thomas precession in the case of a
particle that is moving in a magnetic rather than in an electric field. However,
the non-relativistic correction term for Thomas precession during the motion of an electron in a magnetic field ${\mathbf{B}}$ is given by
 ${\boldsymbol{\omega}}_{T} = {1\over{2 c^{2}}} {\mathbf{v}} \wedge {\mathbf{a}} =
{q\over{2m_{0}c^{2}}} {\mathbf{v}} \wedge ({\mathbf{v}} \wedge {\mathbf{B}})$, which
remains very small in the non-relativistic limit due to the presence of the factor $v^{2}/c^{2}$.

As we want to discuss the anomalous Zeeman effect and the spin-orbit coupling in the rest of the paper, the 
similarities exhibited in Eq. \ref{forces2}
seem to be a good way to make the transition between the two parts of the paper.
 That we can derive the anomalous Zeeman effect and the spin-orbit effect this way from a variant of the calculation leading to Eq. \ref{forces2-split1} and Eq. \ref{forces2} 
is a major upheaval, because these two phenomena are traditionally considered as purely due to the electron spin.
Just as it looked as though we obtained magnetic monopoles from Eq. \ref{forces2-split1} 
without having introduced them in Eq. \ref{forces}, here it looks as though {\em we now obtain
spin-related effects  without having introduced spin} in  the variant of Eq. \ref{forces}.

 It is extremely important to note that the matrix calculations one performs in going from Eq. \ref{definitions} to Eq. \ref{forces2}
 or from the variant to the more correct equation
  {\em  are not quantum mechanical}. 
They are independent of any context of wave equations and entirely classical as all we have 
used is a group-theoretical formalism that automatically accounts for Lorentz symmetry. 
Very obviously,
 these derivations 
also
{\em do not contain spin}. {\em We have only introduced the charge-current four-vector of a single moving point charge.}
The algebraic expressions for the anomalous Zeeman effect and the spin-orbit coupling must thus be considered as purely classical and not spin-related.

We might indeed have been convinced that  these terms are quantum mechanical rather than classical because they fitted nicely into quantum mechanics
after their experimental  discovery, while they were not covered by classical mechanics.  But they become quantum mechanical only by the way we use them in quantum mechanics, where they lead to quantized discrete energy levels,
 rather than to a continuum of energy values. This is indeed a feature
 of the experimental data that we are unable to understand classically.
  Due to this quantization and due to the fact that quantum mechanics looks so inscrutable
  to classical intuition, there was perhaps not too much incentive
 to ask oneself if - at least in principle - the existence  of these supplementary imaginary terms could not have an analogous  counterpart
 within a purely classical relativistic context. As also the normal orbital Zeeman effect is quantized and has a well-known classical
counterpart, it would have been legitimate to ask that question.
In Section \ref{terms} we will discuss the physics of the anomalous Zeeman effect and the spin-orbit coupling in further detail.

\subsection{The Pauli equation} \label{Pauli} 

We must now give the reader a first glimpse of the reason why the fact that the anomalous Zeeman effect and the spin-orbit coupling (without the correction for Thomas precession) seem
to occur in Eq. \ref{forces2} with a factor ${\hbar\over{2m_{0}c}}$ is not a coincidence.
The Dirac matrices just contain the
SL(2,${\mathbb{C}}$) matrices $c{\hat{{\mathbf{p}}}}{\boldsymbol{\cdot\sigma}} = 
-\imath c\hbar{\mathbf{\nabla}}{\boldsymbol{\cdot\sigma}}$ and $-qc{\mathbf{A}}{\boldsymbol{\cdot\sigma}}$
in their block structure. Squaring the Dirac equation will lead to a term $[\,c{\hat{{\mathbf{p}}}}{\boldsymbol{\cdot\sigma}}\,]\,[\,-qc{\mathbf{A}}{\boldsymbol{\cdot\sigma}}\,]$.
And from this product we will obtain a term $-c^{2} \hbar q {\mathbf{B}}{\boldsymbol{\cdot\sigma}}$.
Squaring the Dirac equation also leads to a term $c^{2}{\hat{{\mathbf{p}}}}^{2}$, which has to be divided
by $2m_{0}c^{2}$ to reduce it to the operator ${\hat{{\mathbf{p}}}}^{2}/2m_{0}$ which is used in the Schr\"odinger equation in the non-relativistic limit.
In fact, the transition from relativistic to classical mechanics  is obtained by
putting $E= m_{0}c^{2} + E_{cl}$. Then $E^{2} -c^{2}p^{2} = m_{0}^{2}c^{4}$ leads to $E_{cl}^{2} +2m_{0}c^{2} E_{cl} + m_{0}^{2}c^{4}  -c^{2}p^{2} = m_{0}^{2}c^{4}$. 
The terms $m_{0}^{2}c^{4}$ can be dropped on both sides. After dividing both sides by $2m_{0}c^{2}$ and neglecting the term $E_{cl}^{2}/2m_{0}c^{2}$ based on the observation that
$E_{cl} \ll 2m_{0}c^{2}$, one obtains then $E_{cl} = {p^{2}\over{2m_{0}}}$. The quantum mechanical version of this argument is arguably obtained by
introducing  $E= m_{0}c^{2} + E_{cl}$ in the wave function through $\psi = e^{-\imath m_{0}c^{2}t/\hbar}\psi_{cl}$ in the complete Dirac equation
including the minimal substitution (but we will discover later on that this procedure can contain a major pitfall).

This way one can derive the Pauli  equation from the Dirac equation\footnote{It must be noted that after squaring the Dirac equation, and dividing the result by $2m_{0}c^{2}$, we also recover a term  
$-{q B\over{2m_{0}}}{\hat{L}}_{z}\,\bigone$,
which corresponds to the famous classical term $- {\boldsymbol{\mu\cdot}}{\mathbf{B}}\,\bigone$ and describes the orbital Zeeman effect.\label{Lz-term}}
and due to the division
by $2m_{0}c^{2}$ 
this equation contains now the anomalous Zeeman term $-{\hbar q\over{2m_{0}}} {\mathbf{B\cdot\sigma}}$.
As Eq. \ref{forces2} contains $c{\mathbf{B}}$ this explains the conversion factor ${\hbar\over{2m_{0}c}}$. 
The Pauli equation does not contain a term that looks like the spin-orbit interaction, because it does not use the correct ``minimal'' substitution.
We postpone the discussion of this point to Section \ref{correct-minimal-applied}, because it is rather intricate and raises several subsidiary issues.


\section{Problems with the Traditional Interpretation of  the Anomalous Zeeman Effect} \label{terms}

\subsection{The anomalous Zeeman effect} \label{anomaly-g}

\subsubsection{An inconvenient truth: The physical imagery violates the symmetry} \label{anomaly-g-problem}

The anomalous Zeeman effect  $-{\hbar q\over{2m_{0}}}\, [\,{\mathbf{B}}{\boldsymbol{\cdot\sigma}}\,]$ is traditionally attributed to  a coupling between the magnetic
 field ${\mathbf{B}}$ and the spin ${\hbar\over{2}} {\boldsymbol{\sigma}}$. In reality ${\mathbf{B}}{\boldsymbol{\cdot\sigma}}$ is not a scalar product
 but just the way the vector ${\mathbf{B}}$ is written in the group theory.
Traditionally one interprets
 the term  $-{\hbar q\over{2m_{0}}}\, [\,{\mathbf{B}}{\boldsymbol{\cdot\sigma}}\,]$ indeed as analogous to the orbital Zeeman  term   $-{ qB\over{2m_{0}}}\,{\hat{L}}_{z} \bigone$ 
 $= -{ q\over{2m_{0}}}({\mathbf{B\cdot}}{\hat{{\mathbf{L}}}})\,\bigone$,\footnote{
 Here ${\hat{{\mathbf{L}}}} = (\, -\imath\hbar (y{\partial\over{\partial z}} - z {\partial\over{\partial y}}),$ $-\imath\hbar (z{\partial\over{\partial x}} - x {\partial\over{\partial z}}),$
 $-\imath\hbar (x{\partial\over{\partial y}} - y {\partial\over{\partial x}})\,)$. Hence ${\mathbf{BÊ\cdot}} {\hat{{\mathbf{L}}}} $ stands here
 for the scalar operator $ -\imath\hbar [\, B_{x} (y{\partial\over{\partial z}} - z {\partial\over{\partial y}})$ $+ B_{y} (z{\partial\over{\partial x}} - x {\partial\over{\partial z}})$
 $+B_{z} (x{\partial\over{\partial y}} - y {\partial\over{\partial x}})\,]$, such that  $= -{ q\over{2m_{0}}}({\mathbf{B\cdot}}{\hat{{\mathbf{L}}}})\,\bigone$
 is a scalar operator, which for the case ${\mathbf{B}} \parallel {\mathbf{e}}_{z}$ reduces to $-\imath\hbar B_{z} (x{\partial\over{\partial y}} - y {\partial\over{\partial x}}) 
 \,\bigone$. \label{L-note}}
whereby
 one would have to replace ${\hat{{\mathbf{L}}}}$ by ${\hat{{\mathbf{S}}}} = {\hbar\over{2}}{\boldsymbol{\sigma}}$
 in the shorthand notation 
    $ -{ q\over{2m_{0}}}({\mathbf{B\cdot}}{\hat{{\mathbf{L}}}})$.
 This is done by rewriting $-{\hbar q\over{2m_{0}}}\, [\,{\mathbf{B}}{\boldsymbol{\cdot\sigma}}\,]$ as $-{2 q\over{2m_{0}}}\, [\,{\mathbf{B\cdot}}{\hbar\over{2}}{\boldsymbol{\sigma}}\,]$, where
${\hbar\over{2}}{\boldsymbol{\sigma}}$ is postulated to be the operator ${\hat{{\mathbf{S}}}}$
 corresponding to the  ``spin vector'' ${\mathbf{S}}$, such that the term becomes then $-{\mathbf{B\cdot}} {2q\over{2m_{0}}}{\hat{{\mathbf{S}}}}$ yielding
  an energy eigenvalue
$-{\mathbf{B\cdot}} {2q\over{2m_{0}}}{\mathbf{S}} = - {\boldsymbol{\mu}}_{e}{\mathbf{\cdot B}}$, where  $\mu_{e} = 2\mu_{B}$.
It is because the eigenvalues of $S_{z}$ of ${\hat{S}}_{z}$ are $\pm {\hbar\over{2}}$, that one is then obliged to introduce a factor $g=2$
 into the algebra to recover the correct value  $-{\hbar q\over{2m_{0}}}\, [\,{\mathbf{B}}{\boldsymbol{\cdot\sigma}}\,]$. This interpretation tries thus to represent the term  $-{\hbar q\over{2m_{0}}}\, [\,{\mathbf{B}}{\boldsymbol{\cdot\sigma}}\,]$
as corresponding to the potential
energy ${\boldsymbol{\mu}}_{e}{\mathbf{\cdot B}}$ of a (spin-induced) magnetic dipole ${\boldsymbol{\mu}}_{e}$ within a magnetic field ${\mathbf{B}}$.
But as we have explained in connection with Eq. \ref{Cartan} in Section \ref{intro}, the quantity $-{\hbar q\over{2m_{0}}}\, [\,{\mathbf{B}}{\boldsymbol{\cdot\sigma}}\,]$
represents a {\em pseudo-vector, not a scalar} like  $ -{ q\over{2m_{0}}}({\mathbf{B\cdot}}{\hat{{\mathbf{L}}}})\,\bigone$, whose
scalar character transpires very clearly from the presence of the unit matrix $\bigone$ in it. 
The picture of the magnetic dipole ${\boldsymbol{\mu}}_{e}$ that would be
proportional to a ``spin vector'' ${\mathbf{S}}$
 is based on  the misleading notation ${\mathbf{a}}{\boldsymbol{\cdot\sigma}}$ we warned against above.
The quantity ${\boldsymbol{\sigma}}$
is in reality  the set of the three vectors 
${\mathbf{e}}_{x}$, ${\mathbf{e}}_{y}$, $ {\mathbf{e}}_{z}$.
In the analogy one replaces thus wrongly  the scalar quantity
 $B_{z} {\hat{L}}_{z}\,\bigone  = -\imath\hbar B_{z} (x{\partial\over{\partial y}} - y {\partial\over{\partial x}})\,\bigone$
by the vector quantity $B_{z} {\hat{S}}_{z} = 
B_{z} {\hbar\over{2}}  \sigma_{z}$ (with analogous substitutions for $B_{x} {\hat{L}}_{x}\,\bigone$ and $B_{y} {\hat{L}}_{y}\,\bigone$). 
The notation ${\hbar\over{2}} {\boldsymbol{\sigma}}$ stands for a set of three vectors
while the quantity ${\hat{{\mathbf{L}}}} \bigone$ stands for a set of three scalars. In fact, the three scalars
${\hat{{\mathbf{L}}}} = -\imath\hbar(y {\partial\over{\partial z}} - z  {\partial\over{\partial y}},
z {\partial\over{\partial x}} - x  {\partial\over{\partial z}}, x {\partial\over{\partial y}} - y  {\partial\over{\partial x}})$  are represented in matrix form in the group theory by multiplying them with the unit matrix, 
such that ${\hat{{\mathbf{L}}}}\,\bigone$ becomes a set of three matrices.
But despite their matrix form the quantities  ${\hat{L}}_{j}\,\bigone$ continue to represent  scalars
in the group theory, with an all together
different symmetry than the matrices ${\hbar\over{2}} \sigma_{j}$ which represent vectors.
The term that is interpreted as $-{\boldsymbol{\mu}}_{e} {\mathbf{\cdot B}}$ just cannot be a potential energy as it is a pseudo-vector.
Whereas the algebra used to calculate  the anomalous $g$-factor is exact, such that it correctly reproduces
the experimental results,  the physical  interpretation proposed  is thus mathematically unsustainable, even if it might be intuitively appealing.

Of course these considerations clash ignominiously with accepted notions.
We must therefore insist that the SL(2,${\mathbb{C}}$) and Dirac representations are completely self-consistent formalisms and that their algebra
is a closed system that contains all it needs to contain, such that
it is pointless to attack the conclusion by drawing in considerations that are external to SL(2,${\mathbb{C}}$).
In view of the strong resistance  this conclusion might provoke, we  give further arguments to back it and make a strong case for it (while we will give the correct
interpretation of the term  $-{q\hbar\over{2m_{0}}} {\mathbf{B}}{\boldsymbol{\cdot\sigma}}$ in Section \ref{monopole}).

(1) As we explained above, the quantity ${\hbar\over{2}} {\boldsymbol{\sigma}}$ in the term
$-{\hbar q\over{2m_{0}}}\, [\,{\mathbf{B}}{\boldsymbol{\cdot\sigma}}\,]$ does not represent the spin in SU(2).
It is thus not the spin operator ${\hat{{\mathbf{S}}}}$, 
and it is not correct to transform the term $-{\hbar q\over{2m_{0}}}\, [\,{\mathbf{B}}{\boldsymbol{\cdot\sigma}}\,]$ 
into $- {q\over{m_{0}}} {\mathbf{B\cdot}}{\hat{{\mathbf{S}}}} = -g {q\over{2m_{0}}} {\mathbf{B\cdot}}{\hat{{\mathbf{S}}}}$, with $g=2$.  
 We may note that in the context of the Dirac equation the presence of
the term $-{\hbar q\over{m_{0}}}\, [\,{\mathbf{B}}{\boldsymbol{\cdot\sigma}}\,]$ 
is due to the minimal substitution used to derive the Dirac equation in an electromagnetic field from the free-space
Dirac equation. But this is a  substitution for a point charge, which is why it is called minimal in the first place
(As pointed out in Subsection \ref{correct-minimal}, it  does not even account for the motion of the electron in the laboratory frame).
If we had wanted to account for a potential energy within the magnetic field,  of a magnetic dipole ${\boldsymbol{\mu}}_{e}$
associated with the spin, 
we should have introduced a more complicated substitution, with a term expressing how ${\boldsymbol{\mu}}_{e}$ couples
to the electromagnetic field,  or how it couples to the charge or magnetic dipole moment of another
particle that is also present in its neighbourhood within the magnetic field. As we have not put such spin-related dipole effects 
into the formalism, they cannot come about by magic. 

(2) That there is no spin-related dipole effect in the  term
$-{\hbar q\over{2m_{0}}}\, [\,{\mathbf{B}}{\boldsymbol{\cdot\sigma}}\,]$ can also be appreciated from the fact
that Eqs. \ref{definitions1}-\ref{forces2} and their variant have been derived  without introducing any considerations about spin.
As the terms that occur in these calculations are the same ones as those that occur in the Dirac theory, 
none of the operators in the Dirac theory contains the spin. 
The physical origin of the terms we recollect from the squared Dirac equation is not the presence of spin. The terms only come about
because we treat the problem in full rigor by using relativistic group theory.
Within the Dirac equation,
the spin occurs only implicitly, {\em viz.} inside the spinor wave function (as clearly explained in \cite{Coddens}). It is
 the requirement that the spinor wave function must be an eigenstate of a vector operator ${\mathbf{K}}{\boldsymbol{\cdot\sigma}}$ or a pseudo-vector operator like $-{\hbar q\over{2m_{0}}}\, [\,{\mathbf{B}}{\boldsymbol{\cdot\sigma}}\,]$  (or their four-dimensional analogous expressions in terms of gamma matrices), that forces
the spin to align itself with the vector ${\mathbf{K}}$ or pseudo-vector ${\mathbf{B}}$ in the calculations and is this way responsible for the occurrence of the up and down eigenstates.

(3) We may also appreciate that it would be extremely puzzling if it were true
that we can calculate a magnetic dipole moment ${\boldsymbol{\mu}}_{e}$ produced by the electron spin with fantastic precision in quantum electrodynamics,
without having to specify anything in the calculations about the internal charge-current and
mass distributions  inside the electron. In fact, the presence of such current distributions is
intuitively the only mechanism we know to account for the existence of a magnetic dipole ${\boldsymbol{\mu}}$.
The remark of Lorentz we quoted above shows that using this  mechanism to explain the anomalous Zeeman effect could be wrong despite the fact that it is
intuitively appealing.

(4) As will be pointed out in the Appendix, the interpretation of the orbital Zeeman term $-{\boldsymbol{\mu\cdot}}{\mathbf{B}}$, on which one tries to build here intuition
about the anomalous Zeeman term by analogy, is itself flawed, because there does not exist  such a thing as a potential energy with respect to a magnetic field.
The correct interpretation of  $-{\boldsymbol{\mu\cdot}}{\mathbf{B}}$ will be given in Subsubsection \ref{A-der}.

(5) We may further note that a three-component expression for the spin can never be a complete description within a fully relativistic context, due to the
theorem about the bilinear covariants mentioned in the Introduction. Only covariants with $1$, $4$ or $6$ components exist.
Therefore, either one or three components must be missing in the three-component description. In our work \cite{Coddens}, the
expression for the relativistic generalization of the spin  
operator 
${\hbar\over{2}}\, [{\mathbf{s}}{\boldsymbol{\cdot\sigma}}\,]$ in SU(2), becomes a four-component quantity ${\hbar\over{2}}\, [s_{t}\bigone + {\mathbf{s}}{\boldsymbol{\cdot\sigma}}\,]$ within SL(2,${\mathbb{C}}$).

(6) As we have not introduced spin in Eq. \ref{forces} or its variant, 
all ``dipole'' effects we can expect to obtain from Eq. \ref{forces} or its variant are orbital effects due to moving point charges. 
In fact, we introduce magnetic moments into the formalism
through the term $-{q\over{c}}\,[\,{\mathbf{v}}{\boldsymbol{\cdot\sigma}}\,]$ in Eq. \ref{forces} or its variant.  This can be understood  from  our calculation of a magnetic 
 moment produced by a current  presented in the Appendix. 
The anomalous Zeeman term $-{\hbar q\over{2m_{0}}}\, [\,{\mathbf{B}}{\boldsymbol{\cdot\sigma}}\,]$ can thus not
involve any magnetic dipole moment for the electron that  we are studying: It cannot contain a spin-induced
``intrinsic'' magnetic dipole moment as it can be derived in a context without spin. It also cannot contain any orbital magnetic  ``dipole'' moment as
 it does not contain ${\mathbf{v}}$. 
The anomalous Zeeman effect does indeed not depend on the velocity of the electron. 
In the measurements of the anomalous $g$-factor within a Penning trap \cite{Gabrielse}, we can reduce the velocity of the electron
such that one practically reaches  the limit ${\mathbf{v}} \rightarrow {\mathbf{0}}$. The anomalous Zeeman effect would
still exist.

(7) We may note that the exercise of rewriting $-{\hbar q_{1}\over{ 2m_{0}}}\,[\,{\mathbf{B\cdot}}{\boldsymbol{\sigma}}\,]$ as
 $-({\hat{{\boldsymbol{\mu}}}}_{e}{\mathbf{\cdot B}})\,\bigone$ is not only mathematically flawed but also physically futile.
The wrong interpretation of the algebra is used to  introduce by brute force an intuitive  classical  image,
while the facts of the non-interpreted algebra will eventually force us to give up 
on  that  classical image anyway. 
 The presence of a scalar product  $-{\boldsymbol{\mu}}_{e}{\mathbf{\cdot B}}$
 conjures up the image of an operator
 with  a continuous spectrum, due to the continuum of possible angles between
 ${\boldsymbol{\mu}}_{e}$ and ${\mathbf{B}}$,
 while eventually
 this classical picture of a magnetic dipole in a magnetic field
has to be abandoned as  only the up and down states are observed experimentally. 
 
(8) To  readers who may still feel reluctant to accept the arguments formulated under points (1)-(7) because they fly in the face of the accepted notions, 
we may perhaps  ask  to consider how the inherent profound difficulty of quantum mechanics forces us  to teach it
  a little bit  like a religion. 
A nice example of a quasi-religious mystery
is the concept of  particle-wave duality. With some misplaced irony
one could compare it with the Christian dogma of the mystery of the Holy Trinity. 
Three different persons (the Father, the Son and the Holy Ghost),
 are  claimed to be only one God, 
and one is invited to accept this puzzling postulate as a factual truth, for which will not be  given any further explanation because it is a mystery.
 In complete analogy, an electron is postulated to be both a particle and a wave.
 This is also quite  puzzling a notion, for which we are told that we should
accept it as a quantum mystery \cite{drops}. One could claim sarcastically  that the two mysteries resemble one another as two peas in a pod.
Of course, there is a very essential point that makes all the difference between the quantum mystery and the religious mystery,
and justifies why we can postulate that one has to accept the quantum mystery without further asking and just ``shut up and calculate''.  That point is the agreement
of the theory with experimental evidence.
Quantum mechanics passes the test of the comparison with  experimental data with flying colors by grinding out  all the correct answers with impressive precision.

Asking if a physical theory provides the correct answers is indeed a crucial criterion to assess its value.
But a formalism that turns out the correct answers is a far cry from a theory
if it is mathematically flawed. 
 The points (1)-(7) show that with respect to the criterion of mathematical
self-consistency, the traditional interpretation of the anomalous $g$-factor of the electron in quantum mechanics is not even an option.
What is wrong in this respect   is not the algebra itself, which is correct, such that it
indeed turns out the right answers.
It is the {\em interpretation} routinely given to that algebra that is absurd as it violates the mathematics, even if the algebraic expression 
$-{\hbar q\over{2m_{0}}}\, [\,{\mathbf{B}}{\boldsymbol{\cdot\sigma}}\,]$ correctly accounts for  the anomalous Zeeman effect.
On this and countless other occasions quantum mechanics has time and again
overruled the logically peremptory, true geometrical meaning of the group theory explained in Section \ref{intro} in favour of
a self-cooked  parallel interpretation \cite{Coddens}. 
The argument that we should accept this because
the theory turns out the right answers does not hold sway.
What is confirmed by the experiments is just the algebra, not the interpretation of that algebra.
The real anathema resides in interpreting  $-{\hbar q\over{2m_{0}}}\, [\,{\mathbf{B}}{\boldsymbol{\cdot\sigma}}\,]$ as a scalar, not
in questioning the traditional orthodoxy.

Of course these observations also raise the question how we must define the spin operator
if it is not given by  ${\hat{{\mathbf{S}}}} = {\hbar\over{2}} {\boldsymbol{\sigma}}$.
We may mention that it is possible to preserve the physical image of the spin as a vector ${\hbar\over{2}}\,{\mathbf{s}}$,
by using a different definition $ {\hbar\over{2}}\,[\, {\mathbf{s}}{\boldsymbol{\cdot\sigma}}\,]$ for the spin operator than
  ${\hat{{\mathbf{S}}}} = {\hbar\over{2}} {\boldsymbol{\sigma}}$. This approach can be developed without changing a iota to the results of Pauli's spin calculus, such that
it leads to an identical agreement with experimental data.  It even  presents  less
problems of interpretation,  but the issue entails a whole domino chain of related questions and answers, whose development and discussion
are  beyond the scope of any paper of reasonable length \cite{Coddens}. 

\subsubsection{Difficult questions} \label{issues}

The problem with this discussion of the anomalous Zeeman effect is that
without the traditional interpretation, 
the physics related to the non-relativistic Zeeman term $-{\hbar q\over{2m_{0}}}\, [\,{\mathbf{B}}{\boldsymbol{\cdot\sigma}}\,]$
become mysterious. The problem is two-fold. There is a classical enigma, as the interaction
 exists in principle already within classical
electromagnetism, as the calculations given in Eqs. \ref{definitions1}-\ref{forces2} or their variant clearly illustrate. The question is then
of course what the classical meaning of this  classical effect is supposed to be.
There is also a quantum mechanical enigma. In fact, the two energy levels
observed in the anomalous Zeeman splitting are traditionally  interpreted as corresponding to the spin-up and spin-down states. But how can we
explain the anomalous Zeeman splitting if the derivation of
the term  $-{\hbar q\over{2m_{0}}}\, [\,{\mathbf{B}}{\boldsymbol{\cdot\sigma}}\,]$ 
does not rely on a notion of spin?

\subsubsection{Not a dipole-dipole but a charge-dipole interaction symmetry} \label{q-D}

The first step in our attempts  to make sense of  this puzzling situation is as follows.
The Lorentz transformation for the electromagnetic field under a boost with velocity ${\mathbf{v}}$ can be written as:

\begin{equation} \label{relativity}
{\mathbf{E}}_{\parallel}' = {\mathbf{E}}_{\parallel}, \quad {\mathbf{B}}_{\parallel}' = {\mathbf{B}}_{\parallel}, \quad {\mathbf{E}}'_{\perp} = \gamma({\mathbf{E}}_{\perp} + {\mathbf{v}}\wedge {\mathbf{B}}_{\perp}), \quad {\mathbf{B}}'_{\perp} = \gamma({\mathbf{B}}_{\perp} -{1\over{c^{2}}} {\mathbf{v}}\wedge {\mathbf{E}}_{\perp}).
\end{equation}

\noindent Here the indices $\perp$ and $\parallel$ are with respect to ${\mathbf{v}}$. Starting from a rest frame without magnetic field (${\mathbf{B}} = {\mathbf{0}}$), we
obtain a magnetic field   (${\mathbf{B}}' \neq {\mathbf{0}}$) in a moving frame, such that a magnetic field can be viewed 
(schematically) as a relativistic byproduct of the electric field.
One could introduce the philosophy that only the Coulomb field really exists and that the magnetic field is just an optical illusion.
By adopting this viewpoint, we can see that it is the macroscopic quantity ${\mathbf{B}}$ that contains  
a set of  terms $q{\mathbf{v}}_{j}$ whose sum can be associated with
a macroscopic magnetic dipole moment ${\boldsymbol{\mu}}$.
The whole of Eq. \ref{forces2} would in essence  just be due to a calculation:

\begin{equation*} 
(\, q_{1} \bigone + {q_{1}\over{c}}[\,{\mathbf{v}}_{1}{\boldsymbol{\cdot\sigma}} \,]\,)\,\sum_{j}(\, q_{2j} \bigone - {q_{2j}\over{c}}\,[\,{\mathbf{v}}_{j2}{\boldsymbol{\cdot\sigma}} \,]\,)
= 
\end{equation*}
\begin{equation} \label{charge-current-charge-current-interaction}
\sum_{j} (\, q_{1}q_{2j} \bigone 
                + {q_{1}q_{2j}\over{c}} [\,{\mathbf{v}}_{1}{\boldsymbol{\cdot\sigma}}\,] 
                -  {q_{1}q_{2j}\over{c}} [\,{\mathbf{v}}_{2j}{\boldsymbol{\cdot\sigma}}\,] 
                -   {q_{1}q_{2j}\over{c^{2}}} \,({\mathbf{v}}_{1}{\mathbf{\cdot v}}_{2j})\bigone
                 - \imath {q_{1}q_{2j}\over{c^{2}}} \,[\, ({\mathbf{v}}_{1} \wedge {\mathbf{v}}_{2j}){\boldsymbol{\cdot\sigma}}\,]\,).
\end{equation}

\noindent Here $q_{1}$ and ${\mathbf{v}}_{1}$ are the parameters of the electron we are studying, after
putting it in the electromagnetic field $({\mathbf{E}}, {\mathbf{B}})$, which is generated by a distribution of charged particles
described by the parameters $q_{2j}$ and ${\mathbf{v}}_{2j}$.
The many terms ${\mathbf{v}}_{2j}$ are hidden (with their coupling terms) in the macroscopic quantity
${\mathbf{B}}$, and the terms $q_{2j}$ in ${\mathbf{E}}$ or ${\mathbf{B}}$. 
The macroscopic quantities ${\mathbf{E}}$ and ${\mathbf{B}}$ are only tools to express  interactions of charges and currents with other charges and currents.
We can then ``derive''  the structure of Eq. \ref{forces2}  from the backbone of Eq. \ref{charge-current-charge-current-interaction},
by replacing $\sum_{j}\,q_{2j}{\mathbf{v}}_{2j} \rightarrow c{\mathbf{B}}$ and for the remaining terms 
$q_{2} \rightarrow {\mathbf{E}}$. The term ${q_{1}q_{2j}\over{c}} [\,{\mathbf{v}}_{1}{\boldsymbol{\cdot\sigma}}\,] $ will lead 
to two contributions ${q\over{c}} \,({\mathbf{v\cdot E}})\,\bigone$ and  ${q\over{c}}\, ({\mathbf{v}}\wedge {\mathbf{E}}){\boldsymbol{\cdot\sigma}}$, because ${\mathbf{E}}$ will be defined by ${\mathbf{r}}_{2j}-{\mathbf{r}}_{1}$. 
To make a rigorous derivation one would of course have to weight the various contributions
with their coupling terms.

It can be seen this way that it is much more logical to consider the anomalous Zeeman effect as a term that accounts for the interaction of the charge of the electron 
with the magnetic dipoles produced by the current loop of the moving electrons  which generate
 the magnetic field. That such an effect exists is shown by the classical derivation of Eq. \ref{forces2} wherein the term containing ${\mathbf{B\cdot}}{\boldsymbol{\sigma}}$
 does not allow for any ``intrinsic'' magnetic dipole moment  on behalf of an electron of charge $q$ we subject
 to the magnetic field. It does not make sense to invoke another
 physical mechanism for  the same term in the Dirac equation.
 The  origin of the term containing $q{\mathbf{B}}$ is obvious. 
 We might  initially just consider  the force $[\,q\bigone \,] [\,{\mathbf{E}}{\boldsymbol{\cdot\sigma}}\,]$ instead of Eq. \ref{forces}
 and then generalize both terms independently by Lorentz symmetry.
We  recover then the general expression of Eq.  \ref{forces}, wherein  ${\mathbf{B}}$ is no longer zero. 
We must thus conclude that  there is in general an interaction
  between an electric point charge and a magnetic field.
 The pre-factor $-{\hbar q\over{2m_{0}}}$ in the anomalous Zeeman term
$-{\hbar q\over{2m_{0}}}\,[\,{\mathbf{B\cdot}}{\boldsymbol{\sigma}}\,]$  has  been
used  to attribute a non-classical magnetic dipole moment ${\boldsymbol{\mu}}_{e}$  to the charge $q$, 
while in reality the term $-{\hbar q\over{2m_{0}}}\,[\,{\mathbf{B\cdot}}{\boldsymbol{\sigma}}\,]$ corresponds
to a charge-dipole interaction of a point charge $q$ with the macroscopic dipole ${\boldsymbol{\mu}}$ corresponding to the
current loops that produce ${\mathbf{B}}$.
We can see that the anomalous value $g=2$ is just due to the introduction of 
${\hbar\over{2}}$ with the aim  to recover ${\hbar\over{2}}{\boldsymbol{\sigma}}$ as described above.
 In reality, if we take the liberty to continue to use the textbook misnomer ``magnetic dipoles''
for the magnetic moments of single moving  electrons, then
  all ``magnetic-dipole''  effects in the formalism are orbital, and there is no 
spin-induced magnetic dipole moment in the algebra.
The term in the Dirac equation that gives rise to the anomalous Zeeman term belongs to the symmetric counterpart  
$cq{\mathbf{B}}{\boldsymbol{\cdot\sigma}} + {q\over{c}} ({\mathbf{v}}\wedge {\boldsymbol{E}}){\boldsymbol{\cdot\sigma}}$, 
of the Lorentz force $q({\mathbf{E}} + {\mathbf{v}}\wedge {\mathbf{B}}){\boldsymbol{\cdot\sigma}}$
where the r\^oles of the electric and magnetic fields have been exchanged. 
Within the expression 
$cq{\mathbf{B}}{\boldsymbol{\cdot\sigma}} + {q\over{c}} ({\mathbf{v}}\wedge {\boldsymbol{E}}){\boldsymbol{\cdot\sigma}}$ itself, the term
$cq{\mathbf{B}}{\boldsymbol{\cdot\sigma}}$ is the symmetric counterpart of
 $ {q\over{c}} ({\mathbf{v}}\wedge {\boldsymbol{E}}){\boldsymbol{\cdot\sigma}}$ 
 wherein the r\^oles of charge and ``magnetic dipoles''  have been exchanged. Within $cq{\mathbf{B}}{\boldsymbol{\cdot\sigma}}$
the charge of the electron interacts with the ``magnetic dipoles'' of the moving electrons that generate the magnetic field.
In the term $ {q\over{c}} ({\mathbf{v}}\wedge {\boldsymbol{E}}){\boldsymbol{\cdot\sigma}}$ the ``magnetic dipole''  generated by the moving
electron interacts with the charges of the electric field.
The symmetry of the anomalous Zeeman effect is intrinsically different from that of the orbital
Zeeman effect, as the latter depends on two velocities, rather than one velocity. In other words, the orbital Zeeman effect corresponds to a ``dipole''-dipole interaction,
while the anomalous Zeeman effect stems from a charge-dipole interaction and the spin-orbit coupling to a ``dipole''-charge interaction. 
This shows clearly that  the anomalous $g$-factor
should not be visualized  in terms of a ``dipole''-dipole interaction, as Dirac has done.

\section{The Physical Meaning of the Anomalous Zeeman Effect} \label{monopole}

\subsection{Introduction} \label{Intro-prec}

 Of course, getting a feeling for the anomalous Zeeman effect and explaining why it leads to two Zeeman levels
  is  more complicated, even if it follows very clearly from the algebra. All terms in Eq. \ref{forces2} we ``understood'' classically 
  are in reality facts of life that we had to accept after discovering them experimentally and to which we got used. 
  The anomalous Zeeman effect and the 
  spin-orbit 
  coupling should thus
  be considered on the same footing. However, the spin-orbit term can be understood in terms
  of precession, as explained in Subsection \ref{discussion}. It is a current-charge interaction, and therefore the symmetrical counterpart of the
  anomalous Zeeman effect, which is of the charge-current type. Let us therefore check if we can also interpret the anomalous Zeeman effect in terms of
  precession. To do this we need a more detailed understanding of the magnetic potential and its vorticity.
  
\subsection{The magnetic potential} \label{A-pot}

\subsubsection{Derivation} \label{A-der}

How do we deal with the  kinetic energy of a moving charge  in circular motion when we describe it as a stationary magnetic phenomenon? As we will show, it is done by expressing
the kinetic energy in terms of a fake potential energy $U$.  Potential energy is a scalar. As the moving charge corresponds to a current $q{\mathbf{v}}$,
which is a vector quantity, the only way to create a scalar quantity out of this current  is to combine it with another
vector quantity ${\mathbf{A}}$ into a scalar product, e.g. $U =  -q\,({\mathbf{v \cdot A}})$. 
For the moment we consider ${\mathbf{A}}$ as a general vector that has not yet been specified any further. 
In a constant magnetic field ${\mathbf{B}} = B{\mathbf{e}}_{z}$, with $B > 0$
the moving charge $q < 0$ whose velocity ${\mathbf{v}} = v{\mathbf{e}}_{y}$, $v>0$ corresponds to a current $q{\mathbf{v}}$,  will perform
a uniform circular motion at the cyclotron frequency $\omega_{c} >0 $, which in the non-relativistic limit is given by $\omega_{c} = - {qB\over{m}}$.
The velocity is then $v = - {qBr\over{m}}$.
Let us  rewrite $-q{\mathbf{v \cdot A}}$ as $-q{\mathbf{A \cdot v}} = -{q\over{m_{0}}}{\mathbf{A}}{\mathbf{\cdot p}}$.
To make this correspond to the kinetic energy ${\mathbf{p}}^{2}/2m_{0}$ we must thus have $-{q\over{m_{0}}}{\mathbf{A}} = {\mathbf{p}}/2m_{0} =
 {\mathbf{v}}/2 = -{qB\over{2m_{0}}}r{\mathbf{e}}_{y}$. From this it follows that ${\mathbf{A}} = $  ${1\over{2}} rB {\mathbf{e}}_{y}$ $ = -{1\over{2}} {\mathbf{r}} \wedge {\mathbf{B}}$. This is exactly the magnetic vector potential that corresponds to a constant magnetic
field ${\mathbf{B}}$ as can be checked by calculating ${\mathbf{B}} = {\mathbf{\nabla}} \wedge {\mathbf{A}}$.
Using ${\mathbf{A}} =  -{1\over{2}} {\mathbf{r}} \wedge {\mathbf{B}}$ it is easy to rewrite  $U =  -q\,({\mathbf{v \cdot A}})$ as $U = {\boldsymbol{\mu\cdot}}{\mathbf{B}}$,
where ${\boldsymbol{\mu}} = -{q\over{2m_{0}}}{\mathbf{L}}$.
This derivation does not require any consideration of a current loop.
We see thus that the intuitive picture we use for this term in terms of a true potential energy is wrong, as anticipated under point (4) in Subsubsection
\ref{anomaly-g-problem}.

The sign used in the expression  $-q{\mathbf{v \cdot A}}$ may surprise, but we must remind here the reason why we introduce the 
minimal substitution.
In the absence of a magnetic field, the correct parameters to write the Lorentz transformation would be $(E-qV({\mathbf{r}}), c{\mathbf{p}})$. 
In a first non-relativistic approximation the quantity $E - qV$ we want to obtain 
becomes $E - qV \approx m_{0}c^{2} + {{\mathbf{p}}^{2}\over{2m_{0}}} - qV = m_{0}c^{2} + T - U$, just like in Lagrangian dynamics,
where one justifies  its introduction merely by showing that it makes things work. 
In the case of a purely magnetic field, there is no electric potential, such that this expression becomes then $m_{0}c^{2} + {{\mathbf{p}}^{2}\over{2m_{0}}}$
in a frame  wherein the centre of the orbit is at rest, such that it can be considered
as a ``static'' magnetic situation. The expression we want to obtain requires thus adding the kinetic energy,
and this can be achieved by
 subtracting the ``potential energy'' of the current, which is the negative kinetic energy, as defined above. 
 
 Of course the preceding lines only explain the case when ${\mathbf{v}}\parallel {\mathbf{A}}$, and not the cosine term in $U =  -q\,({\mathbf{v \cdot A}})$.
It is less easy to grasp the meaning of the cosine term in $U =  -q\,({\mathbf{v \cdot A}})$. In principle, in a constant magnetic field
 ${\mathbf{v}}$ must be parallel to ${\mathbf{A}}$. Therefore, if ${\mathbf{v}} \not\parallel {\mathbf{A}}$, the motion of  charged particle must be a forced
motion with respect to the magnetic field.
 Let us therefore consider a charge in uniform motion on a circular orbit with a velocity ${\mathbf{v}}$ in  a constant magnetic field ${\mathbf{B}} = B_{z} {\mathbf{e}}_{z}
 +  B_{x} {\mathbf{e}}_{x}$. This motion takes place in a plane perpendicular to ${\mathbf{B}}$. 
 We will then have
 ${\mathbf{A}} = -{1\over{2}} {\mathbf{r}} \wedge {\mathbf{B}} =  -{1\over{2}} {\mathbf{r}} \wedge B_{z}  {\mathbf{e}}_{z}$  
 $ -{1\over{2}} {\mathbf{r}} \wedge B_{x}{\mathbf{e}}_{x}$. Put ${\mathbf{A}}_{1} =  -{1\over{2}} {\mathbf{r}} \wedge B_{z}  {\mathbf{e}}_{z}$,
  ${\mathbf{A}}_{2} =  -{1\over{2}} {\mathbf{r}} \wedge B_{x}  {\mathbf{e}}_{x}$, such that ${\mathbf{A}} = {\mathbf{A}}_{1} + {\mathbf{A}} _{2}$.
  It is then obvious that ${\mathbf{A\cdot v}} = {\mathbf{A}}_{1}{\mathbf{\cdot v}} +  {\mathbf{A}}_{2}{\mathbf{\cdot v}}$.
This explains why the  part of the kinetic energy corresponding to the velocity ${\mathbf{v}}$ that one can  attribute to  ${\mathbf{B}}_{1} = B_{z} {\mathbf{e}}_{z}$  is given 
  by $ {\mathbf{A}}_{1}{\mathbf{\cdot v}}$. If we consider ${\mathbf{B}}_{1}$ and we observe the circular orbit with velocity ${\mathbf{v}}$,
  then only the part $ {\mathbf{A}}_{1}{\mathbf{\cdot v}}$ of the kinetic energy can be attributed to ${\mathbf{B}}_{1}$. The rest must
  be attributed to the second force which forces the charged particle to follow an orbit in a plane that is not perpendicular to ${\mathbf{B}}_{1}$,
  and which in this analysis is based on ${\mathbf{B}}_{2}$. But in other situations, the origin of the force could be a different kind of
   field than a magnetic field ${\mathbf{B}}_{2}$, e.g.
  an electric field. It is only in such cases that a negative cosine term (leading to a ``negative kinetic energy'') can have physical meaning.
  The field could also be varying with time rather than a constant field as in our example of ${\mathbf{B}}_{2}$.

The magnetic vector potential ${\mathbf{A}}$ is often presented as a meaningless mathematical quantity that has just been introduced to simplify the calculations
and whose use is justified because it makes things work.
After the discovery of the Aharonov-Bohm effect, this viewpoint has been challenged by Feynman \cite{Feynman} who stated that ${\mathbf{A}}$ is for quantum mechanics more
significant than ${\mathbf{B}}$. In relativity, $c{\mathbf{A}}$ builds a four-vector with $V$, such that its meaning should be
as physical as the meaning of $V$, and conceptually related to it.
Despite all this, the vector potential has remained a  concept that is not very intuitive.  Here we have discovered a clear meaning for it.
As often pointed out, ${\mathbf{A}}$ is only
defined up to a constant, just like $V$. Its meaning is thus indeed as clear as the meaning of $V$, and
both quantities are quite intuitive. Several authors \cite{vector-pot}-\cite{vector-pot2} have tried
to make a case for this viewpoint, based e.g. on an experiment by Blondel \cite{Blondel}.

\subsubsection{Larmor frequency} \label{Larmor}

The kinetic energy   ${p^{2}\over{2m_{0}}} = {1\over{2}} m_{0} \omega_{c}^{2}  r^{2}$ of the electron on the circular orbit can also be written in terms of the angular momentum $L= m_{0} \omega_{c}  r^{2} $ as ${p^{2}\over{2m_{0}}}  = {1\over{2}} \omega_{c} L $.
To express the kinetic energy for a particle with an angular momentum $L$ in the form $L \omega$, we must thus not
use  the true cyclotron frequency $\omega_{c}$  for $\omega$, but the fictive Larmor frequency $\omega_{L} = \omega_{c}/2$. This quantity 
pops up in all quantum mechanical calculations. One may feel tempted to infer from 
spotting this quantity in the equations that
the orbital rotational motion in the physical problem studied  is happening at the frequency
$\omega_{L}$ instead of $\omega_{c}$, which is quite puzzling. One might ask oneself why the electron is turning slower on its orbit
than one might expect based on classical mechanics. Is this just one more quantum mystery? The solution to this riddle is thus that $\omega_{L}$ is only
an auxiliary quantity to simplify the notations.

There is another way to highlight this point that the Larmor frequency is just an auxiliary concept.
Larmor's construction starts from the calculation of  the fictitious forces in a rotating frame. In this frame there
will be a fictitious centrifugal force and a fictitious Coriolis force.
The expression for the force in the moving frame is $m_{0}( {d^{2}{\mathbf{r}}\over{dt^{2}}} + 2 {\boldsymbol{\omega}}\wedge {d{\mathbf{r}}\over{dt}} +
 {\boldsymbol{\omega}}\wedge( {\boldsymbol{\omega}}\wedge {\mathbf{r}}))$.
He then expresses that there should be a frequency of rotation where the magnetic field is completely canceled by the fictitious forces.
This is the Larmor frequency. The Larmor frame ``erases'' the magnetic field. In this frame there will be
no position and no velocity that make the particle feel the presence of the magnetic field or the fictitious  Coriolis  and  centrifugal forces
 because all these forces cancel each other exactly.
Of course 
in every point of the inertial  frame the particle is not at rest but moving at a velocity ${\mathbf{v}}({\mathbf{r}})$ rather than being at
rest with a velocity ${\mathbf{v}}' = {\mathbf{0}}$. 
The relation between the velocities
in the two frames is ${\mathbf{v}}({\mathbf{r}}) = {\mathbf{v}}'({\mathbf{r}}) + {\boldsymbol{\omega}} \wedge {\mathbf{r}}$.

It looks contradictory that the orbital motion of a particle in a magnetic field takes place at the cyclotron frequency, 
while we state that the particle does not  feel the magnetic force in the frame that is rotating at the Larmor frequency.
Following the idea that the particle does not feel the magnetic force it would appear that the particle could be at rest in the Larmor frame,
while following from what we know in the inertial frame,
the particle should appear to be moving at a residual frequency $\omega_{L} = \omega_{c} - \omega_{L} = \omega_{c}/2$ in the Larmor frame.
If the particle has a residual velocity in the Larmor frame,
should it then
not have to feel an attractive force that ensures that it stays on the circular orbit at this residual velocity? 
As Larmor's construction shows, it does not erase the centrifugal force  $ m_{0} \,{\boldsymbol{\omega}}\wedge( {\boldsymbol{\omega}}\wedge {\mathbf{r}})$.
It still remains present under the form $-{q^{2}\over{2m_{0}}}{\mathbf{B}}\wedge({\mathbf{r}}\wedge{\mathbf{B}})$
in the rotating frame. When there is a central electric force present in the frame that is much larger, then  we can neglect the term
$-{q^{2}\over{2m_{0}}}{\mathbf{B}}\wedge({\mathbf{r}}\wedge{\mathbf{B}})$. It is this possibility to neglect the term which leads
to Larmor's idea that we erase all fictitious forces. But of course  when there is no electric field we can no longer neglect the  term
$-{q^{2}\over{2m_{0}}}{\mathbf{B}}\wedge({\mathbf{r}}\wedge{\mathbf{B}})$.
In this expression we can substitute ${1\over{2}} {\mathbf{r}}\wedge{\mathbf{B}} = {\mathbf{A}}$. Here   $-{q\over{m_{0}}} {\mathbf{A}} $
 corresponds to ${\mathbf{v}}_{*} = -{\mathbf{r}} \wedge {\boldsymbol{\omega}}_{L}$. We obtain thus $  {q\over{2}} {\mathbf{v}}\wedge {\mathbf{B}}$ which is exactly the missing force, that will yield the missing
frequency $\omega_{L}$. This solves the paradox classically. We can also solve the paradox relativistically.
Relativity shows that there is also an electric field in the Larmor  frame, due to the local Lorenz transformation
(with  boost vector  ${\mathbf{v}}'({\mathbf{r}})$) of the magnetic field in the inertial frame (see Eq. \ref{relativity}).
This electric field exerts a local attractive and central force on the particle.
 It is this electric field that is responsible for the fact that
the particle that is in cyclotron motion in the inertial frame is not at rest in the Larmor frame but moving at the residual frequency 
$\omega_{L} = \omega_{c} - \omega_{L} $. The two solutions of the paradox coincide in the non-relativistic limit.

That the Larmor frequency appears in the equations is due to the fact that we must divide all terms that occur in the squared Dirac equation by
$2m_{0}c^{2}$ to be able to reduce it to the Pauli equation. We see here the same factor of $2$ entering the calculations as
the one that occurs in the derivation of the expression of the magnetic potential. It is the factor $2$ that occurs in the expression ${p^{2}\over{2m_{0}}}$.
And in both cases, the true frequency of the motion is $\omega_{c}$.
An analysis of the original meaning of the Larmor frequency shows that drawing the conclusion that the true frequency would be $\omega_{L}$
rather than $\omega_{c}$, just
because this is the quantity that comes to the fore in the calculation, is wrong because it fails to discern
that the rotating frame introduces an electric field.

\subsubsection{Larmor precession as the vorticity of the magnetic potential} \label{Larmor-vortex}

The traditional minimal substitution is given by:
 
\begin{equation} \label{minimini}
E \rightarrow E - qV, \quad {\mathbf{p}} \rightarrow {\mathbf{p}} - q{\mathbf{A}}.
\end{equation}

\noindent Non-relativistically we can write

\begin{equation} \label{velocity-field}
{\mathbf{p}} - q{\mathbf{A}} = m_{0}({\mathbf{v}} -{q\over{m_{0}}}\,{\mathbf{A}}).
\end{equation}

\noindent From this we can appreciate that $-{q\over{m_{0}}}\,{\mathbf{A}}({\mathbf{r}})$ behaves as a velocity field ${\mathbf{v}}_{*}({\mathbf{r}})$.
For a constant magnetic field ${\mathbf{B}}$ we have ${\mathbf{A}} = -{1\over{2}} {\mathbf{r}} \wedge {\mathbf{B}}$.
On a first contact this looks a bit mysterious, because we can choose the origin at will and calculate ${\mathbf{r}}$ with respect
to this origin, the result will always be correct and independent from the choice of origin. But this fact  is well known under the name of  gauge invariance 
and expresses the fact that the vector potential is defined up to an arbitrary constant.
As the scalar potential and the vector potential are related to each other by Lorentz transformation, it is evident that we cannot choose  both arbitrary constants
simultaneously at will, which is why they are related through a gauge condition.
We will further discuss this arbitrary constant later on in our discussion.
For the constant magnetic field, the velocity field  $-{q\over{m_{0}}}\,{\mathbf{A}}({\mathbf{r}})$ is  of the type:

\begin{equation} \label{velocity-field-2}
{\mathbf{v}}_{*}({\mathbf{r}}) =+ {\mathbf{r}} \wedge {q\over{2m_{0}}}\,{\mathbf{B}} = - {\mathbf{r}} \wedge {\boldsymbol{\omega}}_{L},
\end{equation}

\noindent where we have introduced:

\begin{equation} \label{rotation1}
 {\boldsymbol{\omega}}_{L} =   - {q\over{2m_{0}}}\,{\mathbf{B}}.
\end{equation}

\noindent In other words, in the non-relativistic approximation,  the velocity field is the same as the one we would observe in a frame
that is rotating at an angular velocity ${\boldsymbol{\omega}}_{L}$  corresponding to the Larmor frequency.
According to the thoughts  described above, we can consider this as a fictive auxiliary quantity. The rotating frame we discover here
would then be rotating at a fictive frequency, while the true rotational motion
would be the cyclotron frequency.

Let us now consider the velocity field  ${\mathbf{v}}({\mathbf{r}})$ of a liquid
and imagine that we would put a paddlewheel inside this liquid. The velocity field has vorticity and it will be make the paddlewheel turn as discussed in
\cite{MIT}. In this reference Auroux considers a circular path of radius $r$ on which he calculates the average velocity ${\mathbf{v}}$ of the paddlewheel. 
He then takes the limit $r\rightarrow 0$. The paddlewheel becomes then infinitesimal.
According to this calculation,  the infinitesimal paddlewheel will rotate at a frequency:

\begin{equation} \label{rotation-general}
{\boldsymbol{\omega}} =  {1\over{2}} \nabla \wedge {\mathbf{v}}({\mathbf{r}}).
\end{equation}

\noindent When ${\mathbf{v}} = {\boldsymbol{\omega}} \wedge {\mathbf{r}}$,  such that the liquid bodily rotates with the frequency ${\boldsymbol{\omega}}$,
the paddlewheel will locally spin with the same frequency  as the rotation of the liquid body. We have a daily-life
confirmation of this idea on a merry-go-round. 
Such a merry-go-round is actually a better realization of the idea of a liquid that bodily  rotates at a constant frequency
than a real-life liquid where the angular frequencies might be radius-dependent. Let us associate the merry-go-round with a global rotating frame.
When you are on a merry-go-round your personal local frame will not describe
a circular motion with the directions of its axes $x,y,z$  fixed. The axes of your local  frame are also rotating: Your local frame exhibits precession.
Most of the time we gloss over of this fact by using a reference frame based on a local basis ${\mathbf{e}}_{\phi}, {\mathbf{e}}_{r}$ 
in a symmetry-adapted system of curvilinear coordinates,
which already incorporates this effect.
But the fact that this is a varying basis must be taken into account in the mathematics, and motivates the introduction of covariant derivatives.
The argument of the merry-go-round just expresses Berry's phase \cite{Berry} on a circular orbit which is  a geodesic.

Imagine now that you have a point-like charged particle without spin that is put into circular orbital motion
within a magnetic field. It is then conceivable that the merry-go-round effect could give it  spin, whereby
the angular frequency of that spin  would be the cyclotron frequency. That spinning motion would have a kinetic energy $L \omega_{c}/2= L\omega_{L}$,
where again the Larmor frequency is an auxiliary quantity.
The spinning motion is intrinsic and if we wanted to describe the spinning point particle geometrically, we would have to describe
it as a spinning point. But in geometry, points are not spinning. However, we can describe the spinning motion in terms of vorticity.  
We can treat the Larmor frequency geometrically
by using the same reasoning on ${\mathbf{v}}_{*}$ as on ${\mathbf{v}}$.
We see thus that the ``vorticity''  of  the velocity field of a constant magnetic field is such that a charged point-like particle will spin in the field
at the cyclotron frequency. Any energy calculations that could be based on the equation $E=\hbar\omega$ will have to be done using the Larmor frequency 
and this Larmor frequency can be calculated
from:

\begin{equation} \label{rotation2}
{\boldsymbol{\omega}}_{L} =  {1\over{2}} \nabla \wedge {\mathbf{v}}_{*}({\mathbf{r}}) = -{1\over{2}} \nabla \wedge {q\over{m_{0}}}{\mathbf{A}}({\mathbf{r}}) = -{q\over{2m_{0}}}\,{\mathbf{B}}.
\end{equation}

Of course, in the case of a magnetic field, there is no real liquid that would push on the charge like on a paddlewheel, but we can obtain the results also without
assuming the presence of a liquid. 
The idea of a liquid is only introduced
 to obtain a velocity field. As  we have already a velocity field $-{q\over{m}} {\mathbf{A}}$, we can dispense with the introduction of a  liquid to obtain the mathematical
 result derived. The idea is to describe precession induced by circular motion. Such precession is certainly
 considered in magnetism, and one even calculates relativistic corrections for it in terms of Thomas precession.
 
Within a frame in global rotation, the physical effects of rotation that affect the paddlewheel or the charge are independent from the choice of the origin for the local frame. 
This corresponds to the notion that the magnetic potential
is defined up to a constant. Of course, the force that is needed to counterbalance the ``centrifugal force'' of the rotating frame is provided by the Lorentz force, and
this ``centrifugal force'' will indicate what the true centre of the rotation is.
For the calculation of the physical effects of precession however, it does not matter where the centre of the fictive merry-go-round is.
We could take any point as the centre of rotation,
it would not make any difference.
We may also note that in the limit $r\rightarrow 0$, we reduce the circular current $q{\mathbf{v}}$ to a point-like monopole, exactly according to the idea
introduced above.

Let us further explore these  ideas we used to define monopoles and vector potentials and make the radius of the circular motion
 of the charged particle in the magnetic field shrink to zero. What will happen then?
As the non-relativistic cyclotron frequency is independent from the radius of the circular motion, this will leave us
 with a point-like spinning motion, even for a spin-less charged particle at rest!
    A charge in circular motion on an orbit with a diameter that is so small that we cannot see it with the naked eye will be a hidden motion.
This hidden motion can be treated by the previous arguments, which  lead  to the idea that a magnetic field could make a spin-less particle 
with charge $q$ at 
rest  spin at an angular frequency ${\boldsymbol{\omega}}_{c} = 
  - {q\over{m_{0}}}\,{\mathbf{B}}$. 
The point-like hidden motion 
 can be treated as  the interaction between the  scalar charge $q$ and the vorticity of magnetic potential ${\mathbf{A}}$,
such that we end up with an interaction  of the electric charge with the magnetic field
${\mathbf{B}}$. 
 Energy calculations have to be done however with ${\boldsymbol{\omega}}_{L} =  - {q\over{2m_{0}}}\,{\mathbf{B}}$.
 However, this energy $\hbar\omega_{L}$ will not be the kinetic energy of the orbit shrunk to zero because this is also zero.
 It is the energy due to the precession.
 The moment of inertia that intervenes in the precession is not related to a mass distribution
 that corresponds to a point mass at a distance $r$ from the centre of the circular orbit. It is
 the moment of inertia of the mass distribution inside the spinning top that visualizes the spinning electron (see Subsection \ref{precession}).
Simultaneously, a point-like monopole is not the magnetic charge of a circular current loop whose radius shrinks to zero.
In taking this limit, $v \rightarrow 0$, such that there is no current or magnetic charge left, and no true
magnetic monopole. As also $L  \rightarrow 0$, there is also no magnetic moment left.
 It is the precession that corresponds to the magnetic monopole.
The quantity $cq$ can be symbolically identified with a monopole $q_{m}$, like we have done.
The concept of the magnetic monopole  is useful to distinguish a geometrical point modeling a point charge from a spinning point charge.
We may note that the precession terms which correspond to the anomalous Zeeman effect and the spin-orbit coupling
are thus both related to the hidden rotational energy of the hidden rotational
motion. This explains thus why the terms in Eq. \ref{monopole-force}  
 exist and how they can be associated with magnetic monopoles, such that their classification is correct.
 
 We may finally note why the merry-go-round metaphor fails for the spin-orbit term. In fact, for a circular motion
 in a central Coulomb field ${\mathbf{F}} = {q_{1}q_{2}\over{4\pi\epsilon_{0}r^{3} }} {\mathbf{r}}$, the rotational frequency
 $\omega({\mathbf{r}})$ is not a constant, but depends on ${\mathbf{r}}$, such that the image of a liquid that
 is bodily rotating no longer holds. The consequence hereof is that the paddlewheel will no longer rotate at the same frequency as the liquid.

\subsection{Precession} \label{precession}

Precession changes the rest mass of a particle. 
In \cite{Coddens} we have shown that the Dirac equation can be derived from the Rodrigues equation (Eq. \ref{Rodrigues} in Section \ref{intro}) by
putting $\varphi = \omega_{0} \tau$ in the rest frame of the electron and making the assumption ${\hbar \omega_{0}\over{2}} = m_{0}c^{2}$.
The assumption  was introduced in a hand-waving way by assuming that the rest mass of the electron
would correspond to  the kinetic energy stored in its spinning motion. 

The derivation proposed in \cite{Coddens} of the Dirac equation is entirely classical,
which is a very surprising fact, as the Dirac equation is the core of the whole machinery of quantum mechanics. 
The Schr\"odinger equation can be derived from it. How can  these equations then possibly be classical?
It is explained in \cite{Coddens} that the properties of quantum mechanics that make it different from classical
mechanics are only due to the way we use the Dirac and Schr\"odinger equations in the calculations when we apply them to specific problems. 
In fact, in following the motto ``shut up and calculate''
we introduce unwittingly  features into the algebra that are meaningless from the viewpoint of the classical meaning of the algebra.
One of those features that are puzzling is the superposition principle.
Just as adding rotation matrices does not lead to a new rotation matrix, adding spinors has {\em a priori} not a clear meaning.
Another of the weird things that we cannot understand classically is the quantization of spin and angular momentum.
The algebra of the Dirac equation forces the spin and the angular momentum to align with the magnetic field, as discussed above,
and it is difficult to understand classically why it could not be otherwise.

Let us  here thus just accept  the fact that the spin must be aligned with the magnetic field, 
like quantum mechanics tells us. In fact, when we consider in the calculations  the possibility that the spin or the angular momentum are not aligned,
we find that the spin or angular momentum must precess, but that this does not lead to a constant energy.
To recover a constant energy, one must introduce another force, and the motion becomes then forced with respect to the magnetic field.
If we do not draw this force into the calculations, they are incomplete and it is then vain to try to understand them.
What it would imply that the spin is not aligned without forcing, such that the energy is not constant is not clear, but
this problem is in a sense eluded by the alignment condition.
We may thus assume  that
  the treatment ceases to be classical at the moment we are accepting the alignment condition.

We must now point out that we already know the physical meaning of the anomalous Zeeman effect.  
In fact, it is now time to remember that the Larmor precession term ${\boldsymbol{\omega}}_{L}{\boldsymbol{\cdot\sigma}} = - {q\over{2m_{0}}} {\mathbf{B}} {\boldsymbol{\cdot\sigma}}$ 
we obtained from the discussion in Subsection \ref{A-pot} is identical to the term
obtained from quantum mechanics in the non-relativistic limit
of the Dirac equation for an electron in a magnetic field. 
When the particle is rotating in its rest frame the precession frequency $\omega_{c}$ will add up algebraically to the rotation 
frequency $\omega_{0}$ of the
particle, changing the apparent frequency of its rotation, which is why  it eventually entails a correction
for the energy $E = {\hbar\omega_{0}\over{2}}  \rightarrow E = \hbar(\omega_{0} \pm \omega_{c})/2$ $ = m_{0}c^{2} \pm \hbar \omega_{L}$, where the $\pm$ sign
is due to the existence of spin up and spin down states. 
This gives a very neat explanation 
for the anomalous Zeeman splitting for an electron at rest. The very important point that follows from this argument  is that the amplitude of the anomalous Zeeman 
effect in the Dirac theory must be strictly identical to that of the orbital Zeeman effect, because on orbit  the precession is in phase with the orbital motion
due to the merry-go-round effect, and the precession for the particle at rest is obtained from the precession during the orbital motion by letting the radius
of the orbit shrink to zero.
Moreover the anomalous Zeeman effect is not due to the spin of the electron, but due to {\em the additional spin} given to the electron
by  the interaction of its charge with the magnetic field.

\subsection{Conclusion about the anomalous Zeeman effect} \label{anomaly-g-solve}

We see thus that the theory  is able to calculate the energy values of the two
 equilibrium states $m_{0}c^{2} \pm  {\hbar qB\over{2m_{0}}}$ of the spinning electron in terms of a charge-dipole interaction, 
 without associating a magnetic dipole with the spin.
 The idea is thus that a magnetic field makes any charge turn\footnote{The idea that the charge turns means that it is a magnetic monopole
according to Section \ref{monopole-fuss}. This is in conformity with the fact  that the anomalous Zeeman term $\imath cq{\mathbf{B}}$ in Eq. \ref{forces2} corresponds to the  term that describes the interaction of the electromagnetic field with a monopole at rest in
 Eqs. \ref{forces2-split1}-\ref{monopole-force}. The whole paper with its apparently meandering narrative is thus constructed around proving this connection
 and the underlying physical imagery. 
 The rest of the paper is devoted in a similar way to the connection between the  interaction of the moving magnetic monopole with the electromagnetic field  
 and the spin-orbit coupling.
 \label{overall-consistent}}
  at a frequency
 ${\boldsymbol{\omega}}_{c} = -{\gamma  q\over{m_{0}}} {\mathbf{B}}$, and with a kinetic energy described by the length of the pseudo-vector
  $\hbar {\boldsymbol{\omega}}_{L} = -{\hbar \gamma  q\over{2m_{0}}} {\mathbf{B}}$,
 independently from the issue if it already has spin or otherwise. 
 But when the particle  has spin, only
 the two states wherein the precession vector and the intrinsic-spin vector are aligned give rise to well-defined energies.
 The force that makes the particle spin should not be confused with a torque, already for reasons of dimension only.
The reason for its existence is purely due to relativistic symmetry.

We did not introduce spin in Eqs.  \ref{forces}-\ref{forces2}. 
The only quantity in the Dirac equation
that contains the spin
is the wave function. This is also very clear from reference \cite{Coddens}.
This also explains why we can calculate
 the $g$-factor so accurately in quantum electrodynamics: We just do not use a dipole moment 
 ${\boldsymbol{\mu}}_{e}$ due to the electron spin in the calculation of the equilibrium state. 
Simultaneously this copes with Lorentz's objection that the hypothetical magnetic dipole moment of the electron is too large to 
allow for an explanation in terms of a current loop inside the electron: 
There is simply no spin-associated magnetic dipole moment in the formalism. 
 
 We may  note that the correct theory for ferromagnetism also does  not rely on magnetic dipoles, 
 but on Heisenberg's mechanism of an exchange interaction which is
based on the Pauli principle and the Coulomb interaction.  In the Dirac formalism, charge and spin occur in mere juxtaposition, without
blending into a more complex quantity like a dipole.
The interaction  is just  defined by
the charge, while the formalism shows that the spin ${\mathbf{s}}$ has to line up with the magnetic field ${\mathbf{B}}$, because the wave function must be an eigenstate 
of the pseudo-vector operator ${\mathbf{B}}{\boldsymbol{\cdot\sigma}}$ or ${\mathbf{B}}{\boldsymbol{\cdot\gamma}}$.

The mere juxtaposition of spin and charge in the Dirac equation looks similar to that in the exchange mechanism.
The image of a dipole moment
is just not present in the algebra of the Dirac equation. 
One may speculate about defining a magnetic dipole moment ${\boldsymbol{\mu}}_{e}$, 
because it flatters our intuition of a little magnet,
but this idea that $-{q\over{2m_{0}}} {\mathbf{B}}{\boldsymbol{\cdot\sigma}}$  must willy-nilly correspond to a magnetic dipole moment is  a preconceived notion
that  is just  
flawed mathematically.
As already pointed out above, the appropriate interpretation is that there are three types of
interaction: charge-charge, dipole-dipole, and charge-dipole. The anomalous Zeeman effect is a charge-dipole interaction
 and  one should refrain from identifying this charge-dipole
interaction with a dipole-dipole interaction.
There is thus always a precession energy associated with the presence of a charge $q$ in a 
magnetic field ${\mathbf{B}}$, which is the reason why we find
this term already in Eq. \ref{forces2}, which does not contain spin.

After all this, there is yet another argument in favor of the interpretation of the term $-{q\over{2m_{0}}} {\mathbf{B}}{\boldsymbol{\cdot\sigma}}$
proposed in this paper. Dirac's theory works only
well for the leptons \cite{Greiner}, as the true value of $g$ for the electron is given by: $g_{e} = 2.00231930436153 (53)$.
It works almost as well for the muon $g_{\mu} =  2.0023318416(13)$,
and much less well for the neutron ($g_{n}= 3.82608545(90)$) and the proton ($g_{p}= -5.585694713(56)$). It is tempting to assume that
in the cases of the proton and the neutron, a true dipole moment ${\boldsymbol{\mu}}$ due to internal currents might intervene\footnote{These currents are the currents produced by the quarks. It is impossible to explain the magnetic moment of the neutron whose overall charge is zero, without falling back on its quark structure. \label{quarks}} and give rise to some 
term $-{\boldsymbol{\mu\cdot}}{\mathbf{B}}$, 
while this is just not the case for the electron, which is truly a point particle
or is so small that the effect of such currents, if they exist, can be neglected, and only the charge term  
$-{\hbar q\over{2m_{0}}}\, [\,{\mathbf{B}}{\boldsymbol{\cdot\sigma}}\,]$ plays a significant r\^ole. 

\section{Problems with the traditional derivation of the spin-orbit coupling} \label{correct-minimal-applied}

\subsection{Preliminary remarks}  \label{discussion}

The spin-orbit term  ${1\over{m_{0}c}} \,{1\over{r}} \,{\partial U\over{\partial r}}{\mathbf{L}}{\boldsymbol{\cdot\sigma}}$
  is easily shown to be equal to ${q\over{c}}[\, ({\mathbf{v}}\wedge {\mathbf{E}}){\boldsymbol{\cdot\sigma}}\,])$, where  $U = qV$ is the potential energy.\footnote{
  It may be noted that in the classical Coulomb problem, $v$ can be calculated from ${\mathbf{r}}$ from both conservation laws (for energy and angular momentum). 
The kinetic energy ${1\over{2}} m_{0}v^{2} = E-U$ follows just from the total energy $E$ and the potential energy $U$.
But using the law of conservation of energy ceases to be feasible within the present quantum mechanical context. The problem is that the electron might
  have radiated to settle down on a given orbit. The total energy $E$ of the electron on this orbit is now the quantity we want to determine in the first place.
  If there is no further information available as to the amount of energy that has been radiated away, it is   {\em a priori}  
  impossible to know $E$. The lacking information can however be derived from the conservation of angular momentum.
Specifying ${\mathbf{v}}$ through ${\mathbf{L}} = {\mathbf{r}} \wedge m_{0}{\mathbf{v}}$ in the non-relativistic limit permits to calculate $v$.
From this it is possible to calculate $E = U({\mathbf{r}}) + {1\over{2}} m_{0}v^{2}$.
 It is for these reasons that we express everything in terms of $U$ and $L$.
 But in a purely  relativistic context,
  the equation ${\mathbf{L}} = {\mathbf{r}} \wedge m{\mathbf{v}}$ introduces also $m$, which is related to the
  total energy through $E =mc^{2}$.\label{why-L}}
  Following the discussion in Subsubsection \ref{q-D}
 the spin-orbit interaction is of the ``dipole''-charge type, due to the presence of the term $q{\mathbf{v}} \wedge {\mathbf{E}}$. 
Here $q {\mathbf{v}}$ represents the ``dipole'' and $ {\mathbf{E}}$ contains the charge.
Also here the interpretation of ${\boldsymbol{\sigma}}$ in ${\mathbf{L}}{\boldsymbol{\cdot\sigma}}$ in terms of a ``spin operator'' ${\mathbf{S}} = {\hbar\over{2}}{\boldsymbol{\sigma}}$ is not appropriate. The term ${\mathbf{L}}{\boldsymbol{\cdot\sigma}}$
within ${1\over{m_{0}c}} \,{1\over{r}} \,{\partial U\over{\partial r}}{\mathbf{L}}{\boldsymbol{\cdot\sigma}}$
  just represents the angular momentum. 
Imaging the spin-orbit coupling  as a ``dipole''-dipole interaction containing a 
term ${\mathbf{L\cdot S}}$  is in conflict 
with the symmetry. First of all 
${1\over{m_{0}c}} \,{1\over{r}} \,{\partial U\over{\partial r}}{\mathbf{L}}{\boldsymbol{\cdot\sigma}}$
is a vector rather than a scalar. Secondly,   its  ``dipole''-charge interaction symmetry is not compatible with a ``dipole''-dipole interaction symmetry. 
An analogous problem of an over-interpretation of an operator that violates the symmetry  
occurs in the definition of helicity ${\mathbf{u}}{\boldsymbol{\cdot\sigma}}$, with ${\mathbf{u}} = {\mathbf{p}}/p$ for neutrinos in particle physics.

The 
 frequency $\omega_{T}$ of the Thomas precession correction
is given \cite{Rhodes,Purcell}  by: 

\begin{equation} \label{Wigner}
\omega_{T} = {\gamma^{2}\over{\gamma+1}} {1\over{c^{2}}} {\mathbf{v}} \wedge {\mathbf{a}}.
\end{equation}  

\noindent In an electric field ${\mathbf{E}}$, this can be written as $ {\gamma\over{\gamma+1}} {1\over{m_{0} c^{2}}} {\mathbf{v}} \wedge q {\mathbf{E}}$.
For $\gamma \approx 1$ this corresponds to ${1\over{2m_{0}c^{2}}} {\mathbf{v}} \wedge q {\mathbf{E}}$. To
 correct the spin-orbit term for the Thomas precession  we must subtract $\hbar\omega_{T}/2$ from its absolute value.
This has thus absolutely nothing to do with an electric dipole moment induced by the relativistic motion of a magnetic dipole moment associated with the spin of the electron,
as has often been claimed \cite{electric-dipole}. As we have argued all along, such a hypothetical  magnetic dipole moment  
does not even exist in the context of the Dirac equation. The Thomas precession is a relativistic correction term that contributes to the  global spin-orbit precession
along the orbit.  The global expression is equal to the Thomas precession with the reversed sign.

\subsubsection{Simple derivation of the expressions for the Thomas precession} \label{Thomas-calc}

For what will follow it might be useful to explain carefully what Thomas precession is.
There are actually two equations for the Thomas precession that one can prove. The first one summarizes the conceptual definition of the effect, which is that
the composition of two boosts is a composition of a boost and a rotation:
${\mathbf{B}}({\mathbf{v}}_{2}) {\mathbf{B}}({\mathbf{v}}_{1}) = {\mathbf{R}}({\mathbf{n}},\vartheta_{T}) {\mathbf{B}}({\mathbf{v}})$. The rotation axis ${\mathbf{n}}$ is here perpendicular to both ${\mathbf{v}}_{1}$ and ${\mathbf{v}}_{2}$, such that it is just
the rotation angle $\vartheta_{T}$ and the boost parameter ${\mathbf{v}}$ that have to be calculated. But specifying  the Thomas precession
requires only the calculation of $\vartheta_{T}$.
The second equation formulates the effect of Thomas precession on an orbit, and states that 
${d\vartheta_{T}\over{dt}} = {\gamma^{2}\over{(\gamma +1)c^{2}}} {\mathbf{v}} \wedge {\mathbf{a}}$.

The following derivation of the first equation  follows the argument in Subsection VIII D of \cite{Rhodes}.
We note $\gamma = \cosh w$, $\beta\gamma = \sinh w$, $\beta = \tanh w$. We have then $\cosh{w\over{2}} = \sqrt{{\gamma+1\over{2}}}$,
$\sinh{w\over{2}} = \sqrt{{\gamma-1\over{2}}}$, $\tanh{w\over{2}}=  \sqrt {{\gamma-1\over{\gamma +1}}}$. In SL(2,${\mathbb{C}}$), a boost with velocity ${\mathbf{v}} = v{\mathbf{u}}$ is given by ${\mathbf{B}}({\mathbf{v}}) =
\cosh {w\over{2}} \,\bigone - \sinh {w\over{2}} \,[\,{\mathbf{u}}{\boldsymbol{\cdot\sigma}}\,]$. Here ${\mathbf{u}}$ is the unit vector parallel to ${\mathbf{v}}$.
We must now calculate ${\mathbf{B}}({\mathbf{v}}_{2}){\mathbf{B}}({\mathbf{v}}_{1})$. For simplicity, we can take ${\mathbf{u}}_{1} = {\mathbf{e}}_{x}$,
${\mathbf{u}}_{2} = \cos\alpha\,{\mathbf{e}}_{x} + \sin\alpha\,{\mathbf{e}}_{y}$. We find then:

\begin{align}\label{boost-product}
{\mathbf{B}}({\mathbf{v}}_{2}){\mathbf{B}}({\mathbf{v}}_{1}) & = (\cosh{w_{2}\over{2}}\,\cosh{w_{1}\over{2}} + \sinh{w_{2}\over{2}}\,\sinh{w_{1}\over{2}}\,\cos\alpha)\,\bigone\nonumber\\
& - \imath \, \sinh{w_{2}\over{2}}\,\sinh{w_{1}\over{2}}\,\sin\alpha\,[\,{\mathbf{e}}_{z}{\boldsymbol{\cdot\sigma}}\,]  - \cosh{w_{2}\over{2}}\,\sinh{w_{1}\over{2}} \,[\,{\mathbf{e}}_{x}{\boldsymbol{\cdot\sigma}}\,] \nonumber\\
&- \sinh{w_{2}\over{2}}\,\cosh{w_{1}\over{2}} \,[\,(\cos\alpha \,{\mathbf{e}}_{x}
+\sin\alpha\, {\mathbf{e}}_{y}){\boldsymbol{\cdot\sigma}}\,]. 
\end{align}

\noindent This must be equal to the product  ${\mathbf{R}}(\vartheta_{T},{\mathbf{e}}_{z})\,{\mathbf{B}}({\mathbf{v}})$ $=$
$(\,\cos{\vartheta_{T}\over{2}}\,\bigone - \imath \cos{\vartheta_{T}\over{2}}\,[\,{\mathbf{e}}_{z}{\boldsymbol{\cdot\sigma}}\,]\,)$ $\,(\cosh {w\over{2}} \,\bigone - \sinh {w\over{2}} \,[\,{\mathbf{u}}{\boldsymbol{\cdot\sigma}}\,])$ of a  rotation around the $z$-axis and a boost:

\begin{align}\label{other-product}
{\mathbf{R}}(\vartheta_{T},{\mathbf{e}}_{z})\,{\mathbf{B}}({\mathbf{v}}) & 
= \cos{\vartheta_{T}\over{2}}\, \cosh{w\over{2}} \,\bigone - \imath  \sin{\vartheta_{T}\over{2}}\, \cosh{w\over{2}}\,[\,{\mathbf{e}}_{z}{\boldsymbol{\cdot\sigma}}\,]\nonumber\\
& - \cos{\vartheta_{T}\over{2}}\,\sinh{w\over{2}}  \,[\,{\mathbf{u}}{\boldsymbol{\cdot\sigma}}\,] -  \sin{\vartheta_{T}\over{2}}\,\sinh{w\over{2}}\, [\,({\mathbf{e}}_{z} \wedge {\mathbf{u}}){\boldsymbol{\cdot\sigma}}\,]. 
\end{align}

\noindent By identifying the parts containing the unit matrix $\bigone$ and the parts containing $\imath [\,{\mathbf{e}}_{z}{\boldsymbol{\cdot\sigma}}\,]$
one obtains:

\begin{align}\label{abstract-thomas}
\tan {\vartheta_{T}\over{2}} = { \tanh {w_{1}\over{2}} \, \tanh {w_{1}\over{2}} \, \sin\alpha\over{
1 + \tanh {w_{1}\over{2}} \, \tanh {w_{1}\over{2}}  \cos\alpha}},
\end{align}

\noindent which is Eq. 145 in reference \cite{Rhodes}.

The second equation can be derived from this equation.
But one can calculate the identity ${d\vartheta_{T}\over{dt}} = {\gamma^{2}\over{(\gamma +1)c^{2}}} {\mathbf{v}} \wedge {\mathbf{a}}$ also directly,  by
considering the identity:  ${\mathbf{B}}(d{\mathbf{v}}_{\perp}) {\mathbf{B}}({\mathbf{v}}) = 
{\mathbf{R}}({\mathbf{n}},d\vartheta_{T}) {\mathbf{B}}( d{\mathbf{v}} \oplus{\mathbf{v}}_{\perp} )$, where ${\mathbf{v}} \oplus {\mathbf{w}}$ denotes the
boost vector associated with the composition of boosts $B({\mathbf{w}}) ^{\circ} B({\mathbf{v}})$.
We are considering here only $d{\mathbf{v}}_{\perp}$ as collinear boosts do not lead to Thomas precession.
By using Taylor expansions one can show that to first order ${\mathbf{B}}(d{\mathbf{v}}_{\perp})$ $=$ 
$\bigone - {1\over{2c}}\, dv_{\perp}\,[\,({\mathbf{e}}_{z} \wedge {\mathbf{u}}){\boldsymbol{\cdot\sigma}}\,]$. 
One obtains then ${\mathbf{B}}(d{\mathbf{v}}_{\perp}) {\mathbf{B}}({\mathbf{v}}) = \sqrt{{\gamma+1\over{2}}} \,\bigone $
$- {1\over{2c}}\sqrt{{\gamma+1\over{2}}}\, dv_{\perp}\, [\,({\mathbf{e}}_{z}\wedge {\mathbf{u}}){\boldsymbol{\cdot\sigma}}\,] $
$- \sqrt{{\gamma-1\over{2}}}\, [\,{\mathbf{u}}{\boldsymbol{\cdot\sigma}}\,] $
$- {\imath\over{2c}}\,\sqrt{{\gamma-1\over{2}}}\, dv_{\perp}\, [\,{\mathbf{e}}_{z}{\boldsymbol{\cdot\sigma}}\,]$.
Here  $\gamma$ is the Lorentz factor that corresponds
to $v$.
The identification of this result
with Eq. \ref{other-product} 
yields then  : ${d\vartheta_{T}\over{2}} = {1\over{2c}} dv_{\perp} \sqrt{{\gamma-1\over{\gamma+1}}}$
 (whereby we have replaced of course $\vartheta_{T}$ by $d\vartheta_{T}$). 
From this one obtains ${d\vartheta_{T}\over{d\tau}} = {\gamma\over{\gamma+1}} {1\over{c^{2}}} {\mathbf{v}}\wedge {\mathbf{a}}$  in the co-moving frame, i.e. the rest frame
of the electron. Here $\tau$ is the proper time of the electron, i.e. the time in the co-moving frame. Taking into account the time dilatation
this yields
${d\vartheta_{T}\over{dt}} = {\gamma^{2}\over{\gamma+1}} {1\over{c^{2}}} {\mathbf{v}}\wedge {\mathbf{a}}$ in the laboratory frame.
Here ${\mathbf{a}}_{\perp}$ takes the same value in the laboratory frame and in the co-moving frame
because it is perpendicular to ${\mathbf{v}}$.  This is presumably the simplest derivation possible of the expression for the Thomas precession.
It avoids using hyperbolic geometry to derive this result,
as was done in reference \cite{Rhodes}. We can also see from this derivation why the Dirac equation with the minimal
substitution
cannot be used to derive the Thomas precession. As discussed in Subsection \ref{correct-minimal} (and explained in reference \cite{Coddens}), the minimal substitution accounts for  the instantaneous value of ${\mathbf{B}}({\mathbf{v}})$ in $({\mathbf{r}},t)$, but does not take care
of ${\mathbf{a}}_{\perp}$ or ${\mathbf{R}}(\vartheta_{T},{\mathbf{e}}_{z})$.

\subsection{Generalization of Dirac's minimal  substitution for the case of a moving charge} \label{correct-minimal}

As already stated in Subsection \ref{Pauli}, Dirac's minimal substitution
is not general enough as it does not account for the interactions of the current $q{\mathbf{v}}$ of the moving electron with the electromagnetic potential.
It is explained in \cite{Coddens} that in the context of the Dirac equation, the minimal substitution must not be seen as an abstract rule that one 
copies mechanically from Lagrangian dynamics and that one can justify with hindsight by the fact that it works.
The goal
of the minimal substitution is to introduce the true instantaneous kinetic parameters
 $ (\gamma, \gamma{\mathbf{v}}/c)$, that will permit us to write the true instantaneous
Lorentz transformation for the time. As explained in \cite{Coddens}, the primary idea is that the the transformation of the time 
$ct' = \gamma (ct -{\mathbf{v\cdot x}}/c)$ under a free-space Lorentz boost can also be written using
the parameters $(E,c{\mathbf{p}}) = m_{0}c^{2} (\gamma, \gamma{\mathbf{v}}/c)$ and $m_{0}c^{2}$ instead of $ (\gamma, \gamma{\mathbf{v}}/c)$.
In fact, the Dirac equation is derived from the Rodrigues equation by transforming $({d\over{d c\tau}}, 0,0,0)$ $\rightarrow$ ${\partial \over{\partial ct}},
{\partial \over{\partial x}},{\partial \over{\partial y}},{\partial \over{\partial z}})$ according to the boost that transforms 
$(m_{0}c^{2},0,0,0) \rightarrow (E,c{\mathbf{p}})$.\footnote{This shows that the Dirac equation does thus not account for the Thomas precession by construction.
This may look as a startling claim to the reader, but it will be part of a discussion that we will unfold from Subsection \ref{failure} onwards.
\label{construction}}
But in a potential $V$, the rest energy of a particle is no longer $m_{0}c^{2}$ but  $m_{0}c^{2} + qV$, such that the parameters $(E,c{\mathbf{p}})$ 
of a moving particle are no longer
the correct set of kinetic parameters to describe the instantaneous Lorentz transformation, because $E$ is no longer purely kinetic. 
To be able to write the correct Lorentz transformation, one must know the ``true kinetic energy'', i.e. that part of the total energy that is not potential energy.
This leads to the substitution $E \rightarrow E - qV$, ${\mathbf{p}} \rightarrow {\mathbf{p}} - q{\mathbf{A}}$.
The idea is that for an electron at rest in an electric potential, the equation is $-{\hbar\over{\imath}} {d\over{d\tau}}\gamma_{t} \psi = (m_{0}c^{2} + qV)\,\psi$,
where $\tau$ is the time in the rest frame.
By generalizing this to a general frame by Lorentz covariance we obtain then arguably the general Dirac equation with the minimal substitution 
$E \rightarrow E - qV$, ${\mathbf{p}} \rightarrow {\mathbf{p}} - q{\mathbf{A}}$.

The problem is here that $(qV, cq{\mathbf{A}})$ is not exactly what one would obtain from $(qV,{\mathbf{0}})$ by transforming it to a general frame. 
Whereas the  four-potential $(V, c{\mathbf{A}})$ is Lorentz covariant, the quantity $q$ is not,  because $q$  is part of a charge-current four-vector 
$(\gamma q, \gamma q{\mathbf{v}}/c)$.
Therefore $(qV, cq{\mathbf{A}})$ is not the correct Lorentz covariant generalization for the expression that must be used in the  most general substitution. 
 In the most general substitution we would need
in stead of $qV$ and $q{\mathbf{A}}$  the terms that we can obtain by considering:
$[\,\gamma q\, \bigone +\gamma {q\over{c}} {\mathbf{v}}{\boldsymbol{\cdot \sigma}}\,]\,[\,V\bigone + c{\mathbf{A}}{\boldsymbol{\cdot \sigma}}\,]$.
The result will  lead to two terms $qV$ and $qc{\mathbf{A}}$ with the same sign. This is needed because the terms $qV$ and $qc{\mathbf{A}}$
are used with the same signs
in the traditional minimal substitution. Furthermore, we obtain a term $\gamma q\,( {\mathbf{v\cdot A}})\, \bigone$, which after the substitutions
$E \rightarrow E - qV, {\mathbf{p}} \rightarrow  {\mathbf{p}} - q{\mathbf{A}}$, will lead to a term $-\gamma q\,( {\mathbf{v\cdot A}})\, \bigone$, whose sign is
in conformity with the discussion in Subsubsection \ref{A-der}. It may look confusing that in the derivation of Eq. \ref{definitions} and Eq. \ref{correct-subst} 
 we have not used the strategy of alternating signs as would follow from the rules that prevail for SL(2,${\mathbb{C}}$).
For Eq.  \ref{definitions}  this may be due the way the magnetic potential has been historically  defined. For Eq. \ref{correct-subst}, we  note
that we can make juxtapostions
like $[\,\gamma q\, \bigone +\gamma {q\over{c}} {\mathbf{v}}{\boldsymbol{\cdot \sigma}}\,]\,
[\,V\bigone + c{\mathbf{A}}{\boldsymbol{\cdot \sigma}}\,]$ 
with any number of terms and any choice of signs. The goal is to check to what kind of multi-vectors this leads. 
This is more the case here and the choice made here leads to the correct results for the further use.
These considerations lead to:

\begin{equation*} 
[\, \gamma q \bigone + \gamma q {{\mathbf{v}}\over{c}}{\boldsymbol{\cdot\sigma}}\,] \,[\, V\,\bigone + c{\mathbf{A}}{\boldsymbol{\cdot\sigma}}\,] =
\end{equation*}
\begin{equation} \label{correct-subst}
\gamma q V\, \bigone +
\gamma qc{\mathbf{A}}{\boldsymbol{\cdot\sigma}} 
+ \gamma {q\over{c}} V {\mathbf{v}}{\boldsymbol{\cdot\sigma}}
+  \gamma q\,( {\mathbf{v\cdot A}})\, \bigone 
+ \imath \gamma q ({\mathbf{v}} \wedge {\mathbf{A}}){\boldsymbol{\cdot\sigma}}.
\end{equation}

\noindent These are the quantities that one would have to use in the substitution
that generalizes the minimal substitution.

\subsection{How we obtain the spin-orbit-coupling term from the Dirac equation with the generalized substitution} \label{spin-orbit-algbra}

The generalized substitution was given in Eq. \ref{correct-subst}.
The term $\gamma {q\over{c}} V {\mathbf{v}}{\boldsymbol{\cdot\sigma}}$ in Eq. \ref{correct-subst} is a vector and cannot
contribute to the potential energy. However, after squaring the Dirac equation, which will produce the announced variant
$[\, {\partial\over{\partial ct}}\bigone - {\mathbf{\nabla\cdot}}{\boldsymbol{\sigma}}\,]\,[\, \gamma q \bigone + \gamma q {{\mathbf{v}}\over{c}}{\boldsymbol{\cdot\sigma}}\,] \,[\, V\,\bigone + c{\mathbf{A}}{\boldsymbol{\cdot\sigma}}\,] $,
it will lead to a term containing
$[\,{\mathbf{\nabla}}{\boldsymbol{\cdot\sigma}}\,]\,[\,\gamma {q\over{c}} V {\mathbf{v}}{\boldsymbol{\cdot\sigma}} \,]$. For ${\mathbf{A}} = {\mathbf{0}}$,
 this will lead in the non-relativistic limit to a term ${q\over{c}} ({\mathbf{\nabla\cdot}}(V{\mathbf{v}}))\,\bigone$  and a term
$\imath {q\over{c}}( {\mathbf{\nabla}}\wedge V {\mathbf{v}}){\boldsymbol{\cdot\sigma}}$. 
The first term leads to  $ -{q\over{c}} ({\mathbf{v\cdot E}})\,\bigone$ $+{qV\over{c}}{\mathbf{\nabla\cdot v}}\,\bigone$.
In the former term we  can recognize the power term  in Eq. \ref{forces2}.
The calculation of the term $\imath {q\over{c}}( {\mathbf{\nabla}}\wedge V {\mathbf{v}}){\boldsymbol{\cdot\sigma}}$ yields
$\imath {q\over{c}} [\, V\,({\mathbf{\nabla}}\wedge {\mathbf{v}}) + {\mathbf{v}}\wedge {\mathbf{E}}\,]$.
For uniform circular motion ${\mathbf{\nabla}}\wedge {\mathbf{v}} = 2 {\boldsymbol{\omega}}$. The term $V({\mathbf{\nabla}}\wedge {\mathbf{v}})$
becomes then $2V  {\boldsymbol{\omega}} = -2 {\mathbf{v}}\wedge {\mathbf{E}}$. In total we have thus 
 $ V\,({\mathbf{\nabla}}\wedge {\mathbf{v}}) + {\mathbf{v}}\wedge {\mathbf{E}} =  - {\mathbf{v}}\wedge {\mathbf{E}}$, such that
$\imath {q\over{c}}( {\mathbf{\nabla}}\wedge V {\mathbf{v}}){\boldsymbol{\cdot\sigma}} =- \imath {q\over{c}}  {\mathbf{v}}\wedge {\mathbf{E}}$. 
After multiplication by $-{\hbar\over{2m_{0}c}}$ the term that goes with $\imath$ becomes
$ {\hbar\over{2m_{0}^{2}c^{2}}} ({1\over{r}} {\partial U\over{\partial r}} ) {\mathbf{L}}$,
 which is
 the spin-orbit precession in an electric field, {\em but without the correction for Thomas precession}.
 
\subsection{Apparent failure with respect to the traditional derivation} \label{failure}

If we derive the Pauli equation from the Dirac equation with the generalized substitution, it will thus also contain the spin-orbit term, with the correct sign,
but without the correction for Thomas precession.
Traditionally, the spin-orbit term is derived from the Dirac equation with
the minimal substitution $E \rightarrow E - qV, {\mathbf{p}} \rightarrow  {\mathbf{p}} - q{\mathbf{A}}$. This seems to 
entirely discredit our approach as it makes it
 look as though it contradicts our criticism that this substitution does not take into account the velocity of the electron. By combining the traditional approach with the present approach it may even seem that we get the result for the spin-orbit coupling twice. But as already stated previously,
 all this is not true because the traditional approach is wrong and {\em it is not possible  at all} to derive the spin-orbit term from the Dirac equation
based on the traditional minimal substitution. Let us now explain why.

\subsection{Errors in the traditional derivation} \label{spin-orbit-errors}

\subsubsection{How a change of basis offers a first hint about the existence of an error} \label{first-hint}

Let us point out in which way  the traditional derivation  is   based on  flawed mathematics.
The way we are writing the Dirac equation in this paper  is different from the traditional one. In the Weyl representation we have:

\begin{equation} \label{change-of-basis}
{\mathbf{e}}_{t} \leftrightarrow \gamma^{t} = \left ( 
\begin{tabular}{cc}
& $\bigone$\\
$\bigone$ & \\
\end{tabular}
\right ), \quad 
{\mathbf{e}}_{5} \leftrightarrow \gamma^{5} = \left ( 
\begin{tabular}{cr}
$\bigone$ &\\
 & $-\bigone$ \\
\end{tabular}
\right ),
\end{equation}

\noindent while in the traditional representation we have:

\begin{equation} \label{change-of-basis-2}
{\mathbf{e}}_{5} \leftrightarrow \gamma^{5} = \left ( 
\begin{tabular}{cc}
& $\bigone$\\
$\bigone$ & \\
\end{tabular}
\right ), \quad 
{\mathbf{e}}_{t} \leftrightarrow \gamma^{t} = \left ( 
\begin{tabular}{cr}
$\bigone$ &\\
 & $-\bigone$ \\
\end{tabular}
\right ).
\end{equation}

\noindent The difference just corresponds to a change of basis in ${\mathbb{R}}^{5}$ provided with a metric
$x_{4}^{2}+ x_{5}^{2} - x^{2} - y^{2} -z^{2}$, whereby we change the orientation of the $ct$-axis
 between the axes $Ox_{4}$ and $Ox_{5}$  in the $Ox_{5} x_{4}$ plane, e.g. by a rotation over ${\pi\over{2}}$. 
We can consider space-time as embedded into the five-dimensional space,
and its mathematical properties do not depend on the way it is embedded: Meaningful mathematical properties do not depend on a choice of basis.
The two representations are therefore equivalent by a similarity transformation.
This argument actually indicates  how one can try to prove Pauli's theorem that all choices of gamma matrices are equivalent.
Equivalent choices of gamma matrices correspond just to different choices of a basis.
Now, in the Weyl representation, the Dirac equation for an electron in a Coulomb field will be:

\begin{align} \label{equation-in-Weyl}
 \left (
\begin{tabular}{cc}
& $(-{\hbar\over{\imath}} {\partial \over{\partial ct}} - {qV\over{c}})\, \bigone + {\hbar\over{\imath}} {\mathbf{\nabla}}{\boldsymbol{\cdot\sigma}}$\\
   $(-{\hbar\over{\imath}} {\partial \over{\partial ct}} - {qV\over{c}})\, \bigone - {\hbar\over{\imath}} {\mathbf{\nabla}}{\boldsymbol{\cdot\sigma}}$ & \\
\end{tabular}
\right ) &
\left (
\begin{tabular}{c}
$\psi_{1}$\\
$\psi_{2}$\\
\end{tabular}
\right )  \nonumber\\
=  m_{0} c &
\left (
\begin{tabular}{c}
$\psi_{1}$\\
$\psi_{2}$\\
\end{tabular}
\right ).
\end{align}
 
\noindent We have thus:

\begin{align} \label{equation-in-Weyl-2}
[\,(-{\hbar\over{\imath}} {\partial \over{\partial ct}} - {qV\over{c}})\, \bigone + {\hbar\over{\imath}} {\mathbf{\nabla}}{\boldsymbol{\cdot\sigma}}\,]\, \psi_{2} & = m_{0}c  \,\psi_{1},\nonumber\\
[\,(-{\hbar\over{\imath}} {\partial \over{\partial ct}} - {qV\over{c}})\, \bigone - {\hbar\over{\imath}} {\mathbf{\nabla}}{\boldsymbol{\cdot\sigma}}\,]\, \psi_{1} & = m_{0}c \,\psi_{2},
\end{align}

\noindent which leads to:

\begin{align} \label{equation-in-Weyl-3}
[\,(-{\hbar\over{\imath}} {\partial \over{\partial ct}} - {qV\over{c}})\, \bigone + {\hbar\over{\imath}} {\mathbf{\nabla}}{\boldsymbol{\cdot\sigma}}\,]\,
[\,(-{\hbar\over{\imath}} {\partial \over{\partial ct}} - {qV\over{c}})\, \bigone - {\hbar\over{\imath}} {\mathbf{\nabla}}{\boldsymbol{\cdot\sigma}}\,]\, \psi_{1} & =
m_{0}^{2}c^{2}  \,\psi_{1},
\nonumber\\
[\,(-{\hbar\over{\imath}} {\partial \over{\partial ct}} - {qV\over{c}})\, \bigone - {\hbar\over{\imath}} {\mathbf{\nabla}}{\boldsymbol{\cdot\sigma}}\,]\, 
[\,(-{\hbar\over{\imath}} {\partial \over{\partial ct}} - {qV\over{c}})\, \bigone + {\hbar\over{\imath}} {\mathbf{\nabla}}{\boldsymbol{\cdot\sigma}}\,]\, \psi_{2} & =
m_{0}^{2}c^{2}  \,\psi_{2}.
\end{align}

\noindent The point is now that these equations are completely decoupled. The terms that do not contain $V$ are e.g.
$[\,(-{\hbar\over{\imath}} {\partial \over{\partial ct}} )\, \bigone + {\hbar\over{\imath}} {\mathbf{\nabla}}{\boldsymbol{\cdot\sigma}}\,]\,
[\,(-{\hbar\over{\imath}} {\partial \over{\partial ct}} )\, \bigone - {\hbar\over{\imath}} {\mathbf{\nabla}}{\boldsymbol{\cdot\sigma}}\,]\, \psi_{1} = [\,\hbar^{2} (\Delta -{1\over{c^{2}}}  {\partial^{2} \over{\partial t^{2}}})\,\bigone\,]\,\psi_{1}$,  for the first equation. The terms that combine $(-{\hbar\over{\imath}} {\partial \over{\partial ct}} )\, \bigone$ and 
${qV\over{c}}\, \bigone$ give rise to: ${2\hbar\over{\imath}} {qV\over{c^{2}}} {\partial \psi_{1}\over{\partial t}}$, where we have assumed that $V$ does not vary with time. 
The interesting terms with vector symmetry are those that combine
$ {\hbar\over{\imath}} {\mathbf{\nabla}}{\boldsymbol{\cdot\sigma}}$ and ${qV\over{c}}\, \bigone$. They give rise to ${\hbar\over{\imath}}{q\over{c}}\,[\, {\mathbf{E}}
{\boldsymbol{\cdot\sigma}}\,]\, \psi_{1}$. There is thus a term $[\,{\mathbf{E}}{\boldsymbol{\cdot\sigma}}\,] \psi_{j}$, but no term containing $[\,({\mathbf{E}}\wedge {\mathbf{p}}){\boldsymbol{\cdot\sigma}} \,] \psi_{j}$. There is even not a term
 containing $[\,{\mathbf{p}}{\boldsymbol{\cdot\sigma}}\,] \psi_{j}$, because the two  terms containing ${qV\over{c}} {\hbar\over{\imath}}\,[\, {\mathbf{\nabla}}{\boldsymbol{\cdot\sigma}}\,]$ have opposite signs and cancel in the algebra.
 This is an exact result. There is here no fuss about approximations or series expansions.
The vector operators working on  $\psi_{j}$ are all based on vector quantities in the plane of motion, while $({\mathbf{E}}\wedge {\mathbf{p}}){\boldsymbol{\cdot\sigma}} $
is perpendicular to it. The term $({\mathbf{E}}\wedge {\mathbf{p}}){\boldsymbol{\cdot\sigma}} $ can thus never occur in the algebra
if it is carried out correctly. 

The operator ${2\hbar\over{\imath}} {qV\over{c^{2}}} {\partial \over{\partial t}}\,\bigone$ will admittedly lead to a vector term that is perpendicular
to the plane of motion, that becomes equal to $-\imath \, [\,{\boldsymbol{\omega}}{\boldsymbol{\cdot\sigma}}\,] = -\imath \omega \,[\,{\mathbf{s}}{\boldsymbol{\cdot\sigma}}\,]$ for a particle at rest. Here ${\mathbf{s}}$ is the spin axis, which is in general identified with ${\mathbf{e}}_{z}$, while the plane of motion is identified with the $Oxy$ plane.
After the necessary algebra this term will lead to a change $m_{0} \rightarrow m_{0} +qV/c^{2}$ 
for the rest mass or 
$\hbar \omega_{0} \rightarrow \hbar\omega  = \hbar \omega_{0}  +qV$ for the  rest energy of the particle in a potential. As explained in Subsection \ref{correct-minimal},
the purpose of the minimal substitution was exactly to reproduce this result. Also the term  ${2\hbar\over{\imath}} {qV\over{c^{2}}} {\partial \psi_{1}\over{\partial  t}}\,$
does thus not incorporate the spin-orbit coupling.

As the traditional representation is equivalent to the Weyl representation up to a similarity transformation,
this conclusion must also be valid in the traditional representation. Note that the difference between the Weyl representation and the traditional interpretation
does not affect the coordinates $(x,y,z)$, as it corresponds only to a change of the orientation of the time axis in the $(x_{4},x_{5})$-plane. The two representations must
thus lead to the same results and to the same expressions for the results.
This suggests that something with the traditional derivation
of the spin-orbit term must be wrong. 

We can further argue that the only way to derive the terms for the spin-orbit
coupling correctly from a Dirac equation is  using the extended substitution we introduced in Eq. \ref{correct-subst}. 
In fact, to obtain a term $[\,({\mathbf{E}}\wedge {\mathbf{p}}){\boldsymbol{\cdot\sigma}} \,] \psi_{j}$ one needs
a succession $[\, {\mathbf{\nabla}}{\boldsymbol{\cdot\sigma}} \,]\,V\,  [\,{\mathbf{p}}{\boldsymbol{\cdot\sigma}}\,] \,\psi_{j}$. This can never be obtained
from a succession of two $2 \times 2$ SL(2,${\mathbb{C}}$) operators like the one that occurs in Eq. \ref{equation-in-Weyl-3}. A succession
of three such operators is needed. With the extended substitution one obtains the succession  
$ [\,{\mathbf{\nabla}}{\boldsymbol{\cdot\sigma}}\,] \,V \, [\,{\mathbf{v}}{\boldsymbol{\cdot\sigma}}\,] \,\psi_{j}$ after squaring the Dirac equation, and this
will lead to the required spin-orbit term.
The traditional approach is unable to accomplish this because it misses the presence of $\gamma\,  [\, {qV\over{c}}\, {\mathbf{v}}{\boldsymbol{\cdot\sigma}}\,] $
in the minimal substitution. 

\subsubsection{The origin of the error: the solutions are mixed states} \label{crucial-error1}

Nevertheless, the traditional approach presents a derivation of the spin-orbit term whereby it even seems to account correctly  for the Thomas precession.
This traditional derivation of the spin-orbit coupling is a piece of wrong algebra that by good fortune produces the correct physical result desired. 
As explained in the preceding lines, it starts from the wrong minimal substitution which can never
yield the desired result because it does not contain the interactions with the current. It then introduces a second error to obtain 
the  result that agrees with the experimental observations by brute force.  
What  the second error is can be discovered by considering the Dirac equation for an electron at rest in the absence of any electromagnetic field. 
In the Weyl representation this leads to:

\begin{align} \label{equation-in-Weyl-2b}
[\,(-{\hbar\over{\imath}} {\partial \over{\partial ct}})\, \bigone\,]\, \psi_{2} & = m_{0} c  \,\psi_{1},\nonumber\\
[\,(-{\hbar\over{\imath}} {\partial \over{\partial ct}})\, \bigone \,]\, \psi_{1} & = m_{0}c \,\psi_{2},
\end{align}

\noindent and after ``squaring'' to:

\begin{align} \label{equation-in-Weyl-3b}
[\,(-\hbar^{2} {\partial^{2} \over{\partial c^{2}t^{2}}})\, \bigone\,]\, \psi_{1} & = m_{0}^{2} c^{2}  \,\psi_{1},\nonumber\\
[\,(-\hbar^{2} {\partial^{2} \over{\partial c^{2}t^{2}}})\, \bigone\,]\, \psi_{2} & = m_{0}^{2} c^{2}  \,\psi_{2},
\end{align}

\noindent A viable simultaneous solution of Eqs. \ref{equation-in-Weyl-2b} and \ref{equation-in-Weyl-3b}   is:

\begin{equation} \label{solveWeyl}
 \psi_{1} = \psi_{2}= \, e^{-\imath m_{0}c^{2}t/\hbar} \left (
\begin{tabular}{c}
$1$\\
$0$\\
\end{tabular}
\right ). 
\end{equation}

\noindent In the traditional representation, the Dirac equation becomes:

\begin{align} \label{equation-in-Dirac-2b}
[\,(-{\hbar\over{\imath}} {\partial \over{\partial ct}})\, \bigone\,]\, \psi_{1} & = m_{0} c  \,\psi_{1},\nonumber\\
[\,(-{\hbar\over{\imath}} {\partial \over{\partial ct}})\, \bigone \,]\, \psi_{2} & = - m_{0}c \,\psi_{2}.
\end{align}

\noindent This needs no further squaring as the equations are already decoupled. One possible solution is: 

\begin{equation} \label{solveDirac}
 \psi_{1} = \, e^{-\imath m_{0}c^{2}t/\hbar} \left (
\begin{tabular}{c}
$1$\\
$0$\\
\end{tabular}
\right ), \quad 
 \psi_{2} = \, e^{+\imath m_{0}c^{2}t/\hbar} \left (
\begin{tabular}{c}
$1$\\
$0$\\
\end{tabular}
\right ). 
\end{equation}

\noindent What transpires from the solution in the traditional presentation is that the spinor is a mixed state. It contains  two pure states that lead to
the same rest energy $m_{0}c^{2}$ for the electron, one with the electron spinning counterclockwise and one with the electron spinning clockwise around the $z$-axis:

\begin{equation} \label{solveDiracb}
\psi = 
e^{-\imath m_{0}c^{2}t/\hbar} \left (
\begin{tabular}{c}
$1$\\
$0$\\
$0$\\
$0$\\
\end{tabular}
\right ) +
e^{\imath m_{0}c^{2}t/\hbar} \left (
\begin{tabular}{c}
$0$\\
$0$\\
$1$\\
$0$\\
\end{tabular}
\right ). 
\end{equation}

\noindent The mixed state must of course still be normalized. In the mixed state each pure state has thus the same probability ${1\over{2}}$. 
In fact, the rotation angle $\omega_{0} t$ is an algebraic quantity.  
Feynman found out also that one has to use such mixed states 
composed of pure states with opposite sign and equal probabilities in quantum electrodynamics.
 In following the tradition to associate negative-energy states with anti-particles,
the mixed state contains particles and anti-particles with equal probabilities.

Feynman wondered what the meaning of this would be. The solution of that riddle is that the mixed states describe statistical ensembles of electrons.
Negative frequencies already occur in SU(2), which is just Euclidean geometry and does not contain anti-particles. There is therefore
 absolutely no necessity to associate negative frequencies with anti-particles. It is more appropriate to associate
negative frequencies with particles that spin the other way around than those that give rise to positive frequencies. Both signs of the frequency
will however yield the same kinetic rotational energy, and therefore the same rest mass and energy for the electron.
Considering such a mixed state makes thus perfect physical sense, while the mixed state based on antiparticles is puzzling.

Of course, the interpretation of the negative frequencies we are proposing here clashes once more acrimoniously with currently accepted notions.
But  in  reference \cite{Coddens} we give a derivation of the Dirac equation, that does not rely at all on the existence of anti-particles.
Charge conjugation symmetry is just not part of the picture. As we explain in reference \cite{Coddens} we can consider anti-particles in a later stage,
and then the traditional  arguments can be used to show that we should associate them with negative frequencies. But this is then a different representation than the original one. The situation is a little bit like that with the two different versions of SL(2,${\mathbb{C}}$). Just as the two versions of SL(2,${\mathbb{C}}$) are related one to another by parity transformation and
should not be merged into a single  SL(2,${\mathbb{C}}$) formalism, the two particle and anti-particle representations are related by charge-conjugation
and should not be merged into a single formalism, because a negative frequency would then acquire two different, mutually exclusive interpretations. Moreover in the merged formalism, the rest energies of a positron and an electron
are truly considered as opposite, such that they would add up to zero, while they should add up to twice 511 keV. Of course, the alternative interpretation
of the negative frequencies puts the whole discussion about Majorana and Dirac neutrinos in a different context.\footnote{
Furthermore, it is impossible to derive the Dirac equation for a particle with zero rest mass, because in the derivation given in reference \cite{Coddens} 
it is necessary to consider a particle in its rest frame where $2 m_{0}c^{2} = \hbar\omega_{0}$.
In reality neutrinos do have mass, which further wrong-foots the discussion about Majorana and Dirac neutrinos, because the wave functions of 
the traditional Dirac neutrinos
are considered to be just two-component Weyl spinors, while neutrino wave functions that are solutions of a Dirac equation with non-zero rest mass must be 
four-component mixed states, just like the so-called bi-spinor electron wave functions.\label{Majorana}}

Also in the case of the Weyl representation the states are mixed, even if it is less obvious here because the two pure states carry the same sign
for $\omega_{0}t$. But in the Weyl representation,
the equation  for the pure state of a single electron in its rest frame is:

\begin{equation} \label{one-electronequation-in-Weyl}
 \left (
\begin{tabular}{cc}
& $-{\hbar\over{\imath}} {\partial \over{\partial c\tau}} \, \bigone $\\
   $-{\hbar\over{\imath}} {\partial \over{\partial c\tau}} \, \bigone $ & \\
\end{tabular}
\right ) 
\left (
\begin{tabular}{cc}
$\Psi$ &\\
& $\Psi^{-1\dagger}$\\
\end{tabular}
\right ) = 
 m_{0} c 
 \left (
\begin{tabular}{cc}
& ${\mathbf{s}}{\boldsymbol{\cdot\sigma}} $\\
   ${\mathbf{s}}{\boldsymbol{\cdot\sigma}} $ & \\
\end{tabular}
\right ) 
\left (
\begin{tabular}{cc}
$\Psi$ &\\
& $\Psi^{-1\dagger}$\\
\end{tabular}
\right ).
\end{equation}

\noindent Here the $\Psi$ is a two-column SL(2,${\mathbb{C}}$) matrix representing the spinning electron, while $\Psi^{-1\dagger}$
represents the spinning electron in the SL(2,${\mathbb{C}}$) representation of opposite handedness; $\tau$ is the proper time.
The block-diagonal $4\times 4$ matrix ${\mathbf{D}}(g)$ with the blocks $\Psi$ and $\Psi^{-1\dagger}$ on the diagonal is the Weyl representation 
of a group element $g$ obtained from an even number of reflections and represents the spinning
electron. This equation can never lead to the Dirac equation, because the right-hand side can never be simplified according to:

\begin{equation} \label{one-electronequation-in-Weyl-simplify}
 m_{0} c 
 \left (
\begin{tabular}{cc}
& ${\mathbf{s}}{\boldsymbol{\cdot\sigma}} $\\
   ${\mathbf{s}}{\boldsymbol{\cdot\sigma}} $ & \\
\end{tabular}
\right ) 
\left (
\begin{tabular}{cc}
$\Psi$ &\\
& $\Psi^{-1\dagger}$\\
\end{tabular}
\right ) =
 m_{0} c 
 \left (
\begin{tabular}{cc}
$\Psi$ &\\
& $\Psi^{-1\dagger}$\\
\end{tabular}
\right ).
\end{equation}

\noindent In fact, let us divide out $m_{0}c$ on both sides. Then on the left-hand side the non-zero blocks are off-diagonal, while on the right-hand side they are diagonal. 
The two sides have different symmetries and they can therefore never be identical.
In the traditional representation this kind of observation becomes hidden by the fact that there are no longer vanishing blocks.
Due to the fact that the simplification expressed in Eq. \ref{one-electronequation-in-Weyl-simplify} is not possible, 
one is forced to adopt a linear combination of single-electron states to obtain the Dirac equation. 
This linear combination is (see reference \cite{Coddens}):

\begin{equation} \label{linear-combination-in-Weyl}
{\mathbf{M}} = {\mathbf{D}}(g) +  \left (
\begin{tabular}{cc}
& ${\mathbf{s}}{\boldsymbol{\cdot\sigma}} $\\
   ${\mathbf{s}}{\boldsymbol{\cdot\sigma}} $ & \\
\end{tabular}
\right ) {\mathbf{D}}(g), 
\end{equation}

\noindent and we will have then:

\begin{equation} \label{linear-combination-in-Weyl-b}
 \left (
\begin{tabular}{cc}
& ${\mathbf{s}}{\boldsymbol{\cdot\sigma}} $\\
   ${\mathbf{s}}{\boldsymbol{\cdot\sigma}} $ & \\
\end{tabular}
\right ) {\mathbf{M}} = {\mathbf{M}}. 
\end{equation}

\noindent This shows that also in the Weyl representation the solutions of the Dirac equation are mixed states.

Mixed states do not have a direct obvious meaning in the pure group theory. E.g. the sum of two rotation matrices is not a rotation matrix, and therefore the
sum of two spinors is {\em a priori} not a new spinor.
Group elements cannot be added, they can only be multiplied. Linear combinations of representations of group elements do not represent new group elements.
They belong to the so-called group ring.
One can give such a mixed state however a meaning by considering it as defining a statistical ensemble, just as is done in the traditional interpretation
of such mixed states in quantum mechanics. We can thus consider the free-space solution as degenerate, and the presence of electromagnetic fields
can lift this degeneracy.

We may note that in the simpler formalism of SU(2), the solution of the eigenvalue equation 
${\hbar\over{2}} \,[\,{\mathbf{s}}{\boldsymbol{\cdot\sigma}}\,] \psi = \pm {\hbar\over{2}} \psi$ for a spin aligned with the unit vector ${\mathbf{s}}$ 
is also a mixed state. The not normalized ``spinor'' $\psi$ is the first column of the sum ${\mathbf{S}} = \bigone + {\mathbf{s}}{\boldsymbol{\cdot\sigma}}$
 of the two group elements $\bigone$ and  $ {\mathbf{s}}{\boldsymbol{\cdot\sigma}}$.
For this sum, we will have  $[\,{\mathbf{s}}{\boldsymbol{\cdot\sigma}}\,]\, {\mathbf{S}} = {\mathbf{S}}$. 
In other words: $\psi = (\bigone + {\mathbf{s}}{\boldsymbol{\cdot\sigma}})\, \psi_{1}$,
where $\psi_{1}$ is the first column of ${\mathbf{s}}{\boldsymbol{\cdot\sigma}}$. This leads indeed to $\psi = [\, {\mathbf{s}}{\boldsymbol{\cdot\sigma}}\,]\, \psi$.
When ${\mathbf{s}} = {\mathbf{e}}_{z}$, the mixed nature of $\psi$ becomes  concealed after normalization by the fact that 
$\psi_{1}$ and $[\, {\mathbf{s}}{\boldsymbol{\cdot\sigma}}\,]\, \psi_{1}$ accidentally
take the same numerical values. This is also further discussed in reference \cite{Coddens}.

\subsubsection{Problems in taking the  Schr\"odinger limit of the mixed states} \label{crucial-error2}

Now, the non-relativistic Schr\"odinger limit is obtained by removing the rest mass from the equation.
 If we want to get rid of the rest mass in the Weyl representation,
we can define the classical wave function:

\begin{equation} \label{substitution-in-Weyl}
\left (
\begin{tabular}{c}
$\psi^{cl}_{1}$\\
$\psi^{cl}_{2}$\\
\end{tabular}
\right ) =
e^{\imath m_{0}c^{2}t/\hbar}
\left (
\begin{tabular}{c}
$\psi_{1}$\\
$\psi_{2}$\\
\end{tabular}
\right ). 
\end{equation}

\noindent After such a substitution, the rest mass will be removed from the calculations.
But  using the same trick in the traditional representation 
(see e.g. \cite{Itzykson}, page 65) will lead to a meaningless statistical ensemble
containing states with rest masses  $0$ and $2m_{0}$. The correct substitution to be used in the traditional representation is thus:

\begin{equation} \label{substitution-in-Dirac}
\left (
\begin{tabular}{c}
$\psi^{cl}_{1}$\\
$\psi^{cl}_{2}$\\
\end{tabular}
\right ) =
\left (
\begin{tabular}{cc}
$e^{\imath m_{0}c^{2}t/\hbar}$ &\\
& $e^{-\imath m_{0}c^{2}t/\hbar}$\\
\end{tabular}
\right )
\left (
\begin{tabular}{c}
$\psi_{1}$\\
$\psi_{2}$\\
\end{tabular}
\right ), 
\end{equation} 

\noindent rather than Eq. \ref{substitution-in-Weyl}, because this is the substitution
that leads to a meaningful statistical ensemble. This substitution would look absurd if we thought that the state given by Eq.
\ref{solveDiracb} is a pure state. But for the mixed state, this substitution does make sense.
This difference between the substitutions needed in Eqs. \ref{substitution-in-Weyl} and \ref{substitution-in-Dirac} just reflects the difference
 between the choices $\gamma_{4}$ and $\gamma_{5}$ for the gamma matrix that has to be associated  ${\partial\over{\partial ct}}$ in the algebra.
 It is actually much more logical to write the Dirac equation in the traditional representation as:
 
\begin{equation} \label{logical-Dirac}
\left [\,- {\hbar\over{\imath}}\, {\partial \over{\partial ct}} \left (
\begin{tabular}{rr}
$\bigone$ &\\
& $-\bigone$\\
\end{tabular}
\right )
 + 
\left (
\begin{tabular}{rr}
 & ${\hbar\over{\imath}} {\mathbf{\nabla}}{\boldsymbol{\cdot\sigma}}$\\
$ - {\hbar\over{\imath}} {\mathbf{\nabla}}{\boldsymbol{\cdot\sigma}}$ & \\
\end{tabular}
\right )\,\right ]\,\psi = m_{0} c \,\psi,
\end{equation} 

\noindent thereby considering $- { \hbar\over{\imath}} \, {\partial \over{\partial ct}}\,\gamma_{t} $ rather than
$ - { \hbar\over{\imath}}\, {\partial \over{\partial ct}}$ as the energy operator that generalizes the prescriptions that are valid in the context of the Schr\"odinger equation.
All this  is of course a major watershed because it invalidates the whole philosophy of
 small and large  $2\times 2$ representations in the Dirac equation. However, there are just two SL(2,${\mathbb{C}}$) representations in a
$4\times 4$ representation based on gamma matrices, and these two SL(2,${\mathbb{C}}$) representations are mathematically of ``the same size''. 
 As can be seen from Subsection \ref{Lorentz-group}, they are defined in a completely symmetrical way.
 They are just of different handedness.
 If we refuse to accept this, then the Weyl representation
would have two representations that are equally large, while the traditional representation would have a small and a large one.
Why should such  differences all at once pop up in the application of the mathematics if the physics in the two choices of gamma matrices
must be the same? 

It is further very important to realize that it is absolutely not necessary to make the substitution of Eqs. \ref{substitution-in-Weyl} or \ref{substitution-in-Dirac} 
in order to  obtain the non-relativistic limit, as we showed above.
One can  do the calculations fully relativistically. It suffices then to consider afterwards the limit whereby ${\mathbf{v}}$ becomes small and to subtract $m_{0}c^{2}$
from the final result. But when we do the calculations that way, we will never get a term $2m_{0}$ into the algebra.
 It is the division by the term $2m_{0}$ in the equations that one obtains from the wrong substitution
which leads to the illusion that the algebra can describe the Thomas half correctly. This shows that it is not possible
to obtain the Thomas half from the Dirac formalism, and that this is true both in  the Weyl and in the traditional representation!

Of course it could be argued that we could solve Eq. \ref{equation-in-Dirac-2b} by taking another solution  than Eq. \ref{solveDirac}, whereby
we keep $\psi_{1}$ and put $\psi_{2} = 0$. We would then obtain a pure state. But  this solution will also no longer lead to a term $2m_{0}$ in the algebra, and
it could only be used for an electron at rest.

Finally,  the traditional approach suggests that up to a certain order
 the expression for the Thomas precession  
in terms of the quantity ${\mathbf{v}}\wedge {\mathbf{E}}$
would be generally valid and not depend on the geometry of the orbit, while 
 the further developments will show that it is in many instances necessary to assume that the motion is  uniform and circular.

\subsubsection{Methodological remarks} \label{spin-orbit-method-remarks}

The  conclusion that the traditional calculation of the spin-orbit coupling is wrong is very unsettling.  
The reason why it has not been noticed that this error crept in, is that in the traditional approach the Dirac equation has been guessed. 
This means that we use it without knowing
on what kind of precise assumptions it is built. In the approach described in reference \cite{Coddens}, the equation has been derived.
This enables discerning a number of issues that in the traditional approach are not even suspected. It is very hard to suspect that
the solutions of the Dirac equation have to be mixed states. And it remains difficult to spot that this leads to errors  creating the {\em illusion}
that the spin-orbit coupling would be treated correctly by the Dirac equation,
even when one knows the underlying assumptions. All this reveals the limits of the nonchalant advice that one should just ``shut up and calculate'' 
 because
it would ``work''. But contrary to the claims, it does not work. As seen here, following the advice in a blindfolded way can lead to errors of appreciation. 
Spotting such errors and teasing  out their consequences can prove a very difficult task.

Of course it is not easy to decide which approach is correct and which one is wrong when we pit two different approaches one against another.
It is in this respect always better to give several arguments in support of a conclusion, in order to prove the internal consistency of the logic.
Here we are giving four arguments. 

(1) The first one is based on the fact that the solutions of the Dirac equation are mixed states. This may
be less convincing to the reader because he does not know the results derived in reference \cite{Coddens}, but again these results have an internal
consistence. 

(2) The second  one is based on the fact  that the approach
based on the Weyl representation should yield the same result as the approach based on the traditional representation. 
This is the argument that should be really clear to all readers.

(3) A  third one is based on the fact that the derivation of the Dirac equation ignores instantaneous accelerations.
One should indeed not be surprised that the calculation does not take into account the Thomas precession. As mentioned above and  in 
Subsubsection \ref{A-der}, the Dirac equation is derived by using only 
boost parameters, like e.g. $(E,c{\mathbf{p}})$ in the case of the free-space equation. 
It considers only instantaneous boosts, not instantaneous accelerations. The equation makes sure that we get the instantaneous local boost correct
in every point $({\mathbf{r}},t)$.
Thomas precession can only occur in a sequence of non-collinear boosts. It is thus the result
of an instantaneous acceleration perpendicular to the velocity. An instantaneous boost as used to derive the Dirac equation
can thus  not account for  Thomas precession. 
This is also discussed in further detail in \cite{Coddens}.

(4) One can appreciate the point that the Dirac equation with a four-potential only treats boosts and not Thomas precession 
also as follows. Defining a general Lorentz transformation
requires six independent parameters. The electromagnetic four-potential is defined by four parameters, but they are linked by the Lorentz gauge condition,
such that the four-potential contains three independent parameters. These three parameters define the boost part of the Lorentz transformation, while
the three remaining  independent parameters of the general Lorentz transformation correspond to the rotational part of it. These can thus not be defined
by the Dirac equation with a four-potential, because there are just no parameters in the equation that would
permit defining them.

A weak point  of the Dirac equation is thus that it is  covariant with  respect to boosts, but not with respect to  instantaneous accelerations that are perpendicular
to the instantaneous velocity.

\subsection{A more physical approach to the spin-orbit coupling} \label{spin-orbit-physics}

\subsubsection{Road map} \label{road-map}

Thomas precession is by definition an effect in the co-moving frame. It is a mis-appreciation of the correct clock rate of the electron  due to the fact 
that the co-moving frame is rotating.  According to Purcell's  calculation of the Thomas precession \cite{Purcell} the Thomas precession corresponds
exactly to the difference of the merry-go-round effects in the laboratory frame and in the co-moving frame.
The two merry-go-round effects are different due to Lorentz contraction (or time dilatation).
Of course effects of the motion on the rest mass must be made in the co-moving frame, as it is in this frame
that the electron is at rest. These effects must then be transformed back to the laboratory frame.
But whereas Purcell's calculation gives the correct algebraic result, it does not tell us why we can make the calculation the way he does.
Why do we need to consider the difference between two merry-go-round effects?

A similar problem exists for the other part of the spin-orbit effect, the part that occurs (up to a proportionality factor) in Eq. \ref{forces2}
and is not corrected for Thomas precession. In our derivation of Eq. \ref{forces2} it is mere algebra.
As this is not satisfactory, one would like a physical explanation for it.
Such a classical derivation for it has been proposed in Subsection \ref{similar}. 
But also here the argument developed leads to the correct algebraic result, while it is not completely explained what the idea is behind the calculation.
We may wonder e.g. why  one should  consider the magnetic field
experienced by the electron in the co-moving frame in the first place, while we want to calculate results in the laboratory frame.
 Is it not possible to make a calculation in the laboratory frame right ahead? 
In the calculation only the magnetic part of the electromagnetic field in the co-moving frame intervenes. 
Due to our previous discussion of the anomalous Zeeman effect,
we understand very well the effects of the magnetic field on the electron in its rest frame.  In this calculation we only consider the instantaneous boost. 
This result must of course be back-transformed to the laboratory frame.
What remains to calculate, is the effect of the accelerations perpendicular to the instantaneous velocity,
and this is the Thomas precession. The Thomas precession due to the magnetic field can to a first approximation be neglected.
What remains is thus the effect of the component of the electric field that is perpendicular to the instantaneous velocity, and that will be the Thomas precession.

In the following we will talk a lot about phases corresponding to the merry-go-round effect as opposed to Berry phases. We will however not consider
such phases accumulated over a whole loop, but rather the instantaneous rates of change of them, because these correspond to the idea of precession.
What we want to explain  are the following points:

(1) One can consider two types of precession, a merry-go-round effect in position space ${\mathbb{R}}^{3}$ (which is not curved) 
and a precession in velocity space. The velocity space
is curved and helps actually to visualize  the group manifold. The precession on  this curved space yields the Berry phase after an integration over a closed loop.

(2)  The instantaneous 
rate of change in phase due to the merry-go-round effect in position space corresponds to the part of the spin-orbit effect without Thomas precession. It is
the part that occurs in Eq. \ref{forces2}. 

(3) This instantaneous 
rate of change in phase due  to the merry-go-round  effect in position space 
does not correspond to the exact rate of change of the geometrical phase  due to parallel transport. The exact calculation of the latter can
be made by first considering the merry-go-round effect in position space and then making a correction to it.

(4) The correction we have to carry out to obtain the correct rate of change of geometrical phase is the Thomas precession.
According to Purcell's calculation, it is the difference between the merry-go-round effects  in velocity space in the co-moving frame and in the laboratory frame. 
Perhaps it is worth pointing out that there is no real merry-go-round effect
on the group. When we make a closed orbit on the group, we are getting back to a Lorentz transformation that is identical to the one we started from,
such that the phase difference can only be a multiple of $2\pi$. That we have to worry about the Thomas precession is due to the fact that we are dealing
with a closed orbit in ${\mathbb{R}}^{3}$ rather than on the group and that we have constructed the Dirac equation by only taking into account
the boost part of the Lorentz transformations (see Footnote 
\ref{construction}).

In a merry-go-round calculation one basis vector of the co-moving frame remains always aligned with the instantaneous
velocity.  The problem is thus that such a continuous alignment with the tangent to the orbit in position space  does in general not correspond to parallel transport.
It is possible to get a feeling for this by an analogy with the motion of a triad of basis vectors along a closed loop on the surface of a sphere.
The closed loop could e.g. be a small circle.
  Let us assume we have defined spherical coordinates $(r,\theta,\phi)$
and local frames with basis vectors ${\mathbf{e}}_{r},  {\mathbf{e}}_{\theta}, {\mathbf{e}}_{\phi}$ as usual, such that the $z$-axis corresponds to $\theta = 0$. 
In a trip with uniform velocity $v$ 
along a small circle defined by $\theta = \theta_{0} \neq 0$, the geometrical phase will not be given by $2\pi$ as one could be tempted
to conclude from the permanent alignment of ${\mathbf{e}}_{\phi}$ with the instantaneous velocity ${\mathbf{v}}$.
This permanent alignment corresponds to the merry-go-round effect, and does not account correctly for the value of the geometrical phase.
As shown by the Gauss-Bonnet theorem, the geometrical phase will be given by $\Omega = 2\pi (1-\cos\theta_{0})$, where $\Omega$ is the solid angle
subtended by the small circle. Only when $\theta_{0} = {\pi\over{2}}$ will the procedure of aligning ${\mathbf{e}}_{\phi}$ with ${\mathbf{v}}$
yield the correct result, because then the path is a geodesic. The calculation $\Omega = 2\pi (1-\cos\theta_{0})$ reproduces exactly what happens e.g. in Foucault's pendulum. This will be further developed in Subsubsection \ref{two-merry}.

We may note that the calculation of the global Wigner rotation over a closed path in reference \cite{Rhodes} is based on a theorem of hyperbolic geometry
that just rephrases the Gauss-Bonnet theorem. This shows very clearly that  the angle built up by Thomas precession along a path in velocity space is the Berry phase.
As the hyperbolic geometry is the geometry of velocity space, we are getting here a first clue for the existence
of two precession effects. One in the hyperbolic velocity space, and one in position space. 
Instead of a  path in a curved space of positions, Thomas precession occurs on a path in a curved space of velocities.
This curved space represents actually the parameters that define an element of the Lorentz group. The velocities are the boost parameters and correspond to the boosts, the Berry angles correspond to the rotations.

 In summary, we can state that the Dirac equation is wrong because it does not account
properly for the parallel transport, such that it gets the Berry phase wrong. The reason for this is that the Dirac equation just calculates
 the merry-go-round effect in position space. This idea is already present in reference \cite{Coddens}, but I was just unable to find the 
appropriate words to verbalize it. In reference  \cite{Coddens}, it was also discussed how strange it is to postulate that the wave function is a function.
It means that the Berry angle over a given path must be multiple of $2\pi$ and it is this feature that leads to quantization.
But phases differences that are a multiple of $2\pi$ occur only on geodesics. Hence, quantum mechanics postulates that the
orbits must be geodesics with respect to the spinning motion on the group manifold, and are therefore quantized. General relativity postulates that orbits must be geodesics with respect to displacement motion.
This also implies rotations to a certain extend, but it does not consider the rotation due to the spin. The fully correct picture
would thus imply that the orbits must be geodesics with respect to the combined effect of rotation and translation.

\subsubsection{The Berry phase on the surface of a sphere} \label{two-merry}

The merry-go-round effect on a circle with radius $r$ in the $Oxy$ plane can obtained by considering the infinitesimal angle $\Delta \theta  = \Delta r_{\perp}/r$.
 In vector form this can be rewritten as $\Delta \theta\, {\mathbf{e}}_{z} = {\mathbf{r}}\wedge \Delta {\mathbf{r}}_{\perp} /r^{2}$.
 This leads to ${\boldsymbol{\omega}} = {d\theta\over{dt}}\,{\mathbf{e}}_{z}$ $=$ 
 ${1\over{r^{2}}} {\mathbf{r}}\wedge {\mathbf{v}}$. But such a calculation corresponds only to the correct value for the
 geometrical phase in flat space (which is here the $Oxy$ plane). It is no longer true if we consider the geometrical phase along a small circle with radius
 $r$ on the surface of a sphere with radius $R$, as it does not account for
 the curvature of the surface of the sphere.

How the correct calculation of the geometrical phase runs in curved space can be illustrated
on this example of a motion
along a small circle with radius $r$ on a sphere with radius $R$. This case study corresponds exactly to  description of Foucault's pendulum. 
Let us take the rotation axis of the Earth as the $z$-axis. The locally vertical direction on the surface of the sphere is given by the vector ${\mathbf{e}}_{R}$.
The precession of Foucault's pendulum can be trivially calculated from 
$({\boldsymbol{\omega}}{\mathbf{\cdot e}}_{R})\, {\mathbf{e}}_{R}$, which is the component
of ${\boldsymbol{\omega}} = \omega {\mathbf{e}}_{z} $ along ${\mathbf{e}}_{R}$. This shows that the calculation in curved space cannot be made
by using ${1\over{r^{2}}} {\mathbf{r}} \wedge {\mathbf{v}}$, whereby ${\mathbf{r}}$ would be the position vector of the pendulum with respect
to the centre of the small circle. 

In the calculation
one must take  $d{\mathbf{r}} = dr \,{\mathbf{e}}_{\phi}$ where $r = R\cos\theta$, and
$dr= r \,d\phi$. The term $d{\mathbf{r}}/r$ must then be multiplied by ${\mathbf{e}}_{\phi}Ê\wedge {\mathbf{R}}/R = {\mathbf{e}}_{\theta}$, 
rather than ${\mathbf{r}}/r$, because what one wants to calculate
is the rotation in the local tangent plane spanned by ${\mathbf{e}}_{\phi}, {\mathbf{e}}_{\theta}$. This is a rotation  around the local
vertical defined by the vector ${\mathbf{e}}_{R}$. This calculation leads to ${\mathbf{e}}_{\theta} \wedge {\mathbf{e}}_{\phi} \, d\phi = d\phi \,{\mathbf{e}}_{R}$.
The integral over the circle of this quantity yields $2\pi \cos\theta \,{\mathbf{e}}_{z}$. The de-phasing is then $\Delta \phi = 2\pi  - 2\pi \cos\theta$, or
$2\pi (1 - \cos\theta)\,{\mathbf{e}}_{z}$ in vector form. This is equal to $\iint {1\over{R^{2}}}{\mathbf{e}}_{R} {\mathbf{\cdot}} d{\mathbf{S}}$, where $ d{\mathbf{S}} = R^{2} \sin\theta \,d\theta \,d\phi\,
{\mathbf{e}}_{R}$, in agreement with the Gauss-Bonnet theorem. The quantity $\Delta \phi$ is the Berry phase, which can be expressed as $\oint \,{\mathbf{A\cdot}}d{\mathbf{r}}$,
where the vector quantity ${\mathbf{A}}$ is analogous to a vector potential.
Putting ${1\over{R^{2}}} {\mathbf{e}}_{R} = {\mathbf{\nabla}}Ê\wedge {\mathbf{A}}$ permits to rewrite
$\Delta \phi = \iint \, ( {\mathbf{\nabla}}Ê\wedge {\mathbf{A}})\,{\mathbf{\cdot}}d{\mathbf{S}} = \oint\,{\mathbf{A}}d{\mathbf{r}}$. Here $d{\mathbf{r}} = R\,\sin\theta \,d\phi \,{\mathbf{e}}_{\phi}$, such that one must thus take
${\mathbf{A}} = {1 - \cos\theta\over{R \,\sin\theta}} {\mathbf{e}}_{\phi}$ to obtain the correct result. Using the general expression for ${\mathbf{\nabla}} \wedge {\mathbf{A}}$
in spherical coordinates one can verify that  this value of ${\mathbf{A}}$ leads indeed to ${\mathbf{\nabla}} \wedge {\mathbf{A}} = {1\over{R^{2}}} {\mathbf{e}}_{R}$. As ${\mathbf{A}}$ is a vector, one is tempted to interpret it as a vector potential. 

The vector ${\mathbf{A}}$ shows many similarities with a magnetic vector potential. But of course it is  not a magnetic  vector
potential, because there are no magnetic fields in our problem. The vector ${\mathbf{A}}$ does thus not define a magnetic monopole. 
This can be seen from the theorem of Gauss for an electric field: $\oiint {q\over{4\pi\varepsilon_{0} R^{2}}} {\mathbf{e}}_{R}{\mathbf{\cdot}}d{\mathbf{S}} = {q\over{\varepsilon_{0}}} = \iiint \, \rho({\mathbf{r}})\,d{\mathbf{r}}$. Here $\rho({\mathbf{r}}) = q \delta({\mathbf{r}})$. 
This is the sum of two surface integrals $ \iint_{S_{1}}\, {\mathbf{E\cdot}}d{\mathbf{S}} = {q\over{\varepsilon_{0}}} 2\pi (1-\cos\theta)$ and  $\iint_{S_{2}}\, {\mathbf{E\cdot}}d{\mathbf{S}} = {q\over{\varepsilon_{0}}} 2\pi (1+\cos\theta)$ over the two
areas $S_{1}$ and $S_{2}$ on both sides of the small circle, and which together make up the full sphere.  
The result for one of the two areas is then perfectly analogous with the
calculation we have performed above. As there is by definition only an electric field 
${\mathbf{E}} = {q\over{4\pi\varepsilon_{0} R^{2}}} \,{\mathbf{e}}_{R}$ within this problem,
the quantity ${\mathbf{A}} = {q\over{\varepsilon_{0}}} {1\over{R}} \tan {\theta\over{2}} \, {\mathbf{e}}_{\phi} $, which leads to ${\mathbf{\nabla}}Ê\wedge {\mathbf{A}} = {\mathbf{E}}$, is  not a magnetic potential. We have always
${\mathbf{\nabla\cdot E}} = {\rho({\mathbf{r}})\over{\varepsilon_{0}}}$, with 
 $\rho({\mathbf{r}}) = q\delta({\mathbf{r}})$. For all points different from the origin,
the fact that  ${\mathbf{E}} = {\mathbf{\nabla}} 
\wedge {\mathbf{A}}$  leads to ${\mathbf{\nabla\cdot}}({\mathbf{\nabla}}\wedge {\mathbf{A}}) = 0$. It  can be checked indeed that
 ${\mathbf{\nabla\cdot E}} = 0 $ for ${\mathbf{r}} \neq {\mathbf{0}}$ by using the general expression for the divergence in spherical coordinates.  
 The pitfall is now to conclude from this
 that $\oiint ({\mathbf{\nabla}}\wedge {\mathbf{A}}){\mathbf{\cdot}}d{\mathbf{S}} = 
  \iiint {\mathbf{\nabla\cdot ({\mathbf{\nabla}}\wedge {\mathbf{A}})}}  \,d{\mathbf{r}} = 0$. In fact, one expects ${\mathbf{\nabla\cdot ({\mathbf{\nabla}}\wedge {\mathbf{A}})}}  =0$ by analogy with ${\mathbf{a\cdot}}({\mathbf{a}}\wedge {\mathbf{b}}) =0$, but this is not true at the origin where ${\mathbf{A}}$ has a singularity.
 In putting ${\mathbf{\nabla\cdot ({\mathbf{\nabla}}\wedge {\mathbf{A}})}}  =0$, we forget the singularity in ${\mathbf{\nabla\cdot E}}$ at the origin.
As  $\oiint\,({\mathbf{\nabla}} \wedge {\mathbf{A}}){\mathbf{\cdot}}d{\mathbf{S}} = \iiint {\mathbf{\nabla\cdot E}}  \,d{\mathbf{r}}$, we obtain the correct result when we use
${\mathbf{\nabla\cdot E}} = {\rho({\mathbf{r}})\over{\varepsilon_{0}}}$.
 The same reasoning 
can be applied to the magnetic charge $q_{m}\,\delta({\mathbf{r}}) = cq\,\delta({\mathbf{r}})$ following the analogy 
developed in Footnote \ref{magnetic-charge}. 

The mathematical difficulties that arise here are due to the abstraction of representing the charge distribution by $q\delta({\mathbf{r}})$. The delta ``function''
can be described as a limit
of test charge distributions $\lim_{\lambda \rightarrow \lambda_{0}} \,\rho_{\lambda}({\mathbf{r}})$. The problem is that for integration and differentiation
we are not allowed to assume that
the derivative of the limit will be the limit of the derivative, or the integral of the limit will be the limit of the integral:
$\lim_{\lambda \rightarrow \lambda_{0}} \int \,f({\mathbf{r}}) \rho_{\lambda}({\mathbf{r}})d{\mathbf{r}} $ $\neq$ 
$ \int \, [\, \lim_{\lambda \rightarrow \lambda_{0}} f({\mathbf{r}}) \rho_{\lambda}({\mathbf{r}})\,]\,d{\mathbf{r}} $, and
$\lim_{\lambda \rightarrow \lambda_{0}}\, D  [\, f({\mathbf{r}}) \rho_{\lambda}({\mathbf{r}})\,] $ $\neq$ 
$ D\,[\, f({\mathbf{r}}) \lim_{\lambda \rightarrow \lambda_{0}} \rho_{\lambda}({\mathbf{r}})\,] $.
Changing the order of the operations is in general not a valid procedure. This error occurs in Dirac's definition of the ``delta function'', and now also here
in the differentiation procedure, where it gives rise to the confusion that  ${\mathbf{\nabla\cdot ({\mathbf{\nabla}}\wedge {\mathbf{A}})}}$ would be zero at
${\mathbf{r}} = {\mathbf{0}}$. The limits must be defined in the sense of distributions, and then these difficulties
will disappear.  In fact, ${\mathbf{\nabla\cdot ({\mathbf{\nabla}}\wedge {\mathbf{A}})}}$ is not zero at the origin, even though $\lim_{{\mathbf{r}} \rightarrow {\mathbf{0}}} {\mathbf{\nabla\cdot ({\mathbf{\nabla}}\wedge {\mathbf{A}}({\mathbf{r}}))}} =0$. The correct result for the example of the electric monopole is  
${\mathbf{\nabla\cdot ({\mathbf{\nabla}}\wedge {\mathbf{A}})}} = {\mathbf{\nabla\cdot E}} = {q\over{\epsilon_{0}}}\,\delta({\mathbf{r}} )$.

In complete analogy, there is another solution to the existence of a magnetic monopole than inventing a Dirac string.
It suffices to accept that ${\mathbf{\nabla\cdot B}} = 0$ is no longer generally valid. It is  ${\mathbf{\nabla\cdot B}} \neq 0 $ when there is a monopole,
and  ${\mathbf{\nabla\cdot B}} = 0 $ when there is no monopole. This tallies then exactly with the meaning of the equation ${\mathbf{\nabla\cdot B}} = 0 $,
which expresses that there are no magnetic monopoles. 

The introduction of the Dirac string is illogical and runs contrary to the principle of Occam's razor. 
Based on symmetry reasons one wonders if magnetic monopoles could exist.
As we have shown in Section \ref{monopole-fuss}, the equations contain already for each term its symmetric counterpart. 
The fields and potentials of the electric and the magnetic monopoles
have exactly the same mathematical structure. In the further development one stumbles then on a result ${\mathbf{\nabla\cdot B}} =
 q_{m}\, \mu_{0} \,\delta({\mathbf{r}})  \neq 0$, where it had been
${\mathbf{\nabla\cdot B}}  = 0$ without the monopoles. This result presents a golden opportunity to enhance the symmetry even further by rendering this completely analogous to
${\mathbf{\nabla\cdot E}} = {q\over{\epsilon_{0}}}\,\delta({\mathbf{r}} ) \neq 0$. But instead of accepting this tried and proved solution one rejects it and
 postulates that ${\mathbf{\nabla\cdot B}} = 0 $ should  remain universally valid. This choice is mathematically wrong, it  breaks  the symmetry, and to  cover up for the error, one is then forced to introduce the stunning concept of a  Dirac string. It is all together an egregious procedure 
and boils down to inventing ``new physics'', just to explain away a trivial mathematical paradox inherent to the use of singular distributions.

\subsubsection{Spin-orbit coupling without Thomas precession} \label{steps}

There are several steps in the following demonstration that the merry-go-round effect is equivalent to the part of the spin-orbit coupling that occurs in Eq. \ref{forces2}.
We will neglect here often the relativistic $\gamma$-factors, relativistic effects on the mass and consider only uniform circular motion.
From the treatment it will transpire that for non-circular motion the calculations could be off the mark.

(1) According to Subsubsection \ref{A-der}, ${\mathbf{\nabla}} \wedge {{\mathbf{p_{0}}}\over{2}} = {\mathbf{\nabla}} \wedge q{\mathbf{A}} = q{\mathbf{B}}$.
Furthermore Eq. \ref{rotation-general} shows that in continuing to neglect relativistic modifications in the mass, 
we have ${\mathbf{\nabla}} \wedge {{\mathbf{p_{0}}}\over{2}} = m_{0}{\boldsymbol{\omega}}$.
Combining the two we obtain ${\boldsymbol{\omega}} = {q{\mathbf{B}}\over{m_{0}}}$.

(2) For  circular motion in cylinder coordinates we have ${\mathbf{\nabla}} \wedge {{\mathbf{p_{0}}}\over{2}} = {1\over{2r}} {dL\over{dr}}{\mathbf{e}}_{z}$.

(3) For uniform circular motion we can put ${d\phi\over{dt}} = \omega$. We have then $L = m_{0}\omega r^{2}$.
From this it follows that ${1\over{2r}} {dL\over{dr}} = m_{0} \omega = {L\over{r^{2}}}$.

(4) The calculation of the merry-go-round effect yields ${\boldsymbol{\omega}} = {1\over{r^{2}}} {\mathbf{r}} \wedge {\mathbf{v}}$, such that
$m_{0} {\boldsymbol{\omega}} = {1\over{r^{2}}} {\mathbf{r}} \wedge {\mathbf{p}}_{0}$.

(5) From (1)-(4) it follows that the merry-go-round effect leads to $m_{0} {\boldsymbol{\omega}} = q{\mathbf{B}} = {\mathbf{\nabla}} \wedge q{\mathbf{A}} $.
In deriving the Pauli equation we divide this by $2m_{0}$, such that the frequency that intervenes in the energy calculations
is the Larmor frequency ${q{\mathbf{B}}\over{2m_{0}}}$.

(6) When there is no magnetic field in the laboratory frame, the electric field will yield a magnetic field in the co-moving frame
which  is given by ${\mathbf{B}}' = \gamma {1\over{c}} {\mathbf{v}} \wedge {\mathbf{E}}$. We neglect here the factor $\gamma$.
The calculation of the merry-go-round effect in the co-moving frame for an electron that has a relative velocity $d{\mathbf{w}}$ with respect to it,
does not depend on  $d{\mathbf{w}}$ and is
just given by   $m_{0} {\boldsymbol{\omega}} = q{\mathbf{B}}' =  {q\over{c}} {\mathbf{v}} \wedge {\mathbf{E}}$, as derived above. 
By taking the limit $d{\mathbf{w}} \rightarrow 0$, we obtain then the merry-go-round effect in the moving frame. The corresponding
Larmor frequency is ${q\over{2m_{0}c}} {\mathbf{v}} \wedge {\mathbf{E}}$. Transforming this frequency to the lab frame involves a factor $\gamma$
which we neglect again. This way we have shown that the merry-go-round effect corresponds to the part of the spin-orbit coupling that appears
in Eq. \ref{forces2}. Of course this calculation neglects the radial contribution of $\gamma {\mathbf{E}}$ in the co-moving frame, but as the electron is at rest
in the co-moving frame, this electric field will not lead to precession. This shows that the reason we transform to the co-moving frame
is that we want to eliminate the effects due to the radial part of ${\mathbf{E}}$ from the calculation. The radial part still plays a r\^ole in the Thomas precession.

\subsubsection{Thomas precession as the Berry phase on the velocity space hyperboloid} \label{Thomas-Berry}

We will consider motion in the $Oxy$ plane. We can drop then  the $z$-coordinate from the velocity four-vector
$(\gamma,$ $\gamma v_{x}/c,$ $ \gamma v_{y}/c,$ $ \gamma v_{z}/c)$. We have then $\gamma^{2} - (\gamma v_{x}/c)^{2} -  (\gamma v_{y}/c)^{2} =1$.
This the equation of a hyperboloid in ${\mathbb{R}}^{3}$.
In formal analogy with what has been done for the calculation of the Berry phase on a sphere, we will introduce here hyperbolic coordinates $(R,u,\phi)$
for points $(X,Y,Z)$ in ${\mathbb{R}}^{3}$. The transformation between  the Cartesian coordinates $(X,Y,Z)$ and their corresponding hyperbolic coordinates
is given by:
$Z =  R \cosh u$, $X = R \sinh u \cos\phi$,  $Y = R \sinh u \sin\phi$. This can be considered as an abstract change of coordinates, whereby the real meaning
of $(X,Y,Z)$ is irrelevant.
This change of coordinates  is not a 1-1-mapping between ${\mathbb{R}}^{3}$ and ${\mathbb{R}}^{3}$
as for spherical coordinates, as it implies that $Z^{2} - X^{2} -Y^{2} = R^{2} > 0$. The coordinate transformation is thus only defined for points inside the cone
$Z^{2} - X^{2} -Y^{2} =  0$.
For the special choice
$(X_{0},Y_{0},Z_{0})= (\gamma v_{x}/c, \gamma v_{y}/c, \gamma)$, we have then:
$\gamma = \cosh u$, $\gamma v_{x}/c = \sinh u \cos\phi$, $\gamma v_{y}/c = \sinh u \sin\phi$ such that the velocity space is just the surface $Z^{2} - X^{2} -Y^{2} =  1$ 
(corresponding to $R=1$)
in the vector space ${\mathbb{R}}^{3}$ with coordinates $(X,Y,Z)$. 
 We can define ${\mathbf{e}}_{R}$, ${\mathbf{e}}_{u}$
and ${\mathbf{e}}_{\phi}$ as usual. Note that ${\mathbf{e}}_{R}$ is here no longer orthogonal to  ${\mathbf{e}}_{u}$
and ${\mathbf{e}}_{\phi}$ in the Euclidean sense as the metric of the velocity space is not Euclidean. The basis vectors are rather mutually  orthogonal with respect
to the metric $Z^{2} - X^{2} -Y^{2}$. Calculation of the Jacobian matrix shows that the volume element is given by $R^{2} \sinh u\, du\,d\phi\, dR $.
 The oriented surface element on the revolution hyperboloid $R=1$  is given by: 
 $d{\mathbf{S}} = \sinh u\, du\,d\phi  \,{\mathbf{e}}_{z}$.\footnote{The presence of ${\mathbf{e}}_{z}$ may look here surprising, but we are dealing with
 a problem in four-dimensional space with Minkowski metric, where the coordinates of a point are $(\gamma, \gamma v_{x}/c,  \gamma v_{y}/c,  \gamma v_{z}/c)$.
 We have dropped the coordinate $ \gamma v_{z}/c = 0$ in our approach in order to be able to represent the hyperboloid with a three-dimensional model.
 But as the motion takes place in the $Oxy$-plane, the Thomas precession must be around the $z$-axis, which we dropped out of the description. 
 The vector ${\mathbf{e}}_{z}$
 can no longer be written  as ${\mathbf{e}}_{u} \wedge {\mathbf{e}}_{\phi}$ because we are dealing with  calculations in a four-dimensional
 space. The vector ${\mathbf{e}}_{R}$ is also orthogonal to ${\mathbf{e}}_{u}$ and ${\mathbf{e}}_{\phi}$ in terms of the Minkowski metric but it is the unit vector of the tetrad that is analogous to ${\mathbf{e}}_{t}$ in the space-time tetrad. In our three-dimensional subspace $v_{z} = 0$, we have
 ${\mathbf{e}}_{R} = {\mathbf{e}}_{\phi} \barwedge {\mathbf{e}}_{u}  $,
 where $\barwedge$ is the hyperbolic wedge product.\label{4D-remark}}
The surface surrounded by the closed loop defined by $\gamma = \cosh u_{0}$ is thus $2\pi\,(\cosh u_{0}-1)$.
From $v_{y}/v_{x} = \tan \phi$, one can calculate $(1+\tan^{2}\phi)\,d\phi = (v_{x} dv_{y} - v_{y}d_{x})/v_{x}^{2}$, which leads to
 $d\phi = {\gamma^{2}\over{c^{2} (\gamma^{2} -1)}} {\mathbf{v}}\wedge {\mathbf{a}}$.
 As the Thomas precession along the closed loop $u=u_{0}$ is given by $(\cosh u_{0} -1) \,2\pi$, along a part $d\phi$
of this  closed loop  it is therefore given by $d\vartheta_{T} = (\cosh u_{0} -1) \,d\phi = {\gamma^{2}\over{c^{2} (\gamma +1)}} {\mathbf{v}}\wedge {\mathbf{a}}$.
This is completely analogous to what we did for the small circle on a sphere.
The coordinate lines $\phi = \phi_{0}$ correspond to accelerations ${\mathbf{a}} \parallel {\mathbf{v}}$. The contributions $du$ do therefore not
contribute to the Thomas precession. The integration of  the Thomas precession over a closed loop of any shape will give the total Berry phase according to the
Gauss-Bonnet theorem. No taking  into account the contribution ${\mathbf{a}} \parallel {\mathbf{v}}$ is here just analogous to the way we define 
an integral as the limit of a procedure whereby we use the Simpson rule with ever smaller meshes.\footnote{When this work was finished, A. Gontijo Campos sent
me a copy of a paper by N. Alamo and C. Criado, {\em American Mathematical Monthly},
116 (2009), pp. 439-446, which takes essentially the same route to calculate the Thomas precession.\label{Criado}}

\subsubsection{A critical remark} \label{critical}

After all this a critical question remains. In the fully relativistic solution of the hydrogen problem, the spin-orbit coupling terms do not show up in the calculations.
Where are they? And why do we get  the correct solutions from the minimal substitution if it is incomplete?
The answer is  that
by solving the problem in cylindrical or spherical coordinates, without introducing covariant derivatives\footnote{
The covariant derivative for a tensor is different from the covariant derivative for a vector. For a scalar we do not need to consider covariant derivatives.
For each type of symmetry, their will thus be a different type of covariant derivative. We can thus expect that the covariant derivative for a spinor
will be different from that of a vector. The coordinate transformations we normally use are for vectors, which are rank two in the spinors. Hence, asking for
the covariant derivative of a spinor under a usual coordinate transformation, is like asking for the covariant derivative for vectors when we have defined
a transformation for a rank-two tensor. That is a difficult question that may even be ill-conceived and not always have an answer. 
E.g. not every linear transformation in ${\mathbb{R}}^{5}$ will be compatible with a transformation of the tensor of the five spherical harmonics of degree two,
because a number of symmetry requirements have to be satisfied. Spherical harmonics of degree two are obtained by taking
the tensor product $(x,y,z) \otimes (x,y,z)$ of the spherical harmonic $(x,y,z)$ of degree 1 with itself and taking into account the
symmetry constraint. Presumably the most eloquent illustration of such constraints occurs in elasticity theory, where transformation matrices need to
satisfy a number of symmetry requirements.
The correct approach
would thus be to define first a coordinate transformation for the spinor parameters, and to work further  from these. For SU(2) we could e.g. define a rotation by
the spherical coordinates of its axis $(\theta,\phi)$ and the rotation angle $\varphi$. Incorporating also changes  in $\varphi$ within the transformation corresponds
to a gauge transformation.  The whole problem
of  geometrical phases and precession is just due to the fact that the coordinate transformations we use for vectors are not compatible with
a coordinate transformation for the spinors, since in a coordinate transformation for a spinor the phase will be unambiguously defined on the whole group. We are thus taking a shortcut to a lot of algebra and conceptual problems  in  neglecting covariant derivatives.\label{happy-covariant}}, 
we actually introduce the spin-orbit effect (without the correction for Thomas precession) 
for circular orbits.
 
 As a matter of fact, the transition between Cartesian coordinates and cylindrical coordinates is given by the entirely
 geometric transformation:
 
 \begin{equation}\label{geometrical-transformation-c}
{\partial f\over{\partial r}} {\mathbf{e}}_{r}  +
{1\over{r}} {\partial f\over{\partial \phi}} {\mathbf{e}}_{\phi} +
{\partial f\over{\partial z}}  {\mathbf{e}}_{z}  = 
{\partial f\over{\partial x}} {\mathbf{e}}_{x}  +
{\partial f\over{\partial y}} {\mathbf{e}}_{y} +
{\partial f\over{\partial z}}  {\mathbf{e}}_{z},
\end{equation}

\noindent which shows that $\nabla \equiv {\partial \over{\partial r}} {\mathbf{e}}_{r}  +
{1\over{r}} {\partial \over{\partial \phi}} {\mathbf{e}}_{\phi} +
{\partial \over{\partial z}}  {\mathbf{e}}_{z} $. 
But when we perform dynamical calculations  and calculate temporal derivatives, we must  take into account  the fact that 
$({\mathbf{e}}_{r}(t),  {\mathbf{e}}_{\phi}(t),  {\mathbf{e}}_{z}(t))$ are functions of time. In the dynamical calculations we must
correct for this temporal dependence
 by introducing covariant derivatives.
When we use $({\mathbf{e}}_{r}(t),  {\mathbf{e}}_{\phi}(t), {\mathbf{e}}_{z}(t))$ in dynamical calculations of the position ${\mathbf{r}}(t)$ of a particle,
$({\mathbf{e}}_{r}(t),  {\mathbf{e}}_{\phi}(t), {\mathbf{e}}_{z}(t))$
 will be a precessing frame. 
Let us now consider a co-moving copy of this  frame at the position $P({\mathbf{r}}(t))$ of the particle rather than at the origin $O$.
 By making the calculations in cylindrical coordinates without
introducing the covariant derivatives, this co-moving frame will for uniform circular motion carry out
 exactly the merry-go-round motion in position space. We have seen that this corresponds to the part of the spin-orbit
 coupling that occurs in Eq. \ref{forces2}.
This is fortunate, as the Dirac equation based on the minimal substitution does not correct for the merry-go-round effect. 
The Dirac equation Lorentz transforms  $({\partial\over{\partial\tau}},0)$ to $({d\over{dt}},{\mathbf{\nabla}})$ in $({\mathbf{r}},t)$ by using
the local instantaneous boost with parameter ${\mathbf{v}}({\mathbf{r}},t)$ and without considering the Thomas precession part of the Lorentz transformation. 
In these instantaneous boosts, the triads remain just all the time parallel to the $({\mathbf{e}}_{x}, {\mathbf{e}}_{y}, {\mathbf{e}}_{z})$, such that any kind of spin-orbit
effect remains ignored.
By introducing cylindrical coordinates without introducing covariant derivatives,
 we are thus making up for this shortcoming of the Dirac equation in the case of uniform circular motion,
as we are including the merry-go-round effect in the laboratory frame. But as we already pointed out,
 this fails to account for the Thomas precession and it is thus not completely exact, even for uniform circular motion.

We see thus that the solution of the problem of the hydrogen atom is based on a completely different approach than the one that
tries to calculate the various contributions to the spin-orbit coupling explicitly, as attempted in our approach. 
By not introducing the covariant derivatives we take a short cut
to a part of the calculation. We are making an error by using the minimal substitution in the Dirac equation, with the result that we fail to take into account  the spin-orbit
effect (without the correction for the Thomas precession).  But by making a second error when we forget to use covariant derivatives we finally get
a  result that includes all effects, except the Thomas precession. 
We may finally note that aligning the reference frame with ${\mathbf{v}}$ will not be correctly be accounted for by introducing
symmetry-adapted coordinates when the orbit is no longer circular. There will then be a further error in the calculation of the spin-orbit coupling.
The problem of the spin-orbit coupling could be addressed perhaps more conveniently by a completely different approach 
whereby one tries to write ${d\over{d\tau}}$ in the presence of a potential in a moving frame by introducing Christoffel symbols
and following the approach Einstein used to write dynamical equations of motion in general relativity.

\subsection{Can one correct the Dirac equation for the errors? } \label{spin-orbit-corrections}

Of course one could ask how one should modify the Dirac equation to take into account the Thomas precession correctly.
But treating the Thomas correction with exacting rigor is not a simple matter. The mass does no longer change only by translation (i.e. displacement) but also by rotation
(i.e. spin).
 Let us explain this more in detail.
The expression given in Eq. \ref{Wigner} is correct. 
From $q{\mathbf{E}} = {d{\mathbf{p}}\over{dt}} =
m {\mathbf{a}} + {\mathbf{v}} {dm\over{dt}}$, it follows that ${\mathbf{v}} \wedge q{\mathbf{E}} = m {\mathbf{v}} \wedge {\mathbf{a}}$.
However, due to the Thomas precession and the presence of the potential, $m$ is  no longer given by $m = \gamma m_{0}$. Let us assume that $m = \gamma m_{0} + 
 (\gamma qV \pm {\hbar\omega_{T}\over{2}} )/c^{2}$. As $m>0$ does no longer contain the
 sign of $\omega$, the sign $\pm$ must be taken as negative if the Thomas precession takes place in the same sense as the spin,
 and as positive otherwise.
Hence we have:

\begin{equation} \label{complicated-Thomas}
\omega_{T} = {\gamma^{2}\over{\gamma +1 }}\, \cdot\,{{\mathbf{v}} \wedge q{\mathbf{E}}\over{ \gamma m_{0}c^{2} + \gamma qV\pm {\hbar\omega_{T}\over{2}}  }}
\end{equation}

\noindent We would like to solve  this equation for $\omega_{T}$. To do so, we must first eliminate all reference to ${\mathbf{v}}$ and $v$
from the equation. As explained in Footnote 
\ref{why-L},  $v$ cannot be calculated from a conservation law between kinetic and potential energy like in classical
mechanics, because the electron might have emitted radiation. However, quantized emission of radiation can be taken into account in the value of ${\mathbf{L}}$,
if we admit that the emission of radiation conserves total angular momentum.
One can thus use ${\mathbf{v}} \wedge q{\mathbf{E}} = {1\over{m}} {1\over{r}}
{\partial U\over{\partial r}}{\mathbf{L}}$, where $mc^{2} =  \gamma m_{0}c^{2} + \gamma qV\pm {\hbar\omega_{T}\over{2}}$.
When ${\mathbf{r}} \perp {\mathbf{v}}$ we can also use ${\mathbf{L}} = {\mathbf{r}}Ê\wedge m{\mathbf{v}}$ to put $v = Lc^{2}/(r( \gamma m_{0}c^{2} + \gamma qV \pm {\hbar\omega_{T}\over{2}}))$ and solve this as an equation of $v$ in terms of $\omega_{T}$ and $r$.
This expression for $v$ must then consistently be used to replace all occurrences of $v$ in the equation that results from Eq. \ref{complicated-Thomas} after
the replacement ${\mathbf{v}} \wedge q{\mathbf{E}} = {1\over{m}} {1\over{r}}
{\partial U\over{\partial r}}{\mathbf{L}}$. This way we obtain a complicated equation
 for $\omega_{T}$, with $r$ as a parameter. If the degree of this equation is larger than $4$ it will only be possible to solve it by numerical methods.
 
 It is not obvious that it would be a correct procedure to add this term to the Dirac equation, because after squaring it will lead to new terms.
 The other contributions to the spin-orbit coupling only enter the scene after squaring. This raises the question if $m = \gamma m_{0} + 
 (\gamma qV \pm {\hbar\omega_{T}\over{2}} )/c^{2}$ is actually the correct {\em ansatz}. Should we perhaps not just use $E = mc^{2}$, where $E$ is the total energy,
 to calculate the mass? At first sight, it seems that the equation $m = \gamma m_{0} + 
 (\gamma qV \pm {\hbar\omega_{T}\over{2}} )/c^{2}$ seems to account for all corrections on the rest mass in the co-moving frame after a subsequent
 transformation to the laboratory frame, provided we introduce spherical or cylindrical coordinates  in the solution of the Dirac equation obtained by a minimal substitution.
 When we solve the equation rather in Cartesian coordinates, then we must use the generalized substitution in order
 to get the spin-orbit coupling correct.
 
\subsection{Further remarks on the generalized substitution} \label{further-mag}

The purely magnetic part of Eq. \ref{correct-subst} is:
\vspace{-0.1cm}
\begin{equation} \label{correct-magn-subst}
\gamma cq {\mathbf{A}}{\boldsymbol{\cdot\sigma}} 
+  \gamma q\,( {\mathbf{v\cdot A}})\, \bigone 
+ \imath \gamma q ({\mathbf{v}} \wedge {\mathbf{A}}){\boldsymbol{\cdot\sigma}}.
\end{equation}

\noindent When we use Eq. \ref{correct-subst} in its full generality  to make the correct substitution in the Dirac equation, it will 
 lead to a staggering amount of algebra after squaring
the Dirac equation, even when the electromagnetic fields are not varying with time, because
the quantities ${\mathbf{v}}({\mathbf{r}},t)$ and $\gamma({\mathbf{r}},t)$ will in general still depend on space and time.
There will in any case also be a term with ${\mathbf{a}}{\boldsymbol{\cdot\sigma}}$ due to the combination  $[\,{\partial\over{\partial t}}\bigone\,]\,[\,{\mathbf{v}}{\boldsymbol{\cdot\sigma}}\,]$
in the SL(2,${\mathbb{C}}$) matrices.
We have therefore  just discussed the anomalous Zeeman effect and the spin-orbit coupling in the non-relativistic limit. 
Some further aspects of Eq. \ref{correct-magn-subst}
will be discussed in the Appendix. We will give there e.g. an extremely  simple derivation of the term $-{\boldsymbol{\mu\cdot}}{\mathbf{B}}$ 
that occurs in the literature. This derivation is exact, in contrast with  wishy-washy derivations  based on a treatment of a current loop.


\section{Conclusion} \label{conc2}

The traditional interpretation of the anomalous $g$-factor in the Dirac theory
might be physically  attractive as it corresponds to our macroscopic intuition, but it does not agree with the 
true meaning of the algebra and it violates
the Lorentz symmetry.  We have shown this by developing an alternative approach to the physics of the anomalous $g$-factor by just
respecting the correct geometrical interpretation of the algebra, a feat that the traditional
 approach is not able to accomplish as illustrated by the symmetry violation  mentioned.
We have proposed an interpretation that respects the Lorentz symmetry.
We have thereby stuck to the working philosophy that the algebra should remain strictly the same such that only
 the geometrical interpretation of the algebra can be changed (in such a way that it agrees with the {\em given} geometrical
 meaning of the algebra) and agreement with experiment is automatically preserved.

 However, this working philosophy fails when we try to apply the same methods 
to the algebra
 of the spin-orbit coupling. In searching an explanation for this failure we discover that it is not our approach but
 the traditional approach that contains a number of flaws. A similar analysis of Dirac's theory of the magnetic monopole raises also some 
 troubling issues. 
 The origin of these problems can be traced back to a craze for abstraction whereby  any feeling for the original geometrical 
 meaning of the algebra used in the group representation theory
is lost.

 The whole study presented in this paper just ensues from the natural wish to make sense 
 of Eq. \ref{forces2-split1}/Eq. \ref{forces2} which was obtained from a few lines of algebra based on group
 representation  theory. This group  representation theory provides also all the necessary tools to solve  the  many problems encountered along 
 this search for better insight.
In conclusion we think that this work shows what a powerful tool group theory can be in the quest of trying to make sense of quantum mechanics.

{\em Acknowledgements.}   I wish to thank Prof. Dr. J.-E. Wegrowe for fruitful discussions. 

\newpage

\section{Appendix. Additional Calculations} \label{App}

\subsection{Criticism of the calculation of the energy  $- {\boldsymbol{\mu\cdot}}{\mathbf{B}}$ of a magnetic moment $ {\boldsymbol{\mu}}$  based on a current loop} \label{loop-sect}

Many text books propose that the potential energy of a current loop within a magnetic field
would be given by $U_{dipole} = - {\boldsymbol{\mu\cdot}}{\mathbf{B}}$. This is really problematic for several reasons:

(1) The force of a magnetic field ${\mathbf{B}}$ on a moving charge $q$ with velocity ${\mathbf{v}}$ is
${\mathbf{F}} = q({\mathbf{v}} \wedge {\mathbf{B}})$. As ${\mathbf{F\cdot }}d{\boldsymbol{\ell}} = {\mathbf{F\cdot }}{\mathbf{v}} dt =0$,
a magnetic field cannot do work on a moving charge. It is therefore puzzling how we could define a potential energy $U_{dipole}$ with respect to a magnetic field.

(2)  Some textbooks  argue that the magnetic field exerts a torque on the loop and on the dipole moment ${\boldsymbol{\mu}}$ 
of the loop.  But the torque is calculated in a  special pair of points of the loop 
without discussing what happens in all the other pairs of points.
The special points correspond to the maximal value of the torque. 
By analyzing what happens in other pairs of points, one can find a pair of points where the torque is zero.
This is the minimal possible value 
for the torque. For all other intermediate values of the torque, one can find a corresponding pair of points. 
That one has to select 
the pair of points with the maximal torque to obtain 
the correct value for the ``potential energy''  
of the current loop with dipole moment ${\boldsymbol{\mu}}$ makes the calculation look  hand-waving.
Moreover, a torque is not a force such that it is not clear what its relation with  a potential energy ought to be.

(3) In both cases where one applies these arguments to 
 a single electron in circular orbit around the nucleus of an atom, one has  to assume that the charge $q$ of the electron 
is smeared out over the whole loop. The very definition of ${\boldsymbol{\mu}}$ used in all the atomic calculations  is built on this idea,
which is certainly not correct. Such a single electron is not a true current loop and not a true dipole. Moreover, only one force is acting on it, not a torque. 
One needs thus at least two electrons to define a dipole moment and
 a torque as for a macroscopic current loop. The correct definitions must thus be based on the analysis of a single electron.
 After that, one can put several electrons on the same orbit as one can define for a single electron to obtain the macroscopic quantities
 in terms of dipole moments and torques.
 
\subsection{Magnetic moment of a  current carried by a single electron} \label{current-loop-sect}

We will try to remove here these weird conceptual problems by a better description of the problem.
First we address the recurring  issue that the charge of the electron is not smeared out over a loop.
In an atom, the charge density will not be uniform but singular $\rho(\ell) = q \delta(\ell-
 \ell_{0})$, where $\ell_{0}$ is the position of the charge, and $\ell$ denoted the curvilinear length along the orbit.
This leads to a singular current loop, which we can use to describe
 the real situation of the moving charge. The singular loop will have a singular magnetic moment.
 Relativistically, we will have $I = \gamma I_{0}$.

\begin{equation} \label{singular-loop}
{\boldsymbol{\mu}} 
= \oint \gamma I_{0} {1\over{2}}
\,{\mathbf{r}} \wedge d{\mathbf{r}} = 
\oint I_{0} {1\over{2}} 
\,{\mathbf{r}} \wedge \gamma {\mathbf{v}}\,dt =
\oint {1\over{m_{0}}}  {1\over{2}}\,{\mathbf{r}} \wedge {\mathbf{p}}\,dq \,=\, {q\over{2m_{0}}} {\mathbf{r}}_{0}\wedge {\mathbf{p}}_{0}=  {q\over{2m_{0}}} {\mathbf{L}}_{0}.
\end{equation}

\noindent We have used here $I_{0}\,dt =dq$. We can consider ${\mathbf{r}}_{0}$ as the centre of the circular orbit that defines the loop that can be associated with it.
In an atom, it would be the position vector of the electron with respect to the nucleus.
While it is obvious that $I$ is singular, and that also the charge distribution $dq= \rho(\ell) d\ell =
q \delta(\ell- \ell_{0}) d\ell$ is singular, it is obvious that $\oint dq = \oint q \delta(\ell- \ell_{0}) d\ell =  q$. We obtain then $ {\boldsymbol{\mu}} = - {q\over{2m_{0}}} {\mathbf{L}}_{0}$. We can replace ${\mathbf{L}}_{0}$ by
${\mathbf{L}}$ as ${\mathbf{L}}$ is a constant of motion. 
We may note that it is also possible to define  ${\boldsymbol{\mu}}$ classically such that it does not account for $\gamma$ in ${\mathbf{L}}$. 
One must then write  $\gamma {\boldsymbol{\mu}}$ instead of  ${\boldsymbol{\mu}}$.
The quantity ${\boldsymbol{\mu}}$ is not a dipole moment, because there is only one moving charge. It is a monopole moment.

This calculation of ${\boldsymbol{\mu}}$ is exact, while from most presentations one gets the impression that
it would a back-of-the-envelope calculation that is only a rough estimation. It may be noted that we do not need the loop.
All we have to do is to integrate over a small segment $d{\mathbf{r}}$. 
In fact, in the reasoning followed above, the singular magnetic moment was only defined for a line segment $d{\mathbf{r}}$ and we could choose the loop at will.
The rest of the loop that one may add can be chosen arbitrarily as it will
give a zero contribution to the contour integral. And this is true at any moment. But taking the orbit for the contour
integral has the advantage that we never have to change the arbitrary choice and that the definition will be valid over the whole orbit.
It will then be a definition that suits the description of a stationary situation.
The further manipulations introduce ${\mathbf{L}}$ which is a constant.
And in the end we integrate $dq$.
In conclusion, with the necessary provisos,  the current density of a moving single charged particle is singular and can always be considered as  giving rise to a magnetic charge ${\boldsymbol{\mu}}$ which is a magnetic monopole moment (without hyphen).

 As explained by Griffiths and Hnizdo \cite{Griffiths},
in Gilbert's description  a magnetic dipole moment consists just of two monopole moments. When we consider the current loop from the viewpoint
of a moving Lorentz frame, the symmetry between
the two monopole moments  becomes broken, which leads to a ``hidden momentum''. The conclusions reached in our approach are completely in line
with these ideas and show that a single electron in circular motion corresponds to a magnetic monopole moment.

We still have to address the issue (1). The solution of this conundrum is that $-{\boldsymbol{\mu\cdot}}{\mathbf{ B}}$ does not represent
the potential energy, but the kinetic energy as already pointed out in the main text. The following subsection will show how 
we get it into the formalism by applying the  correct substitution for the coupling of the charge to the free-space Dirac equation.

\subsection{Further discussion of Eq. \ref{correct-magn-subst} and the origin of the term $- {\boldsymbol{\mu\cdot}}{\mathbf{B}}$} \label{loop-sect2}

We consider the terms in Eq. \ref{correct-magn-subst} with the opposite sign, such that the signs will be those that will appear in the
equation after making the substitution $E \rightarrow E -qV, {\mathbf{p}} \rightarrow {\mathbf{p}} - q{\mathbf{A}}$.
 The  term $-\gamma q {\mathbf{A}}{\boldsymbol{\cdot\sigma}} $
has already been discussed in terms of a velocity field with a vorticity. As it is a vector, it can not contribute to the potential energy.
However the vorticity of this term can affect the energy.
The  term $- \gamma q\,( {\mathbf{v\cdot A}})\, \bigone$ can be rewritten as $-{\mathbf{j\cdot A}}$,
such that it can be understood as  the ``potential energy'' of a current, while in reality  it describes its kinetic energy, as verified on a simple example in the main text. 
By introducing  ${\mathbf{A}} = - {1\over{2}} {\mathbf{r}} \wedge {\mathbf{B}}$, the term $- \gamma q\,( {\mathbf{v\cdot A}})\, \bigone$ can also be written as $- {\mathbf{L\cdot}} {q{\mathbf{B}}\over{2m_{0}}} \bigone$, where ${\mathbf{L}} = {\mathbf{r}} \wedge \gamma m_{0}{\mathbf{v}}$. This angular momentum ${\mathbf{L}}$ is defined
with respect to an arbitrary origin, as ${\mathbf{r}}$ is defined with respect to an arbitrary origin. The term 
$ -{\mathbf{L\cdot}} {q{\mathbf{B}}\over{2m_{0}}}$
has the dimension of an energy.
We can write it in terms of the ``potential energy'' term considered above:

\begin{equation} \label{correct-subst2}
- U_{dipole} = - {q\over{2m_{0}}} {\mathbf{L}}{\mathbf{\cdot B}} =  \gamma {\boldsymbol{\mu}}{\mathbf{\cdot B}} = - {\mathbf{L}} {\boldsymbol{\cdot\omega}}_{L}.
\end{equation} 

\noindent The term $\gamma$ associated with ${\boldsymbol{\mu}}{\mathbf{\cdot B}}$ is motivated by the fact
that ${\boldsymbol{\mu}}$ has been defined with ${\mathbf{L}} = m_{0} {\mathbf{r}} \wedge {\mathbf{v}}$.
The minimal substitution will lead to $E \rightarrow E + U_{dipole} = E -  \gamma {\boldsymbol{\mu}}{\mathbf{\cdot B}}$.
 This quantity $U_{dipole}$ is determined up to an arbitrary constant just like a Coulomb potential is determined up to an arbitrary constant.
The arbitrary constant is related to the fact that we can define ${\mathbf{r}}$ in ${\mathbf{A}} =- {1\over{2}} {\mathbf{r}} \wedge {\mathbf{B}}$
with respect to an arbitrarily chosen origin.
The constant can  be fixed by choosing an origin.
When we choose an origin to calculate the Coulomb potential, this will simultaneously determine an origin for the potential $- {\boldsymbol{\mu}}{\mathbf{\cdot B}}$.
If there is no Coulomb potential, then we must find a different criterion to define the origin.
The term ${q\over{2m_{0}}} {\mathbf{L}}$ can be rewritten as ${{\mathbf{L}}\over{2m_{0}c}} q_{m}$. Here  ${{\mathbf{L}}\over{2m_{0}c}}$ contains ${\mathbf{r}}$
as a position vector with respect to the arbitrary origin, such that ${\boldsymbol{\mu}}$ is the monopole moment of the magnetic charge
with respect to this origin.
Magnetic moments can be defined with respect to arbitrary
points, just like angular momentum is in principle defined with respect to an arbitrary origin. This magnetic moment is just the product of the magnetic charge $q{\mathbf{v}}$ with a position vector ${\mathbf{r}}$.
Just as only differences of potential energy make physical sense, only differences of magnetic moments make physical sense.
These differences will occur from the moment on we have two magnetic charges.
It is thus when we have two moving charges in points ${\mathbf{r}}_{1}$ and  ${\mathbf{r}}_{2}$ that 
the arbitrariness disappears because ${\mathbf{r}}_{2}- {\mathbf{r}}_{1}$  does then no longer depend on the choice of the origin, such
that we then can really define a magnetic dipole moment. It is because it consists of many magnetic moments that a macroscopic current loop
can be identified with a magnetic dipole moment. But the singular magnetic moment of a single moving charge
is not a magnetic dipole moment. One could call it an arbitrary magnetic monopole moment (which is not the same as a magnetic-monopole moment). One can split the magnetic dipole moment
into two such magnetic monopole moments, just as we can split a dipole into two monopoles. 

Finally, $- \imath \gamma q ({\mathbf{v}} \wedge {\mathbf{A}}){\boldsymbol{\cdot\sigma}}$ $= $
 $\imath \gamma {q\over{2}} ( ({\mathbf{r\cdot v}}) {\mathbf{B}} -  ({\mathbf{B\cdot v}}) {\mathbf{r}}){\boldsymbol{\cdot\sigma}}$. 
 This term is also determined up to an arbitrary constant. While it has here the dimension of an energy, it's pseudo-vector character
 precludes using it as a potential energy.
 For a circular motion of the charge due to the Lorentz force exerted by the magnetic field, this term reduces zero if we choose the origin
 at the centre of the circle. In fact,
 ${\mathbf{r}}$, ${\mathbf{v}}$,  ${\mathbf{B}}$  are then all mutually orthogonal. Once again, the case ${\mathbf{v}}\not\parallel {\mathbf{A}}$
 can only have meaning in a context of forced motion.
  In the non-relativistic limit, where we can neglect $\gamma$, the force term
  $[\,{\mathbf{\nabla\cdot}}{\boldsymbol{\sigma}}\,]\,  [\,- \imath \gamma q ({\mathbf{v}} \wedge {\mathbf{A}}){\boldsymbol{\cdot\sigma}}\,]$ 
will lead to four terms: two pseudo-scalar terms and two vector terms. One of the vector terms will be the Lorentz force
$q({\mathbf{v}} \wedge ({\mathbf{\nabla}}\wedge {\mathbf{A}})){\boldsymbol{\cdot\sigma}}$. This 
will lead to the torque on a non-aligned  current loop considered by many authors,
 from the moment on we consider two  or more charges rather than one in the current loop.
The other vector  term will be $q( ({\mathbf{\nabla}}\wedge {\mathbf{v}}) \wedge{\mathbf{A}}){\boldsymbol{\cdot\sigma}}$. 
The two pseudo-scalar terms can be written together as $- \imath q {\mathbf{\nabla\cdot}}({\mathbf{v}}\wedge {\mathbf{A}})\,\bigone$ $=$
 $\imath q\,( {\mathbf{v \cdot B}})\,\bigone$ $ - \imath q {\mathbf{A \cdot}}({\mathbf{\nabla}}\wedge{\mathbf{v}})\,\bigone$.
 
We can analyze this non-aligned situation again in terms of two mutually orthogonal fields ${\mathbf{B}}_{1}$ and ${\mathbf{B}}_{2}$.
The motion of a single charged particle in the field ${\mathbf{B}}$ will then just be a circular motion in 
the field ${\mathbf{B}} = {\mathbf{B}}_{1} + {\mathbf{B}}_{2}$. It will take place
in the plane orthogonal to ${\mathbf{B}}$ with velocity ${\mathbf{v}}$. 
We can decompose  ${\mathbf{A}} = {\mathbf{A}}_{1} + {\mathbf{A}}_{2}$.
The global term  $- \imath \gamma q ({\mathbf{v}} \wedge {\mathbf{A}}){\boldsymbol{\cdot\sigma}}$
will be zero, as ${\mathbf{v}} \parallel  {\mathbf{A}}$. However, the two terms $- \imath \gamma q ({\mathbf{v}}_{j} \wedge {\mathbf{A}}){\boldsymbol{\cdot\sigma}}$ will both be different
from zero, with their sum adding up to zero. The  terms $\gamma q ({\mathbf{v}} \wedge {\mathbf{A}}_{1})$ and $\gamma {\mathbf{v\cdot A}}_{1}$ 
have norms  $\gamma vA_{1} \sin\chi$ and $\gamma vA_{1} \cos\chi$, where $\chi$ is the angle between ${\mathbf{v}}$ and ${\mathbf{A}}_{1}$.
The term $\gamma q ({\mathbf{v}} \wedge {\mathbf{A}}_{1})$ accounts thus for that part of the energy that can be associated with ${\mathbf{v}}$, 
other than the kinetic energy  $\gamma {\mathbf{v\cdot A}}_{1}$ which has to be attributed to the motion in the field ${\mathbf{B}}_{1}$.

To conclude, we can state that in the absence of an electric field we obtain in the non-relativistic limit for the substitution required: 
 
\begin{equation} \label{correct-subst3} 
E \rightarrow E  -   {\boldsymbol{\mu}}{\mathbf{\cdot B}}, \quad  c{\mathbf{p}} \rightarrow c{\mathbf{p}}-  cq {\mathbf{A}}. 
\end{equation} 

\noindent This calculation is exact and the picture is clear. We do not understand why the electron spin must be aligned (such
that the energies are quantized and the imaginary terms become zero), because we do not understand the electron spin.
The real vector term $ cq {\mathbf{A}}$ in $ c{\mathbf{p}} \rightarrow c{\mathbf{p}}-  cq {\mathbf{A}}$ corresponds 
in principle to the minimal substitution for a spin-less particle
but its vorticity yields the frequency of a spinning motion that after multiplication by $\hbar/2$ can represent energy pumped into the spin. 
Finally, we can see from all this that in the non-relativistic limit the Zeeman effect is given by: $(L+1+2S) {qB\over{2m_{0}}}$, where $L \ge 0$.

\end{document}